\documentclass[journal]{IEEEtran}
\usepackage{cite}

\newcommand\ident[1]{#1}
\newcommand\rev[1]{#1}

\usepackage{array,multicol,multirow}

\ifCLASSINFOpdf
  \usepackage[pdftex]{graphicx}
\else
  \usepackage[dvips]{graphicx}
  \DeclareGraphicsExtensions{.eps}
\fi
\usepackage{amsmath}
\usepackage{algorithm}
\usepackage[noend]{algorithmic}

\usepackage{array}

\ifCLASSOPTIONcompsoc
  \usepackage[caption=false,font=normalsize,labelfont=sf,textfont=sf]{subfig}
\else
  \usepackage[caption=false,font=footnotesize]{subfig}
\fi
\usepackage{dblfloatfix}

\usepackage{url}
\usepackage[breaklinks]{hyperref}

\usepackage{amssymb,amsmath,amsthm}
\usepackage{xifthen}
\usepackage{graphicx,xcolor}
\graphicspath{{figures/}}

\newcommand{\ing}[1]{\mathsf{#1}}
\newcommand{\Rn}[1]{\ifthenelse{\equal{#1}{}}{\mathbb{R}}{\mathbb{R}^{\ing{#1}}}}
\newcommand{\Cn}[1]{\mathbb{C}^{\ing{#1}}}
\newcommand{\Rset}[2]{{\Rn{\ing{#1} \times \ing{#2}}}}
\newcommand{\Cset}[2]{{\Cn{\ing{#1} \times \ing{#2}}}}
\newcommand{\vect}[1]{\boldsymbol{\mathbf{\MakeLowercase{#1}}}}
\newcommand{\mtrx}[1]{\boldsymbol{\mathbf{\MakeUppercase{#1}}}}
\newcommand{\var}[1]{\mathrm{#1}}

\newcommand{\norm}[2]{\|#1\|_{#2}}

\newcommand{\htransp}[1]{#1^{\mathsf{H}}}

\newcommand{\ind}[2]{\ifthenelse{\equal{#2}{}}{\chi_{#1}}{\chi_{#1}\left( #2 \right)}}

\newcommand{\iter}[2]{#1^{\ing{(#2)}}}

\DeclareMathOperator*{\argmin}{argmin}
\DeclareMathOperator*{\argmax}{argmax}
\DeclareMathOperator*{\minim}{minimize\,}

\DeclareMathOperator*{\sign}{sgn\,}
\newcommand{\abs}[1]{\vert#1\vert}

\DeclareMathOperator*{\subjto}{\,subject\,to\,}

\definecolor{light-gray}{gray}{0.75}

\newtheorem{mydef}{Definition}

\newcommand{\Matlab}{$\text{Matlab}^{\circledR}~$}

\newcommand{\ProcBluesy}{$\text{Intel}^{\circledR}~\text{Xeon}^{\circledR}$}

\newcommand\Id{\mtrx{I}}
\newcommand\lp[1]{{\ell_{#1}}}
\newcommand\lpq[2]{{\ell_{{#1},{#2}}}}
\newtheorem*{remark}{Remark}

\usepackage{tikz,tikz-3dplot,qtree,forest,mathtools,tkz-euclide}
\usetikzlibrary{shapes,external,arrows,calc,intersections,decorations.pathreplacing}
\usetikzlibrary{shapes.geometric,shapes.arrows,decorations.pathmorphing}
\usetikzlibrary{matrix,chains,scopes,positioning,fit}
\usepackage{pgf,pgfplotstable,pgfplots}
\pgfplotsset{compat=newest}
\usepackage{filecontents}
\usepackage[utf8]{inputenc}
\usepackage[croatian, english]{babel}

\begin{document}

%
\title{Sparsity-based audio declipping methods: \ident{selected} overview, new algorithms, and large-scale evaluation}
%
%
%

\author{Clément~Gaultier,
	Sr{\dj}an~Kitić,
	Rémi~Gribonval,~\IEEEmembership{Fellow,~IEEE,}
        and~Nancy~Bertin,~\IEEEmembership{Member,~IEEE}
\thanks{N. Bertin is with Univ Rennes, CNRS, Inria, IRISA. All authors were with Univ Rennes, CNRS, Inria, IRISA when this work started. C. Gaultier and S. Kitić are now with Orange. R. Gribonval is now with Univ Lyon, Inria, CNRS, ENS de Lyon, UCB Lyon 1, LIP UMR 5668, F-69342, Lyon, France.}
}%

\maketitle

\begin{abstract}
Recent advances in audio declipping have substantially improved the state of the art.
Yet, practitioners need guidelines to choose a method, and while existing benchmarks have been instrumental in advancing the field, larger-scale experiments are needed to guide such choices. 
First, we show that the \ident{clipping} 
levels in existing small-scale benchmarks are moderate and call for benchmarks with more perceptually significant \ident{clipping} 
levels. We then propose a general algorithmic framework for declipping that covers existing and new combinations of 
\ident{variants} of state-of-the-art techniques exploiting time-frequency sparsity: synthesis \rev{\emph{vs.}} analysis sparsity, with plain or structured sparsity. Finally, we systematically compare these combinations and \ident{a selection of} state-of-the-art methods. Using a large-scale numerical benchmark and a smaller scale formal listening test, we provide guidelines for various \ident{clipping} 
levels, both for speech and various musical genres. The code is made publicly available for the purpose of reproducible research and benchmarking.
\end{abstract}



%
\IEEEpeerreviewmaketitle

\section{Introduction}

Clipping, also known as saturation, is a common phenomenon that can arise from hardware or software limitations in any audio acquisition pipeline. It results in severely distorted audio recordings. Magnitude saturation can occur at different steps in the acquisition, reproduction or \ident{Analog-to-Digital Conversion (ADC)} process. Restoring a saturated signal is of great interest for many applications in digital communications, image processing or audio. In the latter, while light to moderate clipping causes only some audible clicks and pops, more severe saturation highly affects \ident{the sound quality}, contaminating it with rattle noise \ident{and distortion}. The perceived degradation depends on the \ident{clipping threshold} and the original signal, and it can lead to significant loss in perceived audio quality \cite{tan2003effect}. More recently, studies \cite{tachioka2014speech,harvilla2014least} also showed the negative impact of clipped signals when used in signal processing pipelines for recognition, transcription or classification. 

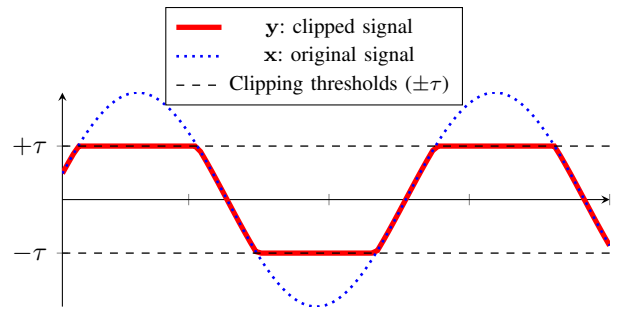
\begin{figure}[htb]
	\centering
	\tikzset{
	point/.style={insert path={node[circle, inner sep=1pt, fill]{}}}
}
\begin{tikzpicture}[
declare function={
	clip(\t,\f)=sin(180*pi*x*\f)*(abs(sin(180*pi*x*\f))<\t) + \t*(sin(180*pi*x*\f)>=\t) -\t*(sin(180*pi*x*\f)<=-\t);
}
]
\begin{axis}[
samples=30,
domain=0.05:2,
y domain=-1:1,
label style={font={\fontsize{8pt}{8pt}\selectfont}},
ytick = {-0.5,0.5},
yticklabels={$-\tau$,$+\tau$},
xticklabels=\empty,
legend style={cells={align=left}, align=left, at={(0.46,1.4)},anchor=north, font={\fontsize{8pt}{8pt}\selectfont}},
axis lines=middle,
width=\columnwidth,
height=0.5\columnwidth
]

\
\addplot[samples=100,line width=2pt,solid,red, name path global=clipped] (x,{clip(0.5,0.5)});
\addplot[samples=100,line width=1pt,blue, dotted,smooth, name path global=sine] (x,{sin(180*pi*0.5*x)});
\addplot[samples=100,line width=0.5pt,dashed,black, name path global=Ptau] (x,0.5);
\addplot[samples=100,smooth,line width=0.5pt,dashed,black, name path global=Ntau] (x,-0.5);
\legend{$\vect{y}$: clipped signal,$\vect{x}$: original signal,Clipping thresholds ($\pm \tau$),}

\end{axis}
\end{tikzpicture}
	\caption{Hard-clipping model \eqref{eq:ClipModel}}
	\label{fig:hardClipModel}
\end{figure}

The idealized hard-clipping model is a simple yet accurate approximation of the magnitude saturation \ident{phenomenon} and allows to easily identify the clipped and reliable samples. Denoting $\vect{x} \in \Rn{L}$ a clean original discrete signal, a \ident{clipped} 
version $\vect{y} \in \Rn{L}$ is modeled \ident{as follows}:
\begin{equation}
\label{eq:ClipModel}
\vect{y}_{\var{i}} = \left\{ 
\begin{array}{l l}
\vect{x}_{\var{i}} & \quad \text{when } \abs{\vect{x}_{\var{i}}} \leq \tau;\\
\sign(\vect{x}_{\var{i}})\tau & \quad \text{otherwise;}\\ \end{array} \right.
\end{equation}
with $\vect{y}_{\var{i}}$ (resp. $\vect{x}_{\var{i}}$) denotes $\var{i}$\textsuperscript{th} sample from $\vect{y}$ (resp. $\vect{x}$) and $\tau$ the hard-clipping threshold. 
This is illustrated \rev{in} Figure~\ref{fig:hardClipModel}.
\ident{In the hard-clipping model, which corresponds for example to digital clipping at the ADC level, it is easy to locate the clipped segments and this is an information used by most \rev{state-of-the-art} declipping methods. In real settings where softer saturation occurs, 
locating the clipped segments is less straightforward, and dedicated methods, such as \cite{laguna2016efficient}, could be used for clipping detection.}
\addtolength\abovecaptionskip{-20pt} 
\addtolength\abovecaptionskip{20pt} 

The terms \textit{declipping} and \textit{desaturation} are equally employed to denote the tentative inversion of this process, aiming at restoring the original signal $\vect{x}$ in its full dynamics and quality, from the sole knowledge of its degraded version $\vect{y}$. This task has gained a lot of attention in the last decade, as effective approaches based on sparse regularization techniques flourished \cite{adler2012audio,harvilla2014least,siedenburg2014audio, ozerov2016multichannel}. After this accumulation of progress, the domain now seems mature enough to address much more severe levels of degradation than those considered in early work. Such an ambition is 
\ident{indeed crucial because} real-world applications require to consider such ranges where saturation is prominently audible, as we will show later. 

In parallel, limited steps have been made towards evaluation on a common data set. Comparisons have been made possible to some extent by the use of sound excerpts from the SMALLbox toolbox \cite{damnjanovic2010smallbox} as experimental material of many works in the field, from the pioneering ones \cite{adler2012audio,siedenburg2014audio} to most recent algorithms \cite{rencker2018consistent,zavivska2019proper}. This freely available data set\footnote{\scriptsize\url{http://small.inria.fr/software-data.html}, ``Audio Inpainting Toolbox''} was itself extracted from the SiSEC evaluation campaign \cite{vincent2011sisec} and is sometimes referred to as ``the SMALL data set''. 
It includes twenty 5-second excerpts (10 speech, 10 music) sampled at 16~kHz (and ten additional 8 kHz speech excerpts unsuited for declipping applications), with a necessarily limited diversity, although effort has been put to cover \ident{various types of audio contents}. On the other hand, some other authors gathered their own data \cite{kitic2013consistent,harvilla2014least,ozerov2016multichannel}. Thus, a large-scale and systematic evaluation of the numerous variants of these algorithms on common data was, to date\footnote{\ident{During the revision process of this paper, the authors became aware of a preprint \cite{Zaviska:2020ua} providing a contribution to the field that is very complementary to the one proposed in this paper. It surveys a wider scope of audio declipping methods, with numerical comparisons on a dataset of 10 single-channel musical excerpts \rev{sampled at 44.1~kHz}.}}, still to be accomplished.

In order to carry out the needed systematic and large-scale benchmarking of \ident{a representative selection of} time-frequency single-channel audio declipping algorithms, this paper introduces a new joint modeling and algorithmic framework, 
mimicking many existing algorithms and encompassing new combinations of the various ideas introduced by them (sparse and cosparse models, plain or structured sparsity). As a prologue, \autoref{sec:quantifysaturation} discusses the objective measures of \ident{clipping} 
levels, which \ident{sheds light on} the incompleteness of previous evaluations, and identifies the perceptually relevant \ident{clipping} 
 levels that real-world applications need to target. After recalling the main ingredients of state-of-the-art algorithms in \autoref{sec:sota}, we present our framework in \autoref{sec:framework} and show how it allows i) to mimic 
existing algorithms, ii) to formulate new combinations and iii) to handily conduct systematic experiments. The versatility of the framework is illustrated in \autoref{sec:experiments} on small-scale and large-scale experiments, evaluated with \ident{signal-based} objective performance measures, \ident{perceptually-motivated objective quality measures}, and a listening test. From these experiments, in the final section, we contribute guidelines for declipping practice and evaluation methodology, together with \ident{ideas on future research}.


\section{How to quantify \ident{clipping}?} 
\label{sec:quantifysaturation}

Quantifying the consequence of \ident{clipping} 
can be itself an interesting problem. Even looking at the degradation from a signal perspective only can lead to various interpretations whether one focuses on the clipping threshold, the distortion or the amount of affected samples. In the following, we compare these usual \ident{clipping level} 
indexes and provide clipping scales where perceptual differences are of interest.

\subsection{\ident{Signal-based} objectives measures of saturation}
Commonly \cite{adler2012audio,harvilla2014least,siedenburg2014audio, ozerov2016multichannel}, 
\ident{clipping} is directly rated from the clipping threshold ($\tau$ on \autoref{fig:hardClipModel}) as it reflects how \ident{strongly} the dynamic range of the signal is affected. The lower $\tau$, the more severe the degradation. Practically, studies usually work with normalized magnitude data for fair comparisons, and the clipping threshold takes values in $[0;1]$. \ident{This value
, which we term \emph{clipping threshold} is also referred to as ``clipping level''.}

\begin{figure*}[htbp]
	\centering
	\subfloat[$\tau$ \rev{\emph{vs.}} SDR\label{fig:ClipvsSDRMusic}]{\includegraphics[trim=1cm 6cm 2cm 7cm,clip,width=0.65\columnwidth]{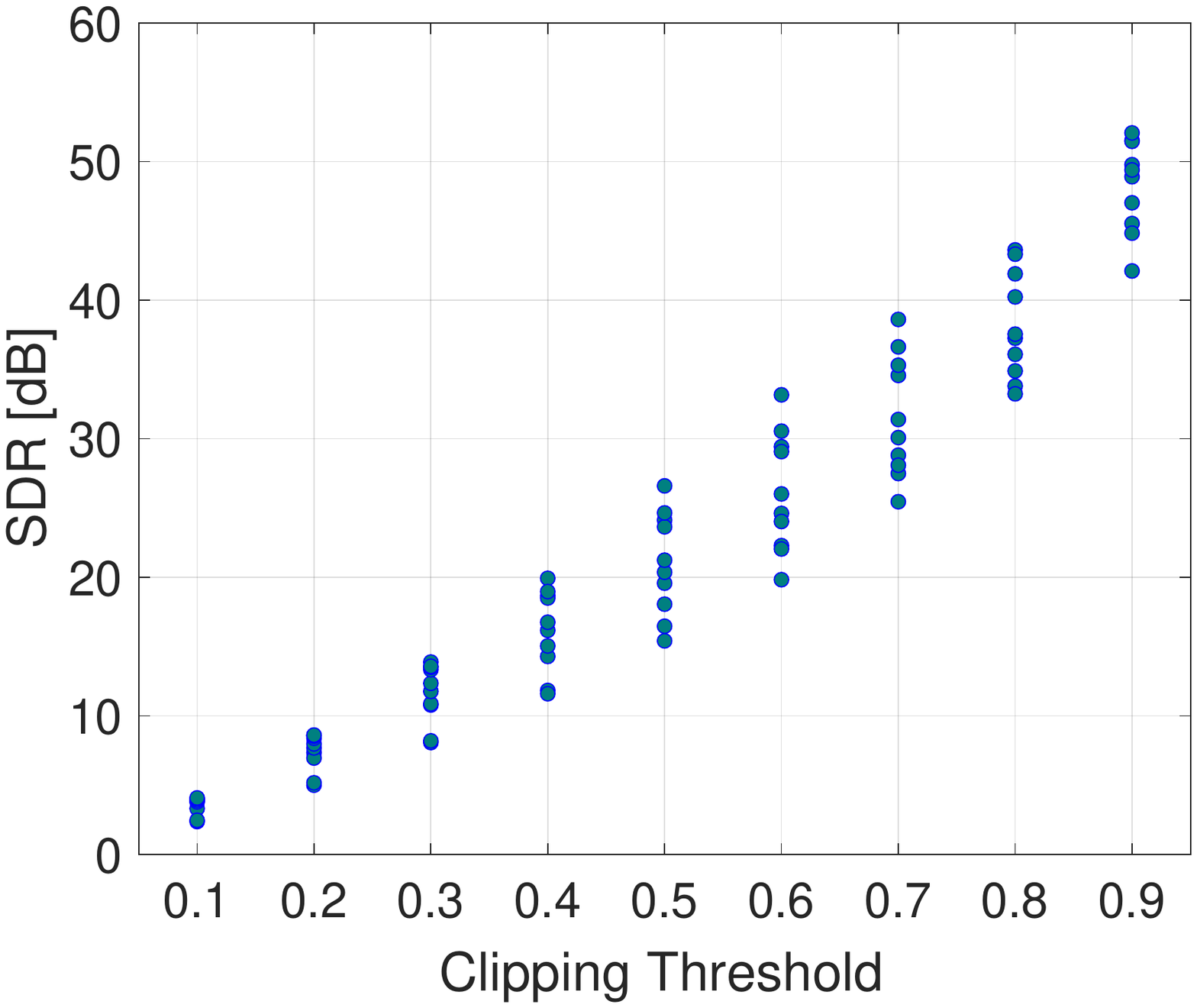}}
	\subfloat[$\tau$ \rev{\emph{vs.}} PEAQ\label{fig:ClipvsPEAQMusic}]{\includegraphics[trim=1cm 6cm 2cm 7cm,clip,width=0.65\columnwidth]{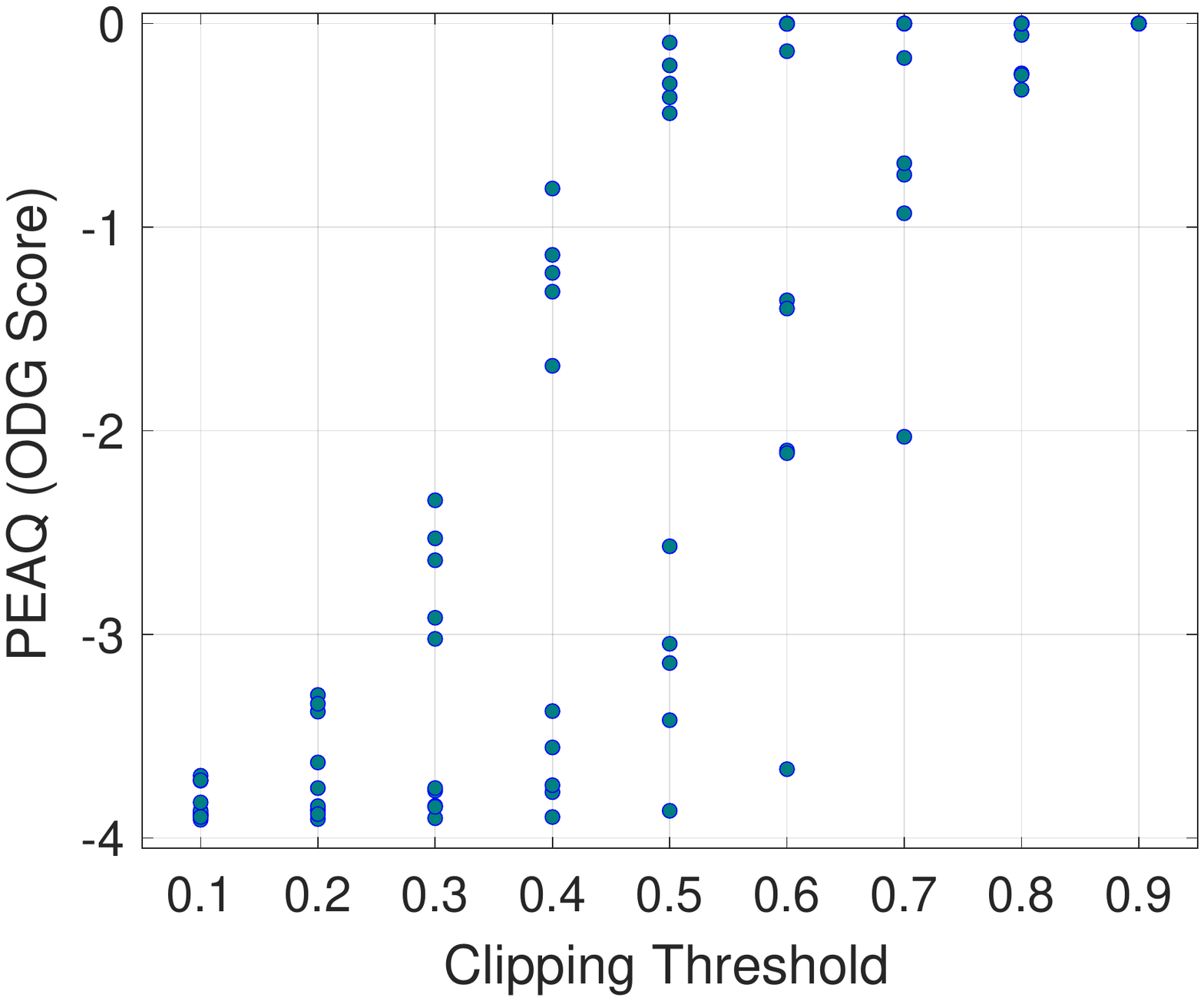}}
	\subfloat[SDR \rev{\emph{vs.}} PEAQ\label{fig:SDRvsPEAQMusic}]{\includegraphics[trim=1cm 6cm 2cm 7cm,clip,width=0.65\columnwidth]{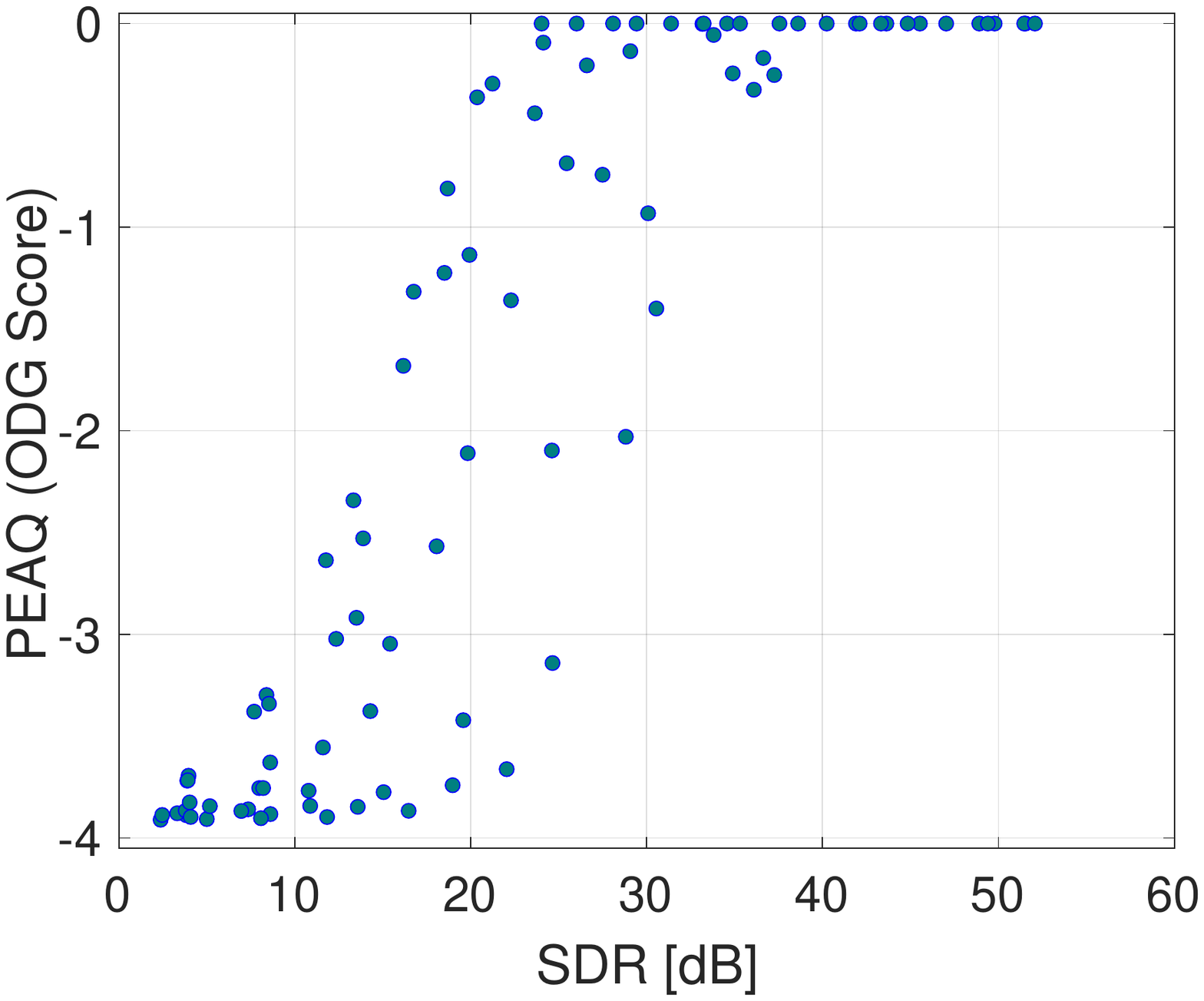}}
	\caption{Clipping measures comparisons (\emph{SMALLbox music examples})\label{fig:MusicQuantifyComp}}
\end{figure*}
Another tool to quantify clipping is the relative level of distortion, measured by the Signal-to-Distortion Ratio in decibels:
\begin{equation}
	\label{eq:SDR}
	\text{SDR} = 10\cdot \log_{10}\left(\frac{\norm{\vect{x}}{2}^2}{\norm{\vect{x}-\vect{y}}{2}^2}\right).
\end{equation}
\ident{Contrary} to the clipping threshold $\tau$, the SDR is highly linked to the content of the initial signal $\vect{x}$ as its computation takes into account the energy of the signal.

In the limit case where the signal is not distorted at all, $\vect{y} = \vect{x}$ and the SDR tends to its maximum value $+\infty$ dB. In the other extreme case, when the \ident{clipping threshold} $\tau$ tends to zero, $\vect{y}$ tends to 
\ident{an all-zero vector} and only retains the sign of the samples of $\vect{x}$. In this limit, the energy of the induced distortion $\norm{\vect{x}-\vect{y}}{2}^{2}$ is as large as the initial signal energy $\norm{\vect{x}}{2}^{2}$ and the SDR tends to its minimum value $0$ dB.

\ident{In order to evaluate the SDR, one needs the original clean signal, while determining the clipping threshold requires knowing the initial dynamic range.} There exists also a mean to estimate the seriousness of the degradation without any reference: by counting the ``proportion'' of a signal that is affected by clipping. In a discrete setting, this boils down to counting the proportion of clipped samples over the complete signal. This ``ratio of clipped samples'' can be denoted 
\begin{equation}
\%_{\text{Clipped}} = \frac{|\{\var{i}\in (1, ... ,\ing{L}_{\textrm{total}})~\mid~\abs{\vect{x}_{\var{i}}}\geq\tau\}|}{\ing{L}_{\textrm{total}}} \cdot 100,
\end{equation}
where here $\ing{L}_{\textrm{total}}$ denotes the total number of samples.
The higher this ratio, the more significant will be the loss as more samples are affected.
Even if this measure can be of great interest to blindly estimate the power of the degradation, it is rarely used in studies, 
as it is not suited for direct comparisons or assessing enhancement.

\subsection{\ident{Perceptually-motivated objective} quality measures}
As for every study involving audio content, numerical indexes used to rate a degradation are relevant if they correlate somehow to 
\ident{perceptual} quality ratings \ident{obtained via listening tests}. Unlike denoising, few studies \ident{have} focused on finding objective numerical descriptors to rate the quality of saturated audio excerpts. Some years ago, \cite{defraene2013declipping} validated the correlation between PEAQ scores~\cite{kabal2002peaq} and clipped audio quality assessments \ident{by using} listening tests. In the following, we will briefly investigate the relationship between clipping thresholds, SDR, PEAQ for music and MOS-PESQ~\cite{rix2001perceptual} for speech to identify a degradation range 
of interest. These preliminary comparisons are held on the SMALL examples. Let us recall that PEAQ scores are adapted to music excerpts and ranges from $-4$ to $0$. The closer the value is \rev{to} $-4$, the more annoying the clipping consequences will be perceived. PESQ scores for speech excerpts \rev{range} from $1$ (bad quality) to $5$ (excellent quality).

\begin{figure*}[htbp]
	\centering
	\subfloat[$\tau$ \rev{\emph{vs.}} SDR\label{fig:ClipvsSDRSpeech}]{\includegraphics[trim=1cm 6cm 2cm 7cm,clip,width=0.65\columnwidth]{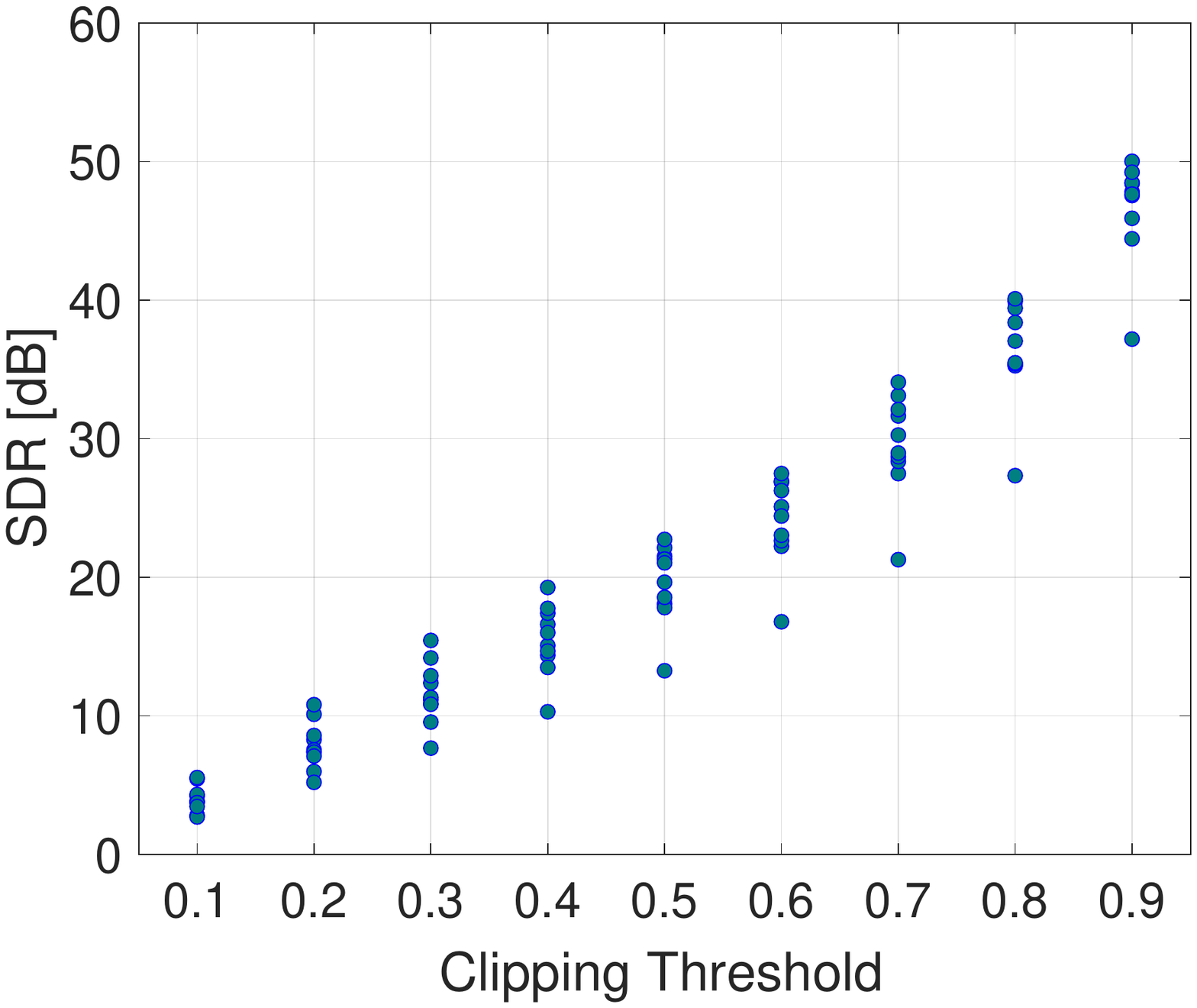}}
	\subfloat[$\tau$ \rev{\emph{vs.}} \ident{PESQ}\label{fig:ClipvsPESQSpeech}]{\includegraphics[trim=1cm 6cm 2cm 7cm,clip,width=0.65\columnwidth]{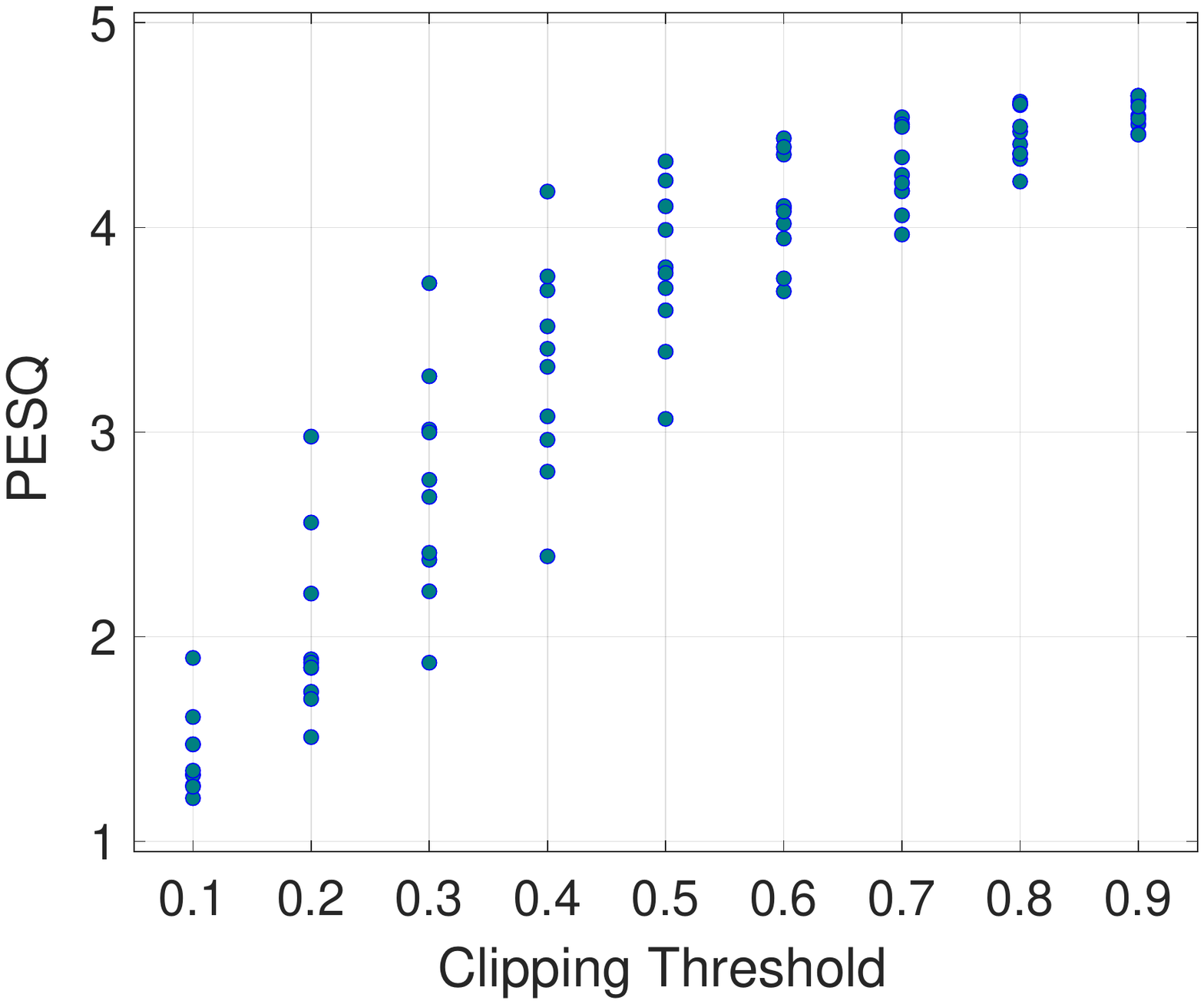}}
	\subfloat[SDR \rev{\emph{vs.}} PESQ\label{fig:SDRvsPESQSpeech}]{\includegraphics[trim=1cm 6cm 2cm 7cm,clip,width=0.65\columnwidth]{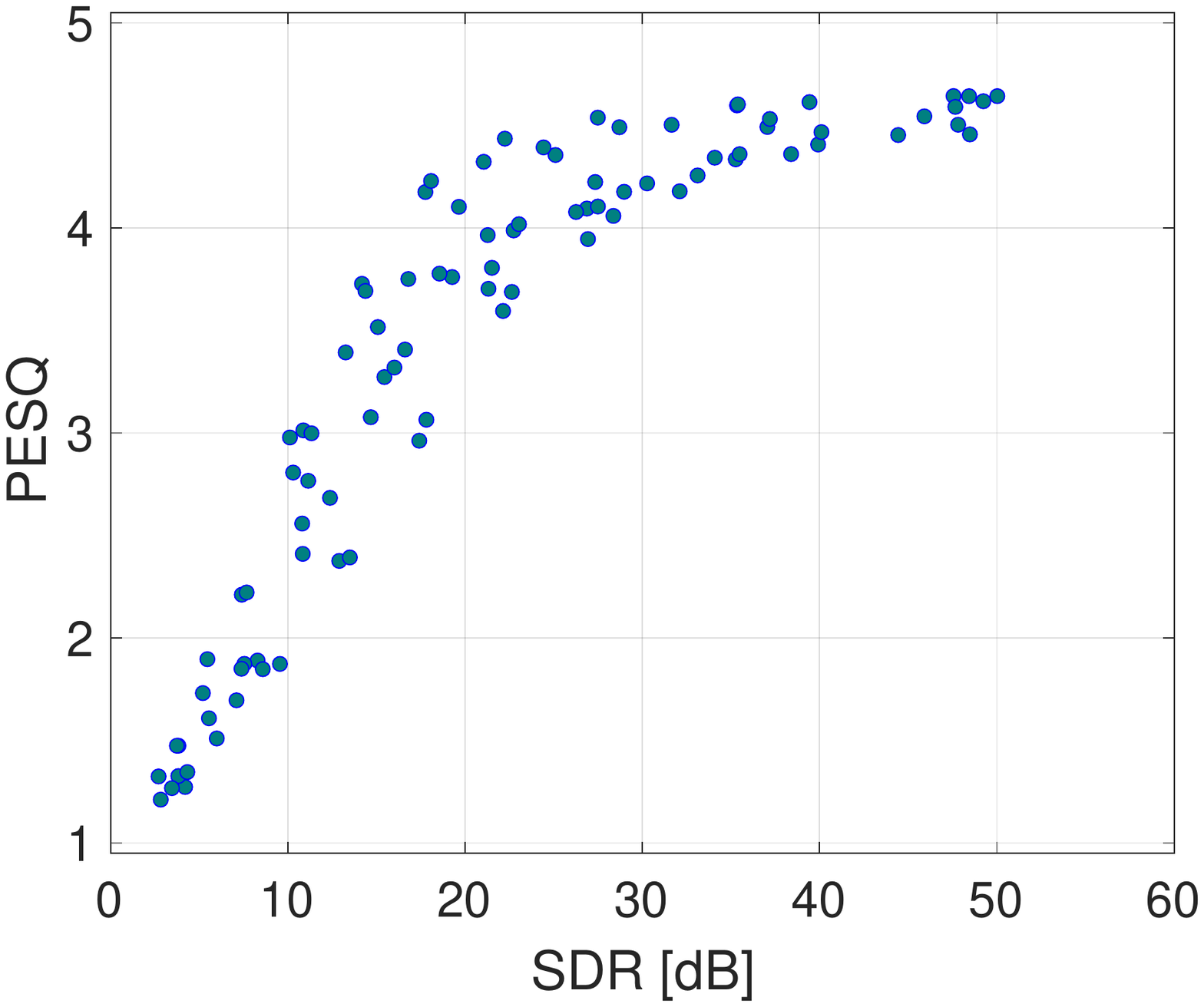}}
	\caption{Clipping measures comparisons (\emph{SMALLbox speech examples})\label{fig:SpeechQuantifyComp}}
\end{figure*}

\subsection{Perceptual relevance:  clipping threshold {\em \rev{vs.}} SDR}
Considering what \ident{signal-based objective} quality measure should be used to describe the strength of the degradation caused by clipping, two main options are available: i) clipping threshold or SDR if a reference signal is available; ii) ratio of clipped samples otherwise. Since we focus in this paper on benchmarking declipping methods in a controlled setting, it is natural to focus on the first option.

As could be expected, and as shown in \autoref{fig:ClipvsSDRMusic} and \autoref{fig:ClipvsSDRSpeech} on the SMALL data set, the SDR is overall an increasing function of the \ident{clipping threshold}. Yet, it remains highly variable as a function of the audio content and its temporal structure. Indeed, for a given \ident{clipping threshold}, a signal with high dynamic will lose a limited number of samples, while more samples will be clipped in a signal whose magnitude values are more uniformly distributed. 
 
To compare the perceptual relevance of the \ident{clipping threshold} and the SDR, \autoref{fig:ClipvsPEAQMusic} and \autoref{fig:SDRvsPEAQMusic} (resp. \autoref{fig:ClipvsPESQSpeech} and \autoref{fig:SDRvsPESQSpeech}) display the PEAQ (resp. PESQ) \ident{perceptually-motivated} objective quality measure as a function of these two criteria, on SMALL music (resp. speech) examples, artificially hard-clipped with \ident{clipping threshold}s between 0.1 and 0.9. As expected, for both speech and music, the estimated rated quality overall increases with the \ident{clipping threshold} and with the SDR. However, as can be observed in \autoref{fig:ClipvsPEAQMusic} and \autoref{fig:ClipvsPESQSpeech}, both PEAQ and PESQ values are highly variable for a given \ident{clipping threshold}, except for the most extreme values. What is even more striking for music is that quality, as measured by PEAQ, stabilizes around its maximum (\emph{imperceptible} degradation) when the SDR roughly exceeds 30 dB. 
To some extent, a similar behaviour is observed for PESQ values on speech. 
\ident{One should be aware that SDR also has weaknesses. Particularly, highly dynamic signals, comprising very short, high-magnitude segments, would score a lower SDR, yet such degradations may not seriously affect 
\ident{the perceived audio quality.}
Nonetheless, in such a case, the clipping threshold measure would perform poorly as well. All this suggests to focus on the SDR rather than the clipping threshold as a primary measure of the \ident{clipping} 
level of the initial signal. }



\subsection{Perceptually relevant degradation ranges}
In light of \autoref{fig:ClipvsSDRMusic}, it appears that most of the SMALL music examples with \ident{clipping threshold} above $0.6$ correspond to an SDR of $30$~dB or more, corresponding to potentially imperceptible degradation. Moreover, as the state of the art in audio declipping shows great potential for the restoration of highly degraded signals, it becomes \ident{desirable} to \ident{focus on} 
test cases involving initial SDRs below $30$~dB. 
\section{Prior work on \ident{sparse} audio declipping}
\label{sec:sota}

\ident{First attempts to address declipping, \textit{e.g.}, with autoregressive models, can be traced back to several decades \cite{janssen1986adaptive}. Significant progress towards efficient \ident{declipping} 
was made in the last ten years by combining ideas from inverse problems and sparse regularization.} \ident{Each of the following subsections described the main ingredients associated to each of these ideas, which encompass the expression of declipping as a linear inverse problem, exploiting consistency \rev{constraints}, and the use of sparse and structured time-frequency models under different 
\ident{variants} (analysis or synthesis). Existing sparse audio declipping methods often combine one or more of these ideas.}

\subsection{Declipping as a linear inverse problem}
The declipping problem can be cast as an \textit{underdetermined, linear inverse problem}, akin to so-called \textit{inpainting} \cite{adler2012audio}. It can be addressed by means of \textit{sparse} regularization in the time-frequency domain using, \textit{e.g.}, a two-stage algorithm based on Orthogonal Matching Pursuit (OMP) \cite{adler2012audio}. In a first stage, the active time-frequency atoms of the sparse representation are greedily identified using OMP on the ``reliable'' samples of the signal (\textit{i.e.}, the samples which have been observed without \ident{clipping}, 
\textit{cf.} Eq.~\eqref{eq:ClipModel}). Then, a declipping constraint is imposed in the transformed domain so the name \emph{Constrained Orthogonal Matching Pursuit}. Besides such greedy algorithms, lines of work involving convex optimization have been investigated, \textit{e.g.}, using $\lp{1}$-norm minimization and a perceptually oriented sparse representation to perform the reconstruction \cite{defraene2013declipping}, or $\lp{2}$-norm regularization for alleviating the effect of clipping in automated speech recognition \cite{harvilla2014least}. More recently, a method based on linearly or quadratically constrained weighted least-squares was introduced \cite{avila2017audio} to tackle a relaxed version of declipping (compression, or soft-clipping, compensation). 

\subsection{\ident{Consistency constraint}}
Besides the exploitation of time-frequency sparsity \ident{(see below)}, a crucial step to improve the efficiency of declipping was the incorporation of the full clipping model \eqref{eq:ClipModel} \emph{within} the iterations of the sparse regularization algorithm \cite{kitic2013consistent}. \rev{The so-called \emph{clipping consistency} constraint aims at doing so, by making sure that the reliable (non-clipped) parts of the observed (clipped) signal exactly match those of the reconstructed signal. At the same time, clipping consistency also ensures that the magnitudes of the reconstructed samples (corresponding to the clipped samples of the original signal) exceed the clipping threshold, with the appropriate sign.} While this idea was already present in \cite{adler2012audio} (but only as a final step), all state-of-the-art declipping methods now apply the consistency constraint at each iteration. This strategy was shown to drastically improve reconstruction performance. Technical details on techniques to enforce such consistency constraints will be given in Section~\ref{sec:framework}.

\subsection{Sparse \& structured time-frequency signal models}
\label{sec:models}

In parallel, a shift from a (now) traditional \textit{sparse synthesis} approach to \textit{sparse analysis} was proposed \cite{kitic2015sparsity}, as well as some model refinements exploiting notions of \textit{structured sparsity}, especially that of \textit{social sparsity} \cite{kowalski2013social} in the time-frequency domain \cite{siedenburg2014audio}. 

\subsubsection{Plain sparsity, analysis \emph{vs.} synthesis}

The \emph{sparse synthesis model} assumes that the signal of interest $\vect{x}$ is built from a linear combination of atoms aggregated in a large dictionary $\mtrx{D}$. We could more precisely write
\begin{equation}
\label{eq:SparseSynthesis}
\vect{x} = \mtrx{D}\vect{z}
\end{equation}

\noindent with $\vect{x}\in\Rn{L}$ the time domain signal (or a windowed frame extracted from this signal), $\mtrx{D}\in\Cset{L}{S}$ the dictionary and $\vect{z}\in\Cn{S}$ a sparse representation of the vector $\vect{x}$ ($\ing{S}\geq\ing{L}$). This models considers that $\norm{\vect{z}}{0}$, the number of non-zero coefficients in $\vect{z}$, is very small compared to the size $\ing{S}$ of the vector. In other words, one needs very few atoms of $\mtrx{D}$ to synthesize $\vect{x}$ from $\vect{z}$.

While synthesis approaches comprise a vast majority of the sparsity-based time-frequency regularization techniques, it has been demonstrated in \cite{nam2011cosparse} and more recently in \cite{kiti:tel-01237323,kitic2015sparsity} that the \emph{analysis sparse model}, also known as the \emph{cosparse} model, can turn out to be advantageous in certain settings, in particular in terms of computational cost. 
Instead of \emph{implicitly} defining a sparse representation $\vect{z}$ of the signal $\vect{x}$ through the sparse synthesis model $\vect{x}=\mtrx{D}\vect{z}$, the rationale of the cosparse model is to \emph{explicitly} assume that
\begin{equation}
\label{eq:cosparseModel}
\vect{z} = \mtrx{A}\vect{x}
\end{equation}
is sparse, where $\mtrx{A}\in\Cset{P}{L}$ is called the analysis operator ($\ing{P}\geq\ing{L}$). Models are equivalent when $\ing{P}=\ing{S}=\ing{L}$ and $\mtrx{A}\mtrx{D} = \Id$. 

\subsubsection{Structured (co)sparse time-frequency models}

\begin{figure} 
	\centering
	\subfloat[Tonal music\label{fig:STFTTonal}]{\includegraphics[trim=1cm 8cm 1cm 8cm,clip,width=0.5\columnwidth]{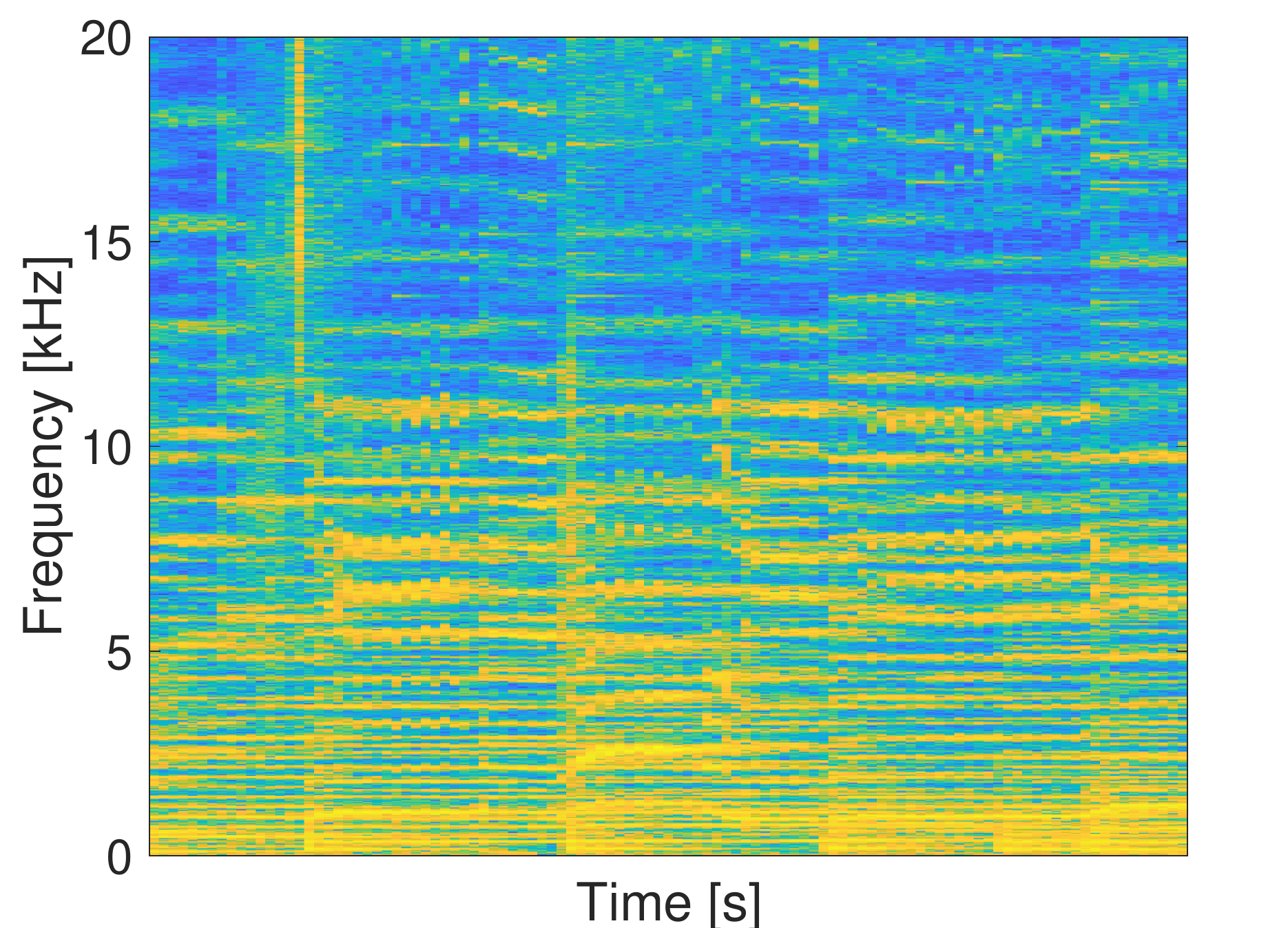}}
	\hfil
	\subfloat[Percussive sounds\label{fig:STFTTransients}]{\includegraphics[trim=1cm 8cm 1cm 8cm,clip,width=0.5\columnwidth]{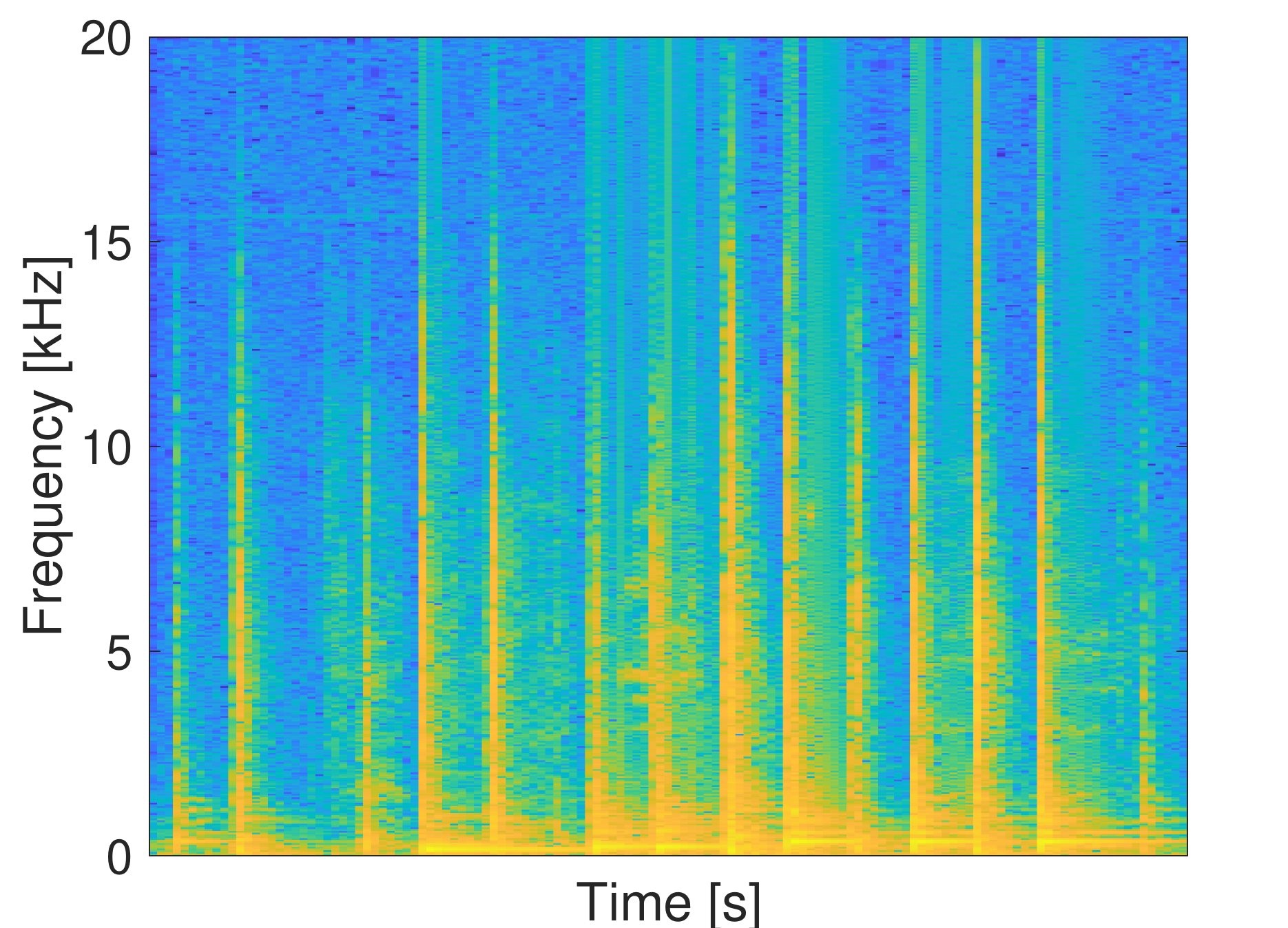}}
	\caption{\ident{Short Time Fourier Transform magnitude of two music signals}\label{fig:STFTstructure}}
\end{figure}

Plain synthesis sparse models as defined in~\eqref{eq:SparseSynthesis} somehow assume that each coefficient in the sparse representation can be active or inactive independently from the others. However, in the context of audio time-frequency modeling one can argue that coefficients are rather arranged in groups as shown in \autoref{fig:STFTstructure}. In a tonal musical excerpt (\autoref{fig:STFTTonal}) high energy coefficients are structured across time in the spectrogram, reflecting the strong presence of harmonics; in a percussive music sample spectrogram (\autoref{fig:STFTTransients}), the dominant coefficients gather across frequency due to transients and beats. Structured forms of sparsity such as group sparsity \cite{kowalski2009structured,jenatton2011structured} or social sparsity \cite{kowalski2009sparsity,kowalski2013social} have emerged as useful refinements 
to take into account such typical time-frequency patterns of audio signals. 

Consider the matrix $\mtrx{x}\in\Rset{L}{G}$ which columns are the windowed frames of an original time-domain audio signal 
and $\mtrx{Z}\in\Cset{S}{G}$ a matrix which columns are a frequency representation of these frames. In other words, this matrix is a time-frequency representation of the underlying audio signal. 
In structured sparse models, the assumed relation between $\mtrx{Z}$ and $\mtrx{X}$ becomes:\\

\noindent
\begin{tabular}{m{0.46\columnwidth}|m{0.46\columnwidth}}
	\multicolumn{1}{c|}{\textbf{Structured analysis model}}&\multicolumn{1}{c}{\textbf{Structured synthesis model}}                                                            \\ 
	$\mtrx{A}\in\Cset{P}{L}, \ing{P} \geq \ing{L}$ & $\mtrx{D}\in\Cset{L}{S}, \ing{S} \geq\ing{L}$                                                                                                                                                            \\
	$\mtrx{Z} \simeq \mtrx{A}\mtrx{X}, \mtrx{Z}\in\Cset{P}{G}$&$\mtrx{D}\mtrx{Z} \simeq \mtrx{X}, \mtrx{Z}\in\Cset{S}{G}$
	\\
	$\norm{\mtrx{z}}{0}\ll\ing{P}\times\ing{G}$&$\norm{\mtrx{z}}{0}\ll\ing{S}\times\ing{G}$
	\\
	$\mtrx{z}$ is ``structured''; &$\mtrx{z}$ is ``structured''.
\end{tabular}

\vspace{0.5cm}
Considering \emph{non-overlapping} groups of indexes in $\mtrx{z}$, \emph{group} sparsity assumes that if some coefficient of the matrix is zero, then all coefficients at indexes belonging to the same group must be also zero, while in ``active'' groups, no sparsity is required. This prior is typically enforced by expressing optimization problems involving mixed-norms such as the $\lpq{2}{1}$ norm \cite{kowalski2009sparseregression}. 
Optimization algorithms to address such minimization problems typically involve structured thresholding operators.
\emph{Social} sparsity extends group sparsity to the case of possibly overlapping groups, and also allows more flexible structures through the use of generic time-frequency patterns. This prior is typically enforced in iterative algorithms with the use of appropriate specific sparsity-promoting operators using dependencies between coefficients. 

\section{Proposed approach}

\label{sec:framework}

In this section, we present a general framework using either simple sparse modeling (analysis or synthesis based) or structured sparse priors to address the audio declipping problem.
Given a matrix $\mtrx{Y}\in\Rset{L}{G}$ \ident{whose} columns are the windowed frames of a clipped time-domain audio signal $\vect{y}$, our goal is to recover an estimate $\hat{\mtrx{x}}$ of the frames $\mtrx{x}$ of the original signal. To this end, one seeks $\hat{\mtrx{x}}$ satisfying:
\begin{itemize}
	\item the modeling constraints described in \autoref{sec:models};
	\item a data fidelity constraint with respect to $\mtrx{Y}$, according to some distortion model (clipping).
\end{itemize}
This is the spirit of the algorithmic framework we develop. It relies on two components: 
\begin{itemize}
	\item a \emph{shrinkage} enforcing (structured) sparsity;
	\item a \emph{generalized projection} onto the clipping-consistent data-fidelity constraint.
\end{itemize}

Before further describing these components in the rest of this section, we formalize some notational conventions.
In the following, lower-case Greek symbols ($\varepsilon$) stands for scalar constant. Lower-case sans serif font ($\ing{i}$) denotes an integer. Lower-case bold font ($\vect{v}$) expresses a vector and upper-case ($\mtrx{V}$) a matrix. $\vect{v}_{\ing{i}}$ is an $\ing{i}$\textsuperscript{th} element of a vector and $\iter{\vect{v}}{i}$ an $\ing{i}$\textsuperscript{th} iterate. $\Theta$ is used for a set. $\mathcal{O}$ stands for a non-linear operator and $F$ a function. \ident{$\mtrx{V}_{\ing{i}\ing{j}}$} represents the component of the matrix $\mtrx{V}$ indexed the $\ing{i}$\textsuperscript{th} row and $\ing{j}$\textsuperscript{th} column. Finally, $\htransp{\mtrx{v}}$ denotes the Hermitian transpose of a matrix $\mtrx{v}$. Curved relation symbols ($\preccurlyeq,\succcurlyeq,\prec,\succ$) are used for entry-wise comparisons between matrices. Other notations will be disambiguated in the text.


\subsection{Sparsity-promoting shrinkage operators}

Shrinkage operators, also known as ``thresholding rules'' \cite{kowalski2014thresholding},  promote sparsity by reducing the magnitude of their input and possibly setting it to zero.
\begin{mydef}[Shrinkage] $\mathcal{S}(\cdot)$, is a shrinkage \ident{if}:
	\begin{enumerate}
		\item $\mathcal{S}(\cdot)$ is an odd function;
		\item $0 \leq \mathcal{S}(x) \leq x$, for all $x \in \mathbb{R}^+$.
		\item $(\mathcal{S}(\cdot))_+$ is nondecreasing on $\mathbb{R}^+$ and $\lim_{x \rightarrow +\infty} (\mathcal{S}(x))_+ = +\infty$, where	$(\cdot)_+:= \max(\cdot,0)$.
	\end{enumerate}
\end{mydef}
\noindent When applied to a (time-frequency) matrix, and written $\mathcal{S}(\mtrx{Z})$, shrinkage is applied entry-wise. Different shrinkage operators have been proposed depending on the sparse prior to account for (\rev{\emph{i.e.}}, plain or structured sparsity). These shrinkage operators are presented below.

\paragraph{Plain sparsity}

The hard-thresholding operator (HT) $\mathcal{H}_{\ing{k}} (\mtrx{Z})$ preserves the $\ing{k}$ coefficients of largest magnitude in $\mtrx{Z}$ and sets the other ones to zero 
\cite{blumensath2009iterative}. It can be used to enforce plain time-frequency sparsity, either analysis or synthesis.

\paragraph{Social (time-frequency) sparsity}

\begin{figure}[!htbp]
	\centering
\newcommand\ColorBox[2][0.5em]{%
	\frame{\textcolor{#2}{\rule{#1}{#1}}}%
}
%
%
%
\definecolor{lred}{rgb}{0.6,0.2,0.}
\definecolor{lgray}{rgb}{0.6,0.6,0.6}

\begin{tikzpicture}[line cap=round,line join=round,>=triangle 45,x=0.2cm,y=0.004cm,scale=2]

\coordinate (A) at (0,0);
\coordinate (B) at (11,0);
\coordinate (C) at (11,300);
\coordinate (D) at (0,300);
\coordinate (a) at (6,50);
\coordinate (b) at (9,50);
\coordinate (c) at (9,150);
\coordinate (d) at (6,150);
\coordinate (i) at (0,100);
\coordinate (j) at (7.5,0);
\coordinate (ij) at (7.5,100);
\coordinate (p1) at (0,50);
\coordinate (p2) at (0,150);
\coordinate (p3) at (6,0);
\coordinate (p4) at (9,0);

\fill [line width=0.6pt,color=lgray,fill=lgray,fill opacity=0.25] (A) -- (B) -- (C) -- (D) -- (A);

\fill [line width=0.6pt,dash pattern=on 1pt off 1pt,color=lred,fill=lred,fill opacity=0.25] (a) -- (b) -- (c) -- (d) -- (a);


\draw [line width=0.3pt, densely dotted,color=lgray] (p1)-- (a);
\draw [line width=0.3pt,densely dotted,color=lgray] (p2)-- (d);
\draw [line width=0.3pt,densely dotted,color=lgray] (p3)-- (a);
\draw [line width=0.3pt,densely dotted,color=lgray] (p4)-- (b);

\draw [line width=0.4pt,densely dotted,color=lgray] (i)-- (ij);
\draw [line width=0.4pt,densely dotted,color=lgray] (ij)-- (j);



\node at (i) [xshift=-5pt]{$\ing{f}$};
\node at (j) [yshift=-10pt]{$\ing{t}$};
\node at (p1) [xshift=-15pt]{$\ing{f}-\ing{F}$};
\node at (p2) [xshift=-15pt]{$\ing{f}+\ing{F}$};
\node at (p3) [yshift=-15pt]{$\ing{t}-\ing{T}$};
\node at (p4) [yshift=-15pt]{$\ing{t}+\ing{T}$};

\node [draw,circle,inner sep=0.5pt,fill] at (ij) {};


\node[anchor=south,scale=0.7] 
at ([xshift=3.25cm,yshift=0.75cm]current bounding box.south west)
{
	\renewcommand\arraystretch{1}
	\begin{tabular}{|cp{0.5cm}|}
	\hline
	\ColorBox{lgray!25} & $\mtrx{Z}$ \\
	 \ColorBox{lred!25} &  $\mtrx{Z}_{\mtrx{P}_{\ing{f}\ing{t}}}$ \\
	$\bullet$ &  $\mtrx{Z}_{\ing{f}\ing{t}}$ \\
	\hline
	\end{tabular}
};

\end{tikzpicture}	
	\caption{Schematic representation of patch extraction from matrix $\mtrx{Z}$\label{fig:Z}}
\end{figure}
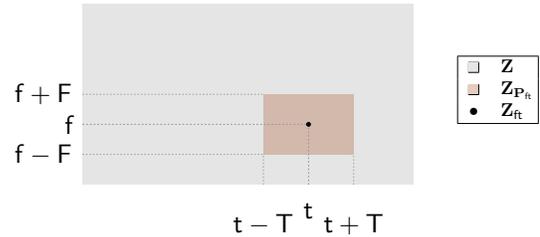

The Persistent Empirical Wiener (PEW) operator \ident{\cite{Siedenburg2012,kowalski2014thresholding}} was successfully used in \cite{siedenburg2014audio} for audio declipping using (synthesis) social sparsity. 
This shrinkage promotes specific local time-frequency patterns around each time-frequency point. Its specification explicitly requires choosing a time-frequency pattern described as a matrix $\Gamma\in\Rset{(\ing{2F+1})}{(\ing{2T+1})}$, \ident{typically} with binary entries. Rows of $\Gamma$ account for the frequency dimension and columns for the time dimension, in \emph{local time-frequency coordinates}. 
Let $\mtrx{Z}$ be a time-frequency representation. 
As illustrated on \autoref{fig:Z}, consider $\ing{ft}$ the coordinates of a time-frequency point in $\mtrx{Z}$ ($\ing{f}$ is a frequency index, $\ing{t}$ the index of a windowed frame) and $\mtrx{P}_{\ing{ft}} := [\ing{f}-\ing{F},\ing{f}+\ing{F}] \times [\ing{t}-\ing{T},\ing{t}+\ing{T}]$ the indices corresponding to a time-frequency patch of size $(\ing{2F+1}) \times (\ing{2W+1})$ centered at time-frame $\ing{t}$ and frequency $\ing{f}$. The matrix $\mtrx{Z}_{\mtrx{P}_{\ing{ft}}}\in\Cset{(\ing{2F+1})}{(\ing{2W+1})}$ is extracted from $\mtrx{Z}$ on these indices, with mirror-padding on the borders if needed. 

\noindent 
Now, given a thresholding parameter $\mu \geq 0$, we can define the PEW shrinkage operator $\mathcal{S}^{\text{PEW}}_{\mu}$. Denoting $\circ$ the (Hadamard) entrywise product and $(\cdot)_+ = \max(\cdot,0)$ the positive part:
\ident{\begin{equation}\label{eqPEW}
\mathcal{S}^{\text{PEW}}_{\mu}(\mtrx{Z} | \Gamma)_{\ing{ft}} := \mtrx{Z}_{\ing{ft}} \cdot \left(1 - \frac{\mu^2}{\norm{ \mtrx{Z}_{\mtrx{P}_{\ing{ft}}} \circ \Gamma}{\text{F}}^2 } \right)_+.
\end{equation}}
Since $\norm{\mtrx{Z}_{\mtrx{P}_{\ing{ft}}} \circ \Gamma}{\text{F}}^2$ is the energy of $\mtrx{Z}$ restricted to a time-frequency neighborhood of $\ing{ft}$ of shape specified by $\Gamma$, the left hand side is zero as soon as this energy falls below $\mu^{2}$. As such, PEW shrinkage effectively promotes structured sparsity.

\begin{remark}
To use similar notations for plain and social sparsity, in the case of analysis (resp. synthesis) plain sparse modeling with $\mtrx{A}\in\Cset{P}{L}$ a forward frequency analysis operator	(resp. $\mtrx{D}\in\Cset{L}{S}$ a dictionary) we denote $\mathcal{S}_{\mu} := \mathcal{H}_{\ing{P} - \mu}$ (resp. $\mathcal{S}_{\mu} := \mathcal{H}_{\ing{S} - \mu}$), for $\mu \in \mathbb{N}^+$, $0 \leq \mu \leq \ing{P}$ (resp. $0 \leq \mu \leq \ing{S}$). Hence, the smaller $\mu$, the less thresholding is performed. The large $\mu$, the sparser the result of both $\mathcal{S}_{\mu}(\cdot)$ and $\mathcal{S}^{\text{PEW}}_{\mu}(\cdot|\Gamma)$. In all cases, $\mu$ is thus a parameter controlling the strength of the performed shrinkage, which is stronger when $\mu$ is larger.
\end{remark}

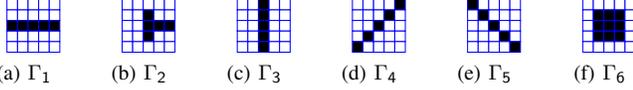
\begin{figure}[!t]
	\centering
	\subfloat[$\Gamma_{\ing{1}}$]{	
	\pgfplotstableread{MusicGammaTonal.cvs}{\Gamma}

		\def\szx{4pt}
		\def\szy{4pt}

	\begin{tikzpicture}
	\pgfplotstableforeachcolumn\Gamma\as\col{
		\pgfplotstableforeachcolumnelement{\col}\of\Gamma\as\colcnt{%
			\draw[draw=blue,very thin,fill=black!\colcnt]($ -\pgfplotstablerow*(0,\szy) + \col*(\szx,0) $) rectangle+(\szx,\szy); 
		}
	}
	\end{tikzpicture}}\hfill
	\subfloat[$\Gamma_{\ing{2}}$\label{fig:GammaPreecho}]{	
	\pgfplotstableread{MusicGammaPreecho.cvs}{\Gamma}

		\def\szx{4pt}
		\def\szy{4pt}

	\begin{tikzpicture}
	\pgfplotstableforeachcolumn\Gamma\as\col{
		\pgfplotstableforeachcolumnelement{\col}\of\Gamma\as\colcnt{%
			\draw[draw=blue,very thin,fill=black!\colcnt]($ -\pgfplotstablerow*(0,\szy) + \col*(\szx,0) $) rectangle+(\szx,\szy); 
		}
	}
	\end{tikzpicture}}\hfill
	\subfloat[$\Gamma_{\ing{3}}$]{	
	\pgfplotstableread{MusicGammaTransient.cvs}{\Gamma}

		\def\szx{4pt}
		\def\szy{4pt}

	\begin{tikzpicture}
	\pgfplotstableforeachcolumn\Gamma\as\col{
		\pgfplotstableforeachcolumnelement{\col}\of\Gamma\as\colcnt{%
			\draw[draw=blue,very thin,fill=black!\colcnt]($ -\pgfplotstablerow*(0,\szy) + \col*(\szx,0) $) rectangle+(\szx,\szy); 
		}
	}
	\end{tikzpicture}}\hfill
	\subfloat[$\Gamma_{\ing{4}}$]{	
	\pgfplotstableread{MusicGammaUp.cvs}{\Gamma}

		\def\szx{4pt}
		\def\szy{4pt}

	\begin{tikzpicture}
	\pgfplotstableforeachcolumn\Gamma\as\col{
		\pgfplotstableforeachcolumnelement{\col}\of\Gamma\as\colcnt{%
			\draw[draw=blue,very thin,fill=black!\colcnt]($ -\pgfplotstablerow*(0,\szy) + \col*(\szx,0) $) rectangle+(\szx,\szy); 
		}
	}
	\end{tikzpicture}}\hfill
	\subfloat[$\Gamma_{\ing{5}}$]{	
	\pgfplotstableread{MusicGammaDown.cvs}{\Gamma}

		\def\szx{4pt}
		\def\szy{4pt}

	\begin{tikzpicture}
	\pgfplotstableforeachcolumn\Gamma\as\col{
		\pgfplotstableforeachcolumnelement{\col}\of\Gamma\as\colcnt{%
			\draw[draw=blue,very thin,fill=black!\colcnt]($ -\pgfplotstablerow*(0,\szy) + \col*(\szx,0) $) rectangle+(\szx,\szy); 
		}
	}
	\end{tikzpicture}}\hfill
	\subfloat[$\Gamma_{\ing{6}}$]{	
	\pgfplotstableread{MusicGammaBlob.cvs}{\Gamma}

		\def\szx{4pt}
		\def\szy{4pt}

	\begin{tikzpicture}
	\pgfplotstableforeachcolumn\Gamma\as\col{
		\pgfplotstableforeachcolumnelement{\col}\of\Gamma\as\colcnt{%
			\draw[draw=blue,very thin,fill=black!\colcnt]($ -\pgfplotstablerow*(0,\szy) + \col*(\szx,0) $) rectangle+(\szx,\szy); 
		}
	}
	\end{tikzpicture}}
	\caption{Extended set of time-frequency neighborhoods used for music\label{fig:Stencils}}
\end{figure}

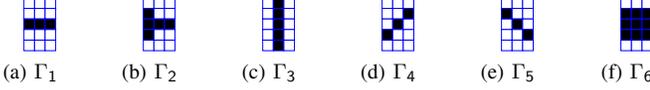
\begin{figure}[!t]
	\centering
	\subfloat[$\Gamma_{\ing{1}}$]{	
	\pgfplotstableread{SpeechGammaTonal.cvs}{\Gamma}

		\def\szx{4pt}
		\def\szy{4pt}

	\begin{tikzpicture}
	\pgfplotstableforeachcolumn\Gamma\as\col{
		\pgfplotstableforeachcolumnelement{\col}\of\Gamma\as\colcnt{%
			\draw[draw=blue,very thin,fill=black!\colcnt]($ -\pgfplotstablerow*(0,\szy) + \col*(\szx,0) $) rectangle+(\szx,\szy); 
		}
	}
	\end{tikzpicture}}\hfill
	\subfloat[$\Gamma_{\ing{2}}$]{	
	\pgfplotstableread{SpeechGammaPreecho.cvs}{\Gamma}

		\def\szx{4pt}
		\def\szy{4pt}

	\begin{tikzpicture}
	\pgfplotstableforeachcolumn\Gamma\as\col{
		\pgfplotstableforeachcolumnelement{\col}\of\Gamma\as\colcnt{%
			\draw[draw=blue,very thin,fill=black!\colcnt]($ -\pgfplotstablerow*(0,\szy) + \col*(\szx,0) $) rectangle+(\szx,\szy); 
		}
	}
	\end{tikzpicture}}\hfill
	\subfloat[$\Gamma_{\ing{3}}$]{	
	\pgfplotstableread{SpeechGammaTransient.cvs}{\Gamma}

		\def\szx{4pt}
		\def\szy{4pt}

	\begin{tikzpicture}
	\pgfplotstableforeachcolumn\Gamma\as\col{
		\pgfplotstableforeachcolumnelement{\col}\of\Gamma\as\colcnt{%
			\draw[draw=blue,very thin,fill=black!\colcnt]($ -\pgfplotstablerow*(0,\szy) + \col*(\szx,0) $) rectangle+(\szx,\szy); 
		}
	}
	\end{tikzpicture}}\hfill
	\subfloat[$\Gamma_{\ing{4}}$]{	
	\pgfplotstableread{SpeechGammaUp.cvs}{\Gamma}

		\def\szx{4pt}
		\def\szy{4pt}

	\begin{tikzpicture}
	\pgfplotstableforeachcolumn\Gamma\as\col{
		\pgfplotstableforeachcolumnelement{\col}\of\Gamma\as\colcnt{%
			\draw[draw=blue,very thin,fill=black!\colcnt]($ -\pgfplotstablerow*(0,\szy) + \col*(\szx,0) $) rectangle+(\szx,\szy); 
		}
	}
	\end{tikzpicture}}\hfill
	\subfloat[$\Gamma_{\ing{5}}$]{	
	\pgfplotstableread{SpeechGammaDown.cvs}{\Gamma}

		\def\szx{4pt}
		\def\szy{4pt}

	\begin{tikzpicture}
	\pgfplotstableforeachcolumn\Gamma\as\col{
		\pgfplotstableforeachcolumnelement{\col}\of\Gamma\as\colcnt{%
			\draw[draw=blue,very thin,fill=black!\colcnt]($ -\pgfplotstablerow*(0,\szy) + \col*(\szx,0) $) rectangle+(\szx,\szy); 
		}
	}
	\end{tikzpicture}}\hfill
	\subfloat[$\Gamma_{\ing{6}}$]{	
	\pgfplotstableread{SpeechGammaBlob.cvs}{\Gamma}

		\def\szx{4pt}
		\def\szy{4pt}

	\begin{tikzpicture}
	\pgfplotstableforeachcolumn\Gamma\as\col{
		\pgfplotstableforeachcolumnelement{\col}\of\Gamma\as\colcnt{%
			\draw[draw=blue,very thin,fill=black!\colcnt]($ -\pgfplotstablerow*(0,\szy) + \col*(\szx,0) $) rectangle+(\szx,\szy); 
		}
	}
	\end{tikzpicture}}
	\caption{Extended set of time-frequency neighborhoods used for speech\label{fig:SpeechStencils}}
\end{figure}

Examples of time-frequency patterns $\Gamma$ chosen for music are given in \autoref{fig:Stencils} and for speech in \autoref{fig:SpeechStencils}. %
The unit associated to individual squares is the time-frequency index of the DFT. In the experimental sections, we will take 64 ms time-frames for music and 32~ms time-frames for speech; the total time span for each pattern $\Gamma$ is 320~ms for music and 96~ms for speech. While similar, the patterns for music and for speech are at different time scales, given the different scales of stationarity in speech and music. The structures of these patterns have various properties: $\Gamma_{\ing{1}}$, with a frequency localized and time-spread support, will emphasize tonal content; vice-versa, $\Gamma_{\ing{3}}$ will emphasize transients and attacks; $\Gamma_{\ing{2}}$ is designed \cite{siedenburg2012audio} to avoid pre-echo artifacts; patterns $\Gamma_{\ing{4}}$ and $\Gamma_{\ing{5}}$ are introduced to stress tonal transitions; finally, $\Gamma_{\ing{6}}$ can serve as a default pattern 
when no particular structure is identified.

\subsection{Clipping-consistent generalized projections} We present below the generalized projection 
that will be crucial to enforce the clipping-consistent data-fidelity constraint along the iterations of the proposed algorithm family. 
\label{subec:SolveDeclip}

Denote $\Omega_+$ (resp. $\Omega_-$) the collection of indices 
of the samples in matrix $\mtrx{Y}$ affected by positive (resp. negative) magnitude clipping. Similarly denote $\Omega_r$ the indices of the reliable samples (not affected by clipping). For any of these sets $\Omega$, given a generic matrix $\mtrx{V}$ of the same dimensions as $\mtrx{Y}$, define $\mtrx{V}_{\Omega}$ the matrix formed by keeping unchanged the entries of $\mtrx{V}$ indexed by $\Omega$ and setting the rest to zero.

The set of clipping-consistent time-domain estimates $\mtrx{W}$ of  $\mtrx{X}$ can be expressed
for the analysis setting by $$\Theta_{\textrm{time}}(\mtrx{Y}) := \left\{ 
\mtrx{W}\ |
\begin{array}{l}
\mtrx{W}_{\Omega_r} = \mtrx{Y}_{\Omega_r};\\
\mtrx{W}_{\Omega_+} \succcurlyeq \mtrx{Y}_{\Omega_+}; \\ 
\mtrx{W}_{\Omega_-} \preccurlyeq \mtrx{Y}_{\Omega_-}.\end{array} \right\}
$$
where $\mtrx{W}$ is of the same size as $\mtrx{Y}$. For the synthesis setting, the set of clipping-consistent time-frequency estimates $\mtrx{W}$ of $\mtrx{X}$ is defined as
$$\Theta_{\textrm{time-freq}}(\mtrx{Y})  := \left\{ 
\mtrx{W}\ |
\begin{array}{l}
(\mtrx{D}\mtrx{W})_{\Omega_r} = \mtrx{Y}_{\Omega_r};\\
(\mtrx{D}\mtrx{W})_{\Omega_+} \succcurlyeq \mtrx{Y}_{\Omega_+}; \\ 
(\mtrx{D}\mtrx{W})_{\Omega_-} \preccurlyeq \mtrx{Y}_{\Omega_-}.\end{array} \right\}.$$
Here $\mtrx{W}$ is a time-frequency estimate gathering as many frames as in $\mtrx{Y}$ and $\ing{S}$ frequency points, so that $\mtrx{D}\mtrx{W}$ is a clipping-consistent time-domain estimate of $\mtrx{X}$. These choices hold for both plain and structured versions, and the set $\Theta$ is convex in all cases.

\begin{mydef}[Generalized projection]
	Let $\Theta$ be a nonempty \emph{convex} set, and $\mtrx{M}$ be a full column rank matrix. Given a time-frequency matrix $\mtrx{Z}$, we denote $\mathcal{P}_{\Theta,\mtrx{M}}(\mtrx{Z})$ the (unique) solution of the following optimization problem:
	\begin{equation}\label{eq:ConstrainedProjection}
	\minim_{\mtrx{W} \in \Theta} \norm{\mtrx{M} \mtrx{W} - \mtrx{Z}}{\mathrm{F}}. 
	\end{equation} 
\end{mydef}

In the analysis setting, seeking a clipping consistent time-domain estimate $\mtrx{W}$ of $\mtrx{X}$ whose time-frequency representation approximates a given $\mtrx{Z}$ corresponds to computing $\mathcal{P}_{\texttt{ana}}(\mtrx{Z}):= \mathcal{P}_{\Theta,\mtrx{M}}(\mtrx{Z})$ with $\Theta = \Theta_{\textrm{time}}(\mtrx{Y})$ and $\mtrx{M}:=\mtrx{A}$. Considering an analysis operator $\mtrx{A}$ such that $\htransp{\mtrx{A}}\mtrx{A} = \mtrx{I}$,
minimizing  $\norm{\mtrx{A}\mtrx{W}-\mtrx{Z}}{\text{F}}^{2}$
with $\mtrx{W} \in \Theta$ is equivalent to 
\[
\minim_{\mtrx{W}} \norm{\mtrx{W} - \htransp{\mtrx{A}}\mtrx{Z}}{\text{F}}^{2} 
\subjto \mtrx{W} \in \Theta.
\]
\noindent As the constraint is written component-wise, 
the desired projection 
can be expressed accordingly for the $\ing{n}$-th sample of the $\ing{t}$-th time-frame as:
$$
[\mathcal{P}_{\texttt{ana}}(\mtrx{Z})]_{\ing{nt}}
\!=\!\left\{ 
\begin{array}{l l}
(\htransp{\mtrx{M}}\mtrx{Z})_{\ing{nt}} & \!\text{if}\!\left\{\begin{array}{l}\ing{nt}\in\Omega_+,(\htransp{\mtrx{M}}\mtrx{Z})_{\ing{nt}} \geq \tau;\\ \text{or}\\\ing{nt}\in\Omega_-,(\htransp{\mtrx{M}}\mtrx{Z})_{\ing{nt}} \leq -\tau;\end{array}\right.\\
\mtrx{Y}_{\ing{nt}}
& \text{otherwise.}\\
\end{array} \right.$$
In this case, matrix-vector products with $\htransp{\mtrx{M}}$ dominate the computational cost of the generalized projection. When this can be done with a fast transform, the analysis 
\ident{variant} has low complexity.

For the synthesis case, seeking a clipping consistent time-frequency estimate $\mtrx{W}$ of $\mtrx{X}$ which is close to a given time-frequency representation $\mtrx{Z}$ corresponds to computing $\mathcal{P}_{\texttt{syn}}(\mtrx{Z}):= \mathcal{P}_{\Theta,\mtrx{M}}(\mtrx{Z})$ with $\Theta = \Theta_{\textrm{time-freq}}(\mtrx{Y})$ and $\mtrx{M}:=\mtrx{I}$. Such a projection step can be approximated with a nested iterative procedure \cite{kiti:tel-01237323}. Even if this can help building an efficient projection algorithm, the overall computation cost for the synthesis 
\ident{variant} remains substantially higher than the analysis version (making it almost intractable). More recently, a closed-form solution for the declipping projection in the synthesis case was obtained \cite{Rajmic:2019dd} when $\mtrx{D}$ is a Parseval tight frame (\rev{\emph{i.e.}}, $\mtrx{D}\htransp{\mtrx{D}} = \mtrx{I}$). 
As detailed in \hyperref[app:ProxDeclipping]{appendix} this projection can then be expressed as:
\begin{align}\label{eq:SythesisDeclipClosedForm}
\mathcal{P}_{\texttt{syn}}(\mtrx{Z})
&	= \mtrx{Z} - \htransp{\mtrx{D}}(\mtrx{D}\mtrx{Z} - 
\Pi(\mtrx{Z})),
\end{align}

\noindent with
$$
[\Pi(\mtrx{Z})]_{\ing{nt}}
\!=\!\left\{ 
\begin{array}{l l}
(\mtrx{D}\mtrx{Z})_{\ing{nt}} & \!\text{if}\left\{\begin{array}{l}\ing{nt}\in\Omega_+,(\mtrx{D}\mtrx{Z})_{\ing{nt}} \geq \tau;\\ \text{or}\\\ing{nt}\in\Omega_-,(\mtrx{D}\mtrx{Z})_{\ing{nt}} \leq -\tau;\end{array}\right.\\
\mtrx{Y}_{\ing{nt}}
 & \text{otherwise.}\\
\end{array} \right.$$

\noindent When fast products with $\mtrx{D}$ and $\htransp{\mtrx{D}}$ can be achieved, the synthesis 
projection also has low complexity. 

\subsection{Generic declipping algorithm}

A generic algorithm for declipping exploiting sparsity, either in its synthesis or analysis 
\ident{variant,}  with and without structure, is described in \autoref{alg:AbstractAlgo}. It takes as input parameters:
	\begin{itemize}
		\item a convex set $\Theta$ and a matrix $\mtrx{M}$ embodying the declipping data fidelity constraint and the domain (time or frequency) in which it is specified;
		\item a parameterized family of shrinkages $\{\mathcal{S}_{\mu}(\cdot)\}_{\mu}$, where the 
		strength of the 
		shrinkage is controlled by $\mu$: in the extreme cases $\mathcal{S}_0(\mtrx{Z}) = \mtrx{Z}$ and $\mathcal{S}_{\infty}(\mtrx{Z}) = \mtrx{0}$; 
		\item a rule $F: \mu \mapsto F(\mu)$ to update the shrinkage strength across iterations, and an initial strength $\iter{\mu}{\ing{0}}$; 
		\item an initial time-frequency estimate $\iter{\mtrx{Z}}{0}$;
		\item stopping parameters $\beta$ (a threshold on the relative error) and $\ing{i}_{\max}$ (a maximum number of iterations).
	\end{itemize}

\begin{algorithm}[!t]
	
	\caption{Generic Algorithm: $\mathcal{G}$}
	\label{alg:AbstractAlgo}
	
	\begin{algorithmic} 
		\REQUIRE $\Theta, \mtrx{M}, \{\mathcal{S}_{\mu}(\cdot)\}_{\mu}, \iter{\mu}{\ing{0}}, F(\cdot), \iter{\mtrx{Z}}{0}, \beta, \ing{i}_{\max}$
		\STATE \underline{\textit{Initialization step:}}
		\STATE $\iter{\mtrx{U}}{0}=\mtrx{0}$;
		\FOR{$\ing{i} = 1 \text{ to } \ing{i}_{\max}$}
		\STATE \underline{\textit{Projection step on the declipping constraint:}}
		\STATE $\iter{\mtrx{w}}{i} =  \mathcal{P}_{\Theta,\mtrx{M}}(\iter{\mtrx{z}}{i-1} - \iter{\mtrx{u}}{i-1})$ 
		\STATE \underline{\textit{Projection step on the modeling constraint:}}
		\STATE $\iter{\mtrx{z}}{i} =  \mathcal{S}_{\iter{\mu}{i-1}} \left( \mtrx{M} \iter{\mtrx{w}}{i} + \iter{\mtrx{u}}{i-1} \right)$
		\STATE \underline{\textit{Update step:}}
		\IF{$\frac{\norm{\mtrx{M}\iter{\mtrx{w}}{i} - \iter{\mtrx{Z}}{i}}{\text{F}} }{\norm{\mtrx{M}\iter{\mtrx{w}}{i}}{\text{F}}} \leq \beta$}
		\STATE $\text{terminate}$
		\ELSE
		\STATE $\iter{\mtrx{u}}{i} = \iter{\mtrx{u}}{i-1} +  \mtrx{M} \iter{\mtrx{w}}{i} - \iter{\mtrx{z}}{i}$
		\STATE $\iter{\mu}{i} = F(\iter{\mu}{i-1})$
		\ENDIF
		\ENDFOR
		\RETURN $\iter{\mtrx{w}}{\ing{i}}$\ [and optionally $\iter{\mu}{i}$, $\iter{\mtrx{z}}{i}$] 
	\end{algorithmic}
\end{algorithm}

\noindent The notation $\iter{\mtrx{Z}}{i}$ highlights that the corresponding variable is in any use-case a sparse/structured time-frequency representation. The variable $\iter{\mtrx{u}}{i}$ is an intermediate time-frequency ``residual'' variable typical of approaches inspired by ADMM (Alternating Direction Method of Multipliers) \cite{boyd2011distributed}. At iteration $\ing{i}$, an estimate of $\mtrx{Z}$ is $\iter{\hat{\mtrx{Z}}}{i} := \iter{\mtrx{z}}{i-1} - \iter{\mtrx{u}}{i-1}$. The interpretation of the other variables is use-case dependent:
\begin{itemize}
	\item {\bf analysis \rev{variant}:} $\mtrx{M} := \mtrx{A}$ is the frequency analysis operator; $\iter{\mtrx{w}}{i}$ is an estimate of the time frames $\mtrx{X}$, that satisfies the time-domain data-fidelity constraint $\Theta_{\textrm{time}}(\mtrx{Y})$ while being closest to $\iter{\hat{\mtrx{Z}}}{i}$  \emph{in the time-frequency domain}; the algorithm outputs a time-domain estimate.
	\item {\bf synthesis \rev{variant}:}  $\mtrx{M} := \mtrx{I}$; $\iter{\mtrx{w}}{i}$ is a time-frequency estimate of $\mtrx{Z}$; the data-fidelity constraint $\Theta_{\textrm{time-freq}}(\mtrx{Y})$ is \emph{expressed in the time-frequency domain}; the algorithm outputs a time-frequency estimate, from which we get a time-domain estimate by synthesis $\hat{\mtrx{x}} := \mtrx{D}\iter{\mtrx{W}}{i}$ with $\mtrx{D}$ the inverse frequency transform operator.
\end{itemize}

\noindent Due to the expression of $\Theta$ respectively in the time domain and the time-frequency domain, the analysis and synthesis 
\ident{variants} can have different computational properties as will be further studied. We can summarize \autoref{alg:AbstractAlgo} as a generalized procedure:

\begin{equation}
\mathcal{G}(\Theta, \mtrx{M}, \{\mathcal{S}_{\mu}\}_{\mu}, \iter{\mu}{\ing{0}}, F, \iter{\mtrx{Z}}{0}, \beta, \ing{i}_{\max}).
\end{equation} 

For reproducibility, the \Matlab code for all variants of this generic algorithm 
\rev{is available \cite{gaultier:hal-03024116v1} under a BSD-3 licence. 
Audio examples are also available on the web\footnote{\url{https://spade.inria.fr/selected-declipping-overview}}.}

\subsection{Plain (co)sparse audio declippers}\label{sec:SparseDeclippers}

In practice, the algorithms are built to work on a frame-based manner (see further details at the end of this section). In the plain (co)sparse cases, $\mtrx{Y} \in \Rset{L}{1}$ is a vector corresponding to a single windowed frame extracted from the clipped signal $\vect{y}$. We instantiate the general algorithm $\mathcal{G}$ (\autoref{alg:AbstractAlgo}) by choosing the operators described in \autoref{tab:SparseDeclip}. The update rule $F$ for $\mu$ is set to gradually decrease $\mu$ by $1$ at each iteration, starting from $\iter{\mu}{0} = \ing{P} - 1$ for the analysis case (resp. $\iter{\mu}{0} = \ing{S} - 1$ for the synthesis case). This way, we relax the sparse constraint along with the iterations.

\begin{table}[h!]
	\caption{Parameters of \autoref{alg:AbstractAlgo} for the Plain Sparse Declippers \label{tab:SparseDeclip}}
	\centering
	\begin{tabular}{m{0.45\columnwidth}|m{0.55\columnwidth}}
		
		\multicolumn{1}{c|}{\textbf{Analysis}}                                                                                                                                                                & \multicolumn{1}{c}{\textbf{Synthesis}}                                                                                                                                                                   \\ 
		\\
		$
		\Theta = \Theta_{\textrm{time}}(\mtrx{Y})
		$
		& 
		$
			\Theta = \Theta_{\textrm{time-freq}}(\mtrx{Y})
		$
		\\
		\\
		$\mtrx{M} = \mtrx{A} \in \Cset{P}{L}, \ing{P} \geq \ing{L}$                                                                                                                                                                                                                   &$\mtrx{M} = \mtrx{I} \in \Cset{L}{L}$,                                                                                                                                                           \\
		$\mathcal{S}_{\mu} (\cdot)= \mathcal{H}_{\ing{P} - \mu} (\cdot)$                                                                                                                                                                &$\mathcal{S}_{\mu} (\cdot)= \mathcal{H}_{\ing{S} - \mu} (\cdot)$,                                                                                                                                        \\
		$\iter{\mu}{0} = \ing{P}-1$
		&
		$\iter{\mu}{0} = \ing{S}-1$
		\\
		$F : \mu \mapsto \mu - 1$
		&
		$F : \mu \mapsto \mu - 1$
		\\
		$\iter{\mtrx{z}}{0} = \mtrx{A}\mtrx{y},\;\mtrx{A} \in \Cset{P}{L}$
		&
		$\iter{\mtrx{z}}{0} = \htransp{\mtrx{D}}\mtrx{y},\; \mtrx{D} \in \Cset{L}{S}$
	\end{tabular}%
\end{table}

Iterating \autoref{alg:AbstractAlgo} with the parameters described above gives a declipped estimate $\hat{\mtrx{w}}$ such that:
\[
\hat{\mtrx{w}} := 
\mathcal{G}(
\Theta, \mtrx{M}, \{\mathcal{S}_{\mu}(\cdot)\}_{\mu}, \iter{\mu}{\ing{0}}, F, \iter{\mtrx{Z}}{0}, \beta, \ing{i}_{\max}
).
\]
The declipped frame is $\hat{\mtrx{x}} := \hat{\mtrx{w}} \in \Rset{L}{1}$ for the analysis version, while for the synthesis version $\hat{\mtrx{x}} := \mtrx{D}\hat{\mtrx{w}} \in \Rset{L}{1}$.\\

\subsection{Social (co)sparse audio declippers}\label{sec:SocSparseDeclippers}

These algorithms are built to work on a frame based manner. Their input is a matrix $\mtrx{Y} \in \Rset{L}{2b+1}$ made of $2\ing{b}+1$ windowed frames of the clipped signal $\vect{y}$. Their goal is to estimate the central frame $\vect{x} = \mtrx{X}(:,\ing{b}+1)$ of the matrix $\mtrx{X} \in \Rset{L}{(2b+1)}$ of frames of the original signal. 

The sparsifying operator is set to $\mathcal{S}_{\mu}^{\text{PEW}}(\cdot\mid\Gamma)$~\eqref{eqPEW} and the update rule is now set to $F_{\alpha}: \mu \mapsto \alpha \mu$. Here $\mu$ plays a different role compared to the plain (co)sparse declippers. Indeed, $\mu$ does not directly tune a sparsity level but an energy. 
The resulting parameters are summarized in \autoref{tab:SocialSparseDeclip}.

\begin{table}[h!]
	\caption{Parameters of \autoref{alg:AbstractAlgo} for the Social Sparse Declippers \label{tab:SocialSparseDeclip}}
	\centering
	\begin{tabular}{m{0.45\columnwidth}|m{0.55\columnwidth}}
		\multicolumn{1}{c|}{\textbf{Analysis}}                                                                                                                                                                & \multicolumn{1}{c}{\textbf{Synthesis}}                                                                                                                                                                   \\ 
		\\
		$
			\Theta = \Theta_{\textrm{time}}(\mtrx{Y})
		$
		& 
		$
				\Theta = \Theta_{\textrm{time-freq}}(\mtrx{Y})
		$
		\\
		\\
		$\mtrx{M} = \mtrx{A} \in \Cset{P}{L}, \ing{P} \geq \ing{L}$                                                                                                                                                                                                                   & $\mtrx{M} = \mtrx{I} \in \Cset{L}{L}$, 
		\\
		$\mathcal{S}_{\mu} (\cdot)= \mathcal{S}^{\text{PEW}}_{\mu}(\cdot|\Gamma)$                                                                                                                                                                & $\mathcal{S}_{\mu} (\cdot)= \mathcal{S}^{\text{PEW}}_{\mu}(\cdot|\Gamma)$,                                                                                                                                        \\
		$\iter{\mu}{0}$: see \autoref{sec:experiments}
		&
		$\iter{\mu}{0}$: see \autoref{sec:experiments}
		\\
		$F = F_{\alpha}: \mu \mapsto \alpha \mu$
		&
		$F = F_{\alpha}: \mu \mapsto \alpha \mu$
		\\
		$\iter{\mtrx{z}}{0} = \mtrx{A}\mtrx{y},\;\mtrx{A} \in \Cset{P}{L}$
		&
		$\iter{\mtrx{z}}{0} = \htransp{\mtrx{D}}\mtrx{y},\; \mtrx{D} \in \Cset{L}{S}$
	\end{tabular}%
\end{table}
\noindent Social declippers with a \emph{predefined} time-frequency pattern $\Gamma$ can be compactly written using \autoref{alg:AbstractAlgo} as:
\begin{eqnarray*}
	\begin{bmatrix}
		\hat{\mtrx{w}}(\Gamma)\\ 
		\mu(\Gamma)\\ 
		\mtrx{Z}(\Gamma)
	\end{bmatrix} := \mathcal{G}(
	\Theta, \mtrx{M}, \{\mathcal{S}^{\text{PEW}}_{\mu}(\cdot|\Gamma)\}_{\mu}, \iter{\mu}{\ing{0}}, F_{\alpha}, \iter{\mtrx{Z}}{0}, \beta, \ing{i}_{\max}
	),
\end{eqnarray*}
For the analysis version the declipped time-domain matrix is $\hat{\mtrx{x}} := \hat{\mtrx{w}} \in \Rset{L}{1}$, while for the synthesis version $\hat{\mtrx{x}} := \mtrx{D}\hat{\mtrx{w}} \in \Rset{L}{1}$. The central frame of the original signal is estimated as $\hat{\vect{x}} = \hat{\mtrx{x}}(:,\ing{b}+1)$.

\subsection{Adaptive social (co)sparse declippers}

In light of recent work on adaptive social denoising \cite{gaultier2017audascity}, a more adaptive social declipper uses the above described social declipper version as a building \ident{block} to select the ``optimal'' pattern $\Gamma$ within a prescribed collection $\left\{ \Gamma_{\ing{j}}\right\}_{\ing{j}=1}^{\ing{J}}$ for the processed signal region, by running few iterations of the algorithm (typically \ident{$\ing{i}_{\text{init}} = 10$}) for each $\Gamma_{\ing{j}}$. We select the pattern $\Gamma_{\ing{j}^{\star}}$ yielding a residual with time-frequency representation of highest entropy. The higher the entropy, the less structured the residual. Hence, this criterion tends to put most of the structured content in the estimate.

For a given $\ing{j}$, we can define the resulting time-frequency residual: $\mtrx{R}_{\ing{j}}  := \mtrx{M}\hat{\mtrx{w}}_{\ing{j}} - \iter{\mtrx{Z}}{0}$. Computing a $\ing{Q}$-bin histogram of the modulus of its entries yields $\hat{\var{p}}^{(\ing{j})}$, an empirical probability distribution, which (empirical) entropy is
\begin{equation}
\label{eq:entropy}
\var{e}_{\ing{j}} = - \sum_{\ing{q=1}}^{\ing{Q}} \hat{\var{p}}_{\ing{q}}^{(\ing{j})} \log_2  (\hat{\var{p}}_{\ing{q}}^{(\ing{j})}).
\end{equation}
A heuristic to choose $\ing{Q}$ is the Herbert-Sturges rule \cite{sturges1926choice}.

The choice of the first value $\iter{\mu_{\ing{j}}}{0}$ and of the update rule $F_{\alpha}$ as well as the choice of the collection of allowed time-frequency patterns $\left\{ \Gamma_{\ing{j}}\right\}_{\ing{j}=1}^{\ing{J}}$, are essential for the algorithm performance. These will be specified in \autoref{sec:experiments}.

\noindent Once the best time-frequency pattern $\Gamma_{\ing{j}^{\star}}$ is selected, we run \autoref{alg:AbstractAlgo} with the parameters listed in \autoref{tab:SocialSparseDeclip} and ``warm-started'' with $\mu^{(0)}$, $\mtrx{Z}^{(0)}$ and a sufficiently large $\ing{i}_{\max}$ (typically \ident{$\ing{i}_{\text{max}} =10^6$}) to get
\[
\hat{\mtrx{w}}:=\mathcal{G}(
\Theta, \mtrx{M}, \{\mathcal{S}^{\text{PEW}}_{\mu}(\cdot|\Gamma_{\ing{j}^{\star}})\}_{\mu}, 
\mu_{\ing{j}^{\star}}, F_{\alpha}, \mtrx{Z}_{\ing{j}^{\star}},  \beta, \ident{\ing{i}_{\text{max}}}
).
\] 
By ``warm-starting'', we mean that the initialization parameters are taken as $\mu^{(0)} =\mu_{\ing{j}^{\star}} $ and $\mtrx{Z}^{(0)}=\mtrx{Z}_{\ing{j}^{\star}}$ where the $\star$ denotes the values at the end of the $\Gamma_{\ing{j}^{\star}}$ selection procedure.


The pseudo-code of the adaptive social declippers for a given block of adjacent frames $\mtrx{Y} \in \Rset{\ing{L}}{\ing{(2b+1)}}$ is given in \autoref{alg:AUDASCITYDeclip}. Again, for the analysis version $\hat{\mtrx{x}} := \hat{\mtrx{w}}$, while for the synthesis version $\hat{\mtrx{x}} := \mtrx{D}\hat{\mtrx{w}}$,  and the central frame of the original signal is estimated as $\hat{\vect{x}} = \hat{\mtrx{x}}(:,\ing{b}+1)$.

\begin{algorithm}[!t]
	\caption{Adaptive Social Sparse Declippers 	\label{alg:AUDASCITYDeclip}}
	
	\begin{algorithmic} 
		\REQUIRE $\mtrx{Y}$, $\varepsilon$, $\mtrx{A}$ or $\mtrx{D}$, $\left\{\Gamma_{\ing{k}}\right\}_{\ing{k}}$, $\{\iter{\mu_{\ing{k}}}{0}\}_{\ing{k}}$, $\alpha,\beta$, \ident{$\ing{i}_{\text{init}}$, $\ing{i}_{\text{max}}$}
		\STATE set parameters from \autoref{tab:SocialSparseDeclip}, $\alpha=1$
		\FORALL{$\ing{k}$}
		\STATE
		\STATE $\begin{bmatrix}
		\hat{\mtrx{w}}_{\ing{k}}\\ 
		\mu_{\ing{k}}\\ 
		\mtrx{Z}_{\ing{k}}
		\end{bmatrix} {\small \gets} 
		\mathcal{G}(\Theta, \mtrx{M},\{\mathcal{S}_{\mu}^{\text{PEW}}(\cdot|\Gamma_{\ing{k}})\}_{\mu}, \iter{\mu_{\ing{k}}}{\ing{0}}, F_{\alpha}, \iter{\mtrx{z}}{0}, \beta,\ident{\ing{i}_{\text{init}}})$
		\STATE
		\STATE Compute $\var{e}_{\ing{k}}$ as in~\eqref{eq:entropy}
		\ENDFOR
		\STATE $\ing{k}^{\star} := \argmax_{\ing{k}} \var{e}_{\ing{k}}$, $\alpha=0.99$
		\STATE $\hat{\mtrx{w}}:=\mathcal{G}(
		\Theta, \mtrx{M}, \{\mathcal{S}^{\text{PEW}}_{\mu}(\cdot|\Gamma_{\ing{k}^{\star}})\}_{\mu}, 
		\mu_{\ing{k}^{\star}}, F_{\alpha}, \mtrx{Z}_{\ing{k}^{\star}},  \beta, \ident{\ing{i}_{\text{max}}}
		).$
		\RETURN $\hat{\mtrx{w}}$
	\end{algorithmic}
	
\end{algorithm}

\subsection{Overlap-add processing}
As the algorithms work on a frame-based manner, the declipped signal is obtained by overlap-add synthesis with a synthesis window satisfying the Constant OverLap-Add (COLA) constraint with respect to the analysis window. \ident{Particularly, we use the square root \emph{periodic} Hamming window both as an analysis and synthesis window, which satisfies COLA for the given $75$\% overlap between successive frames \cite{smithHamming, smithWOLA}. Hamming window is known for its strong suppression of the neighbouring sidelobes, hence it is an intuitive generic choice for the window function. Nevertheless, other types of windows could be more appropriate, depending on the spectral content of the target signal.}


\section{Experimental validation}
\label{sec:experiments}

To assess the performance of the 
\ident{variants} (analysis \rev{\emph{vs.}} synthesis, plain \rev{\emph{vs.}} structured, adaptive or not) of the framework described in the previous section, and to compare it with state-of-the-art approaches at various degradation levels, we conduct several sets of experiments. After describing the data sets and performance measures, we describe and analyze a first small scale experiment on the SMALL data set. The lessons from this experiment lead to the design of a larger-scale experiment which is analyzed in further details.

\subsection{Datasets}
\label{subsec:Datasets}

\begin{table}
	\centering
	\caption{Test data summary. All data are single-channel with 16 kHz sampling rate.}
	\label{tab:testData}
	\resizebox{\columnwidth}{!}{%
		\begin{tabular}{l|c|c|c|c|c|c|c|}
			\cline{2-8}
			& SMALL & \multicolumn{5}{|c|}{RWC} & TIMIT     \\ 
			\cline{3-7}
			&  & \emph{Pop} & \emph{Jazz} & \emph{Chamber} & \emph{Orchestra} & \emph{Vocals} &  \\ \hline
			\multicolumn{1}{|l|}{Duration [s]} & 100 &  3000  &   3000    &   3150  &  2700  & 1320  &   600      \\ \hline
			\multicolumn{1}{|l|}{Excerpts Nb.}  &  20   & 100 &  50  &    35     &    9    &   6    &  135    \\ \hline
		\end{tabular}
	}
\end{table}

\begin{figure*}[!htbp]
	\centering
	\subfloat[SMALL: Music ($\Delta$SDR)\label{fig:SMALLMusicSDR}]{\includegraphics[trim=1cm 8.5cm 2cm 9.5cm,clip,height=3.2cm]{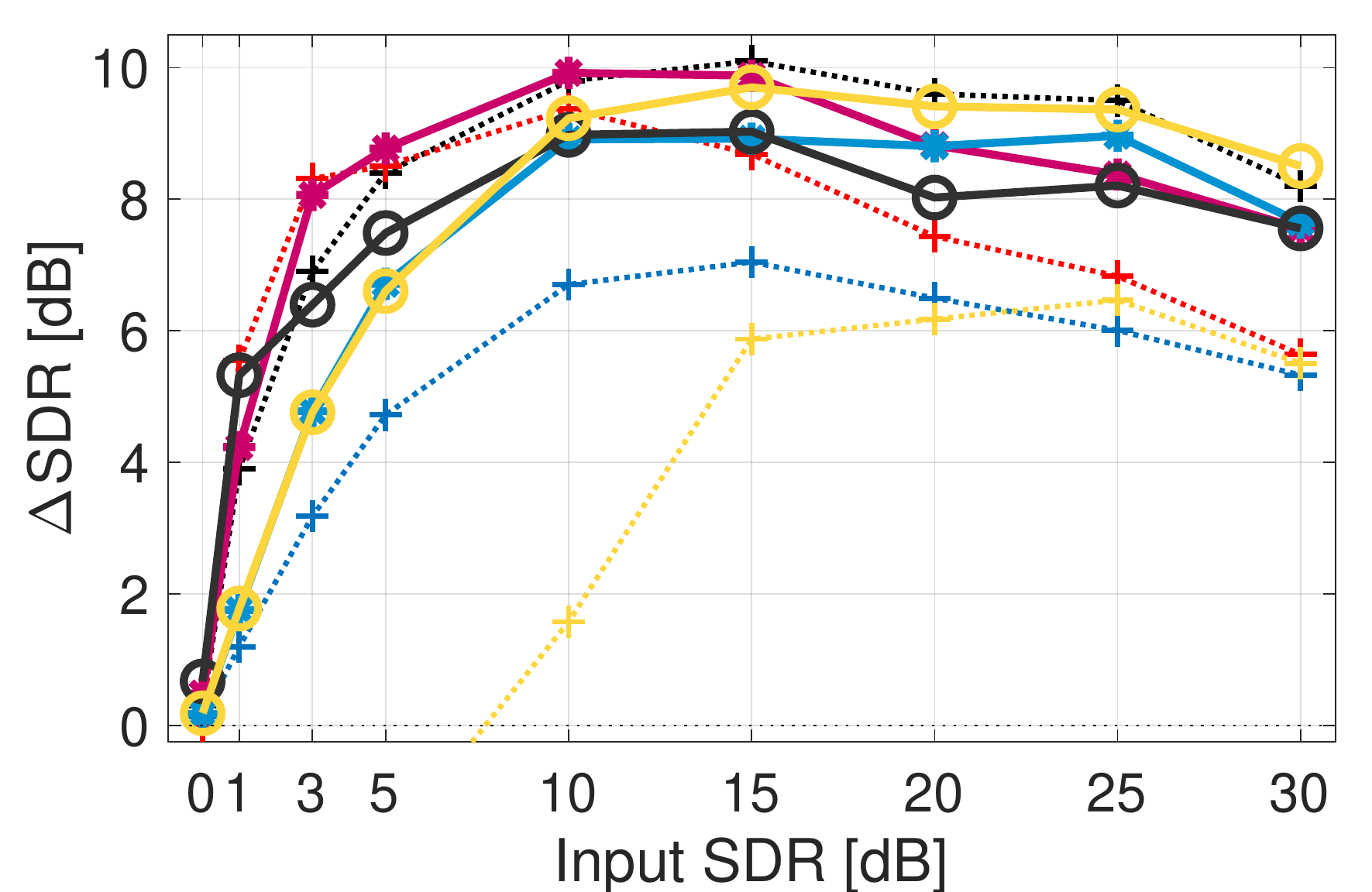}}\hfill\hspace{-1.1cm}
	\subfloat[SMALL: Music ($\Delta$PEAQ)\label{fig:SMALLMusicSOAPEAQDec}]{\includegraphics[trim=0.5cm 8.5cm 0.5cm 9.5cm,clip,height=3.2cm]{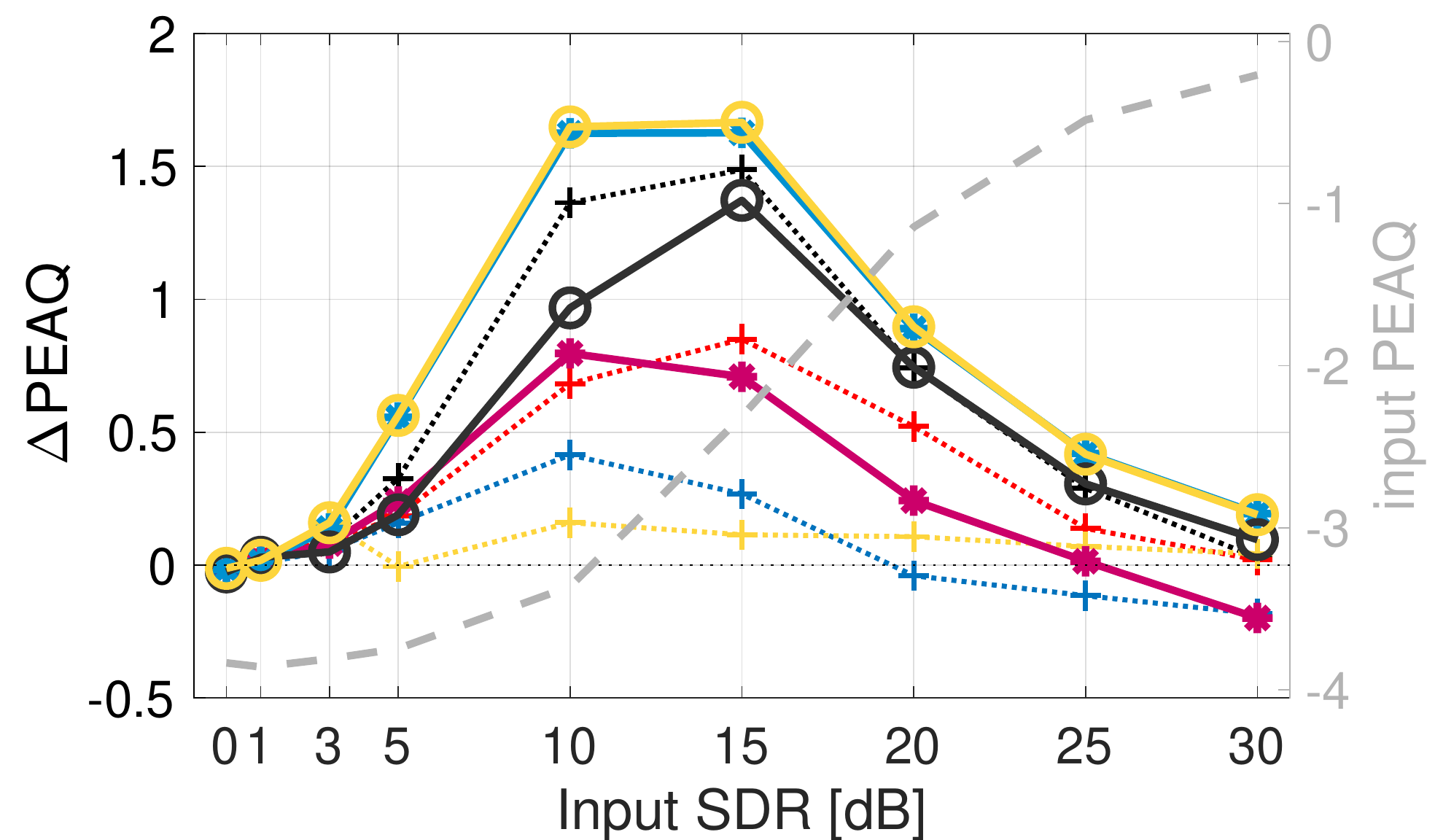}}\hfill
	\subfloat[SMALL: Legend \label{fig:SMALLLegend}]{\includegraphics[trim=5cm 10cm 1cm 10cm,clip,height=3cm]{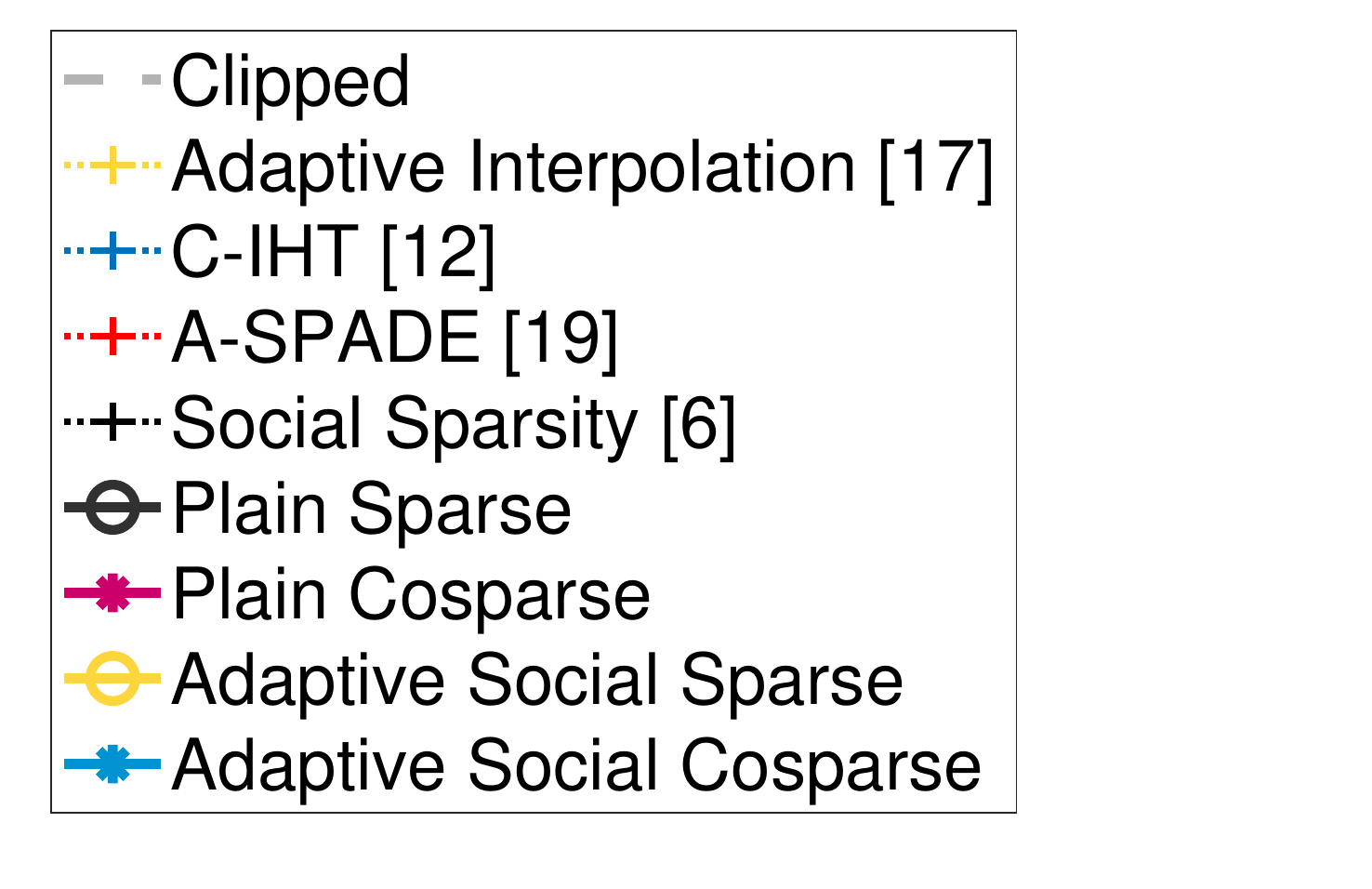}}\\
	\subfloat[SMALL: Speech ($\Delta$SDR) \label{fig:SMALLSpeechSDR}]{\includegraphics[trim=1cm 8.5cm 2cm 9.5cm,clip,height=3.2cm]{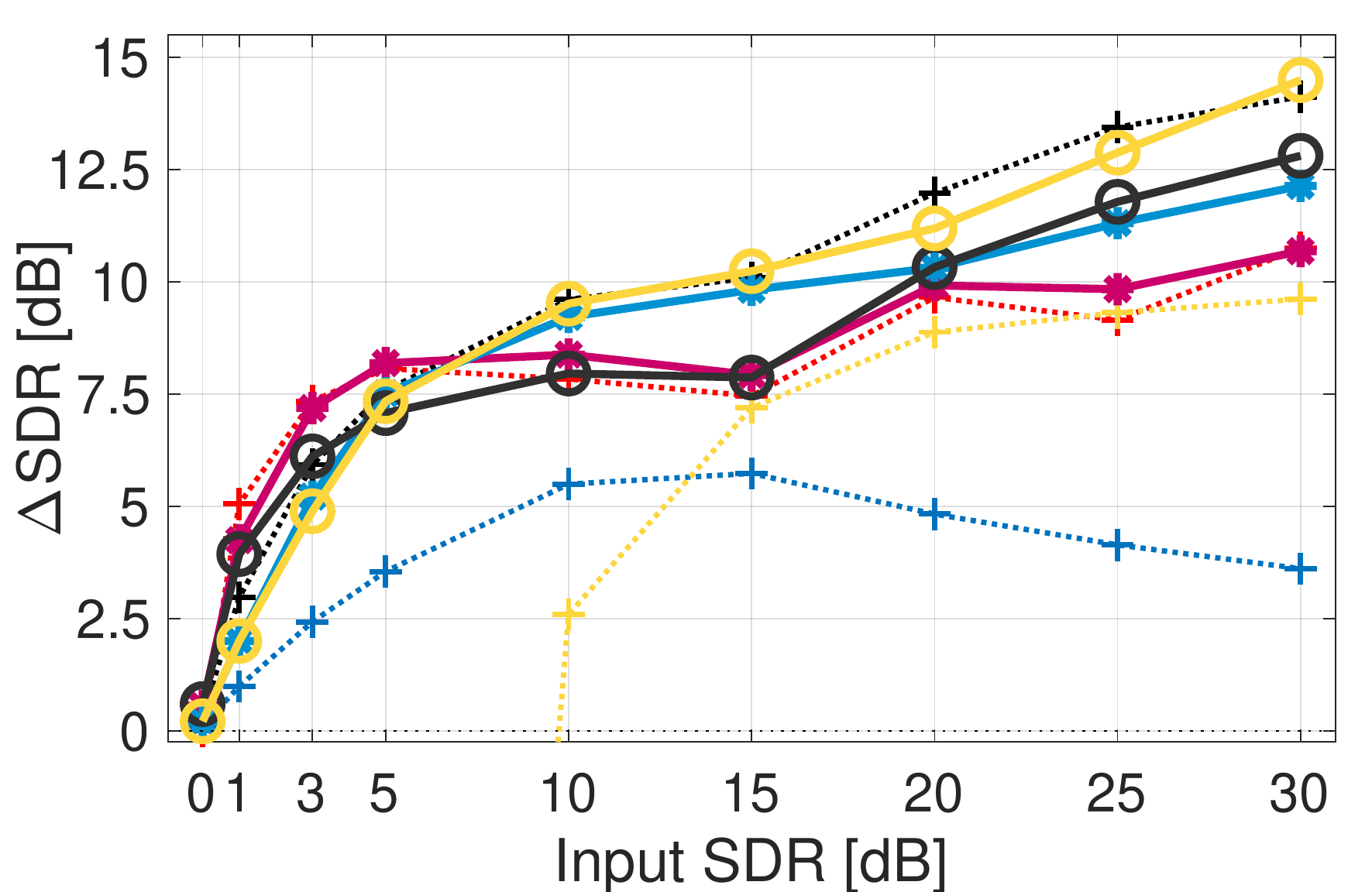}}\hfill
	\subfloat[SMALL: Speech ($\Delta$PESQ)\label{fig:SMALLSpeechSOAPESQDec}]{\includegraphics[trim=0.5cm 8.5cm 0.5cm 9.5cm,clip,height=3.2cm]{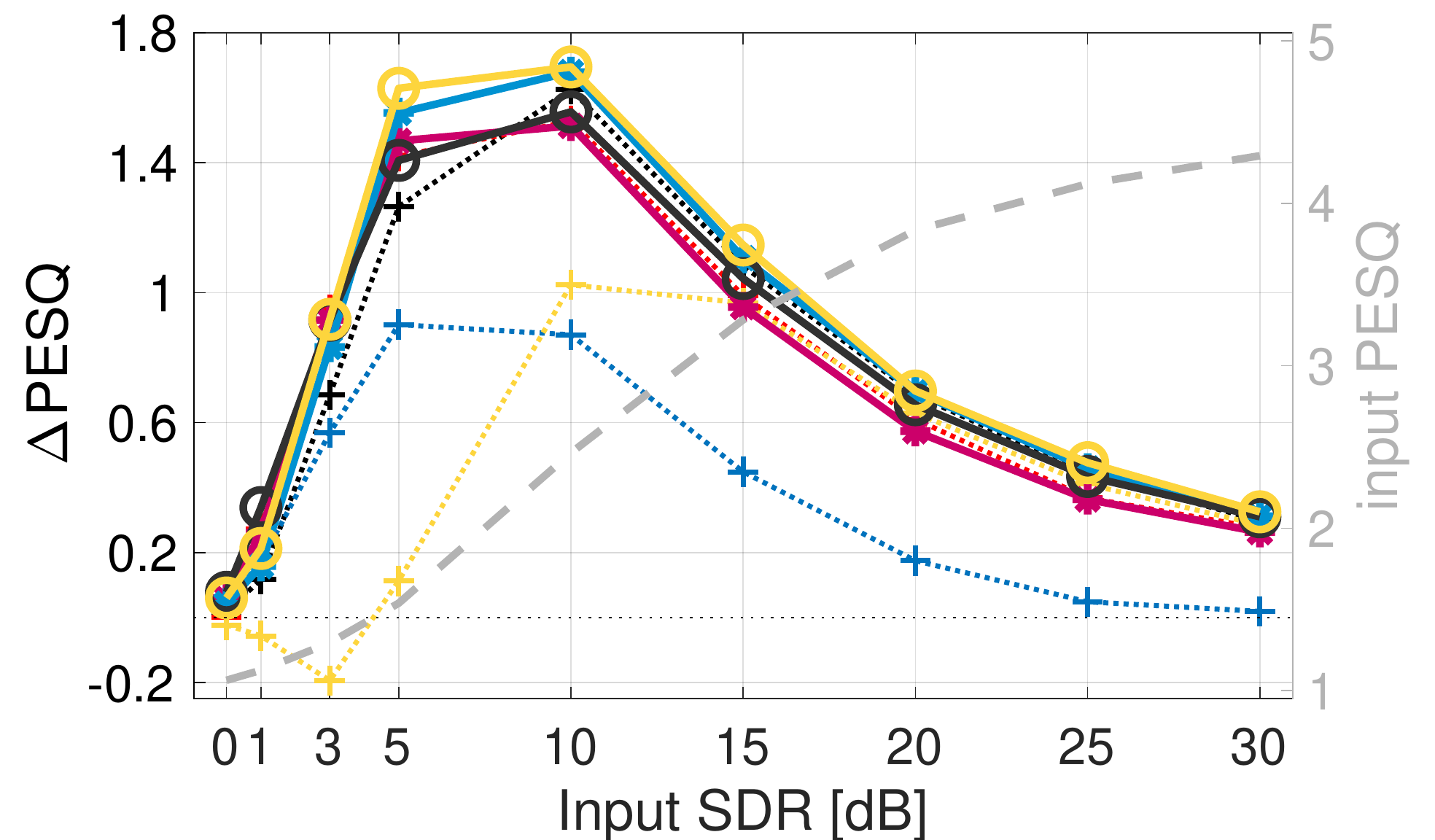}}\hfill
	\subfloat[SMALL: Speech ($\Delta$STOI)\label{fig:SMALLSpeechSOASTOIDec}]{\includegraphics[trim=0.5cm 8.5cm 0.5cm 9.5cm,clip,height=3.3cm]{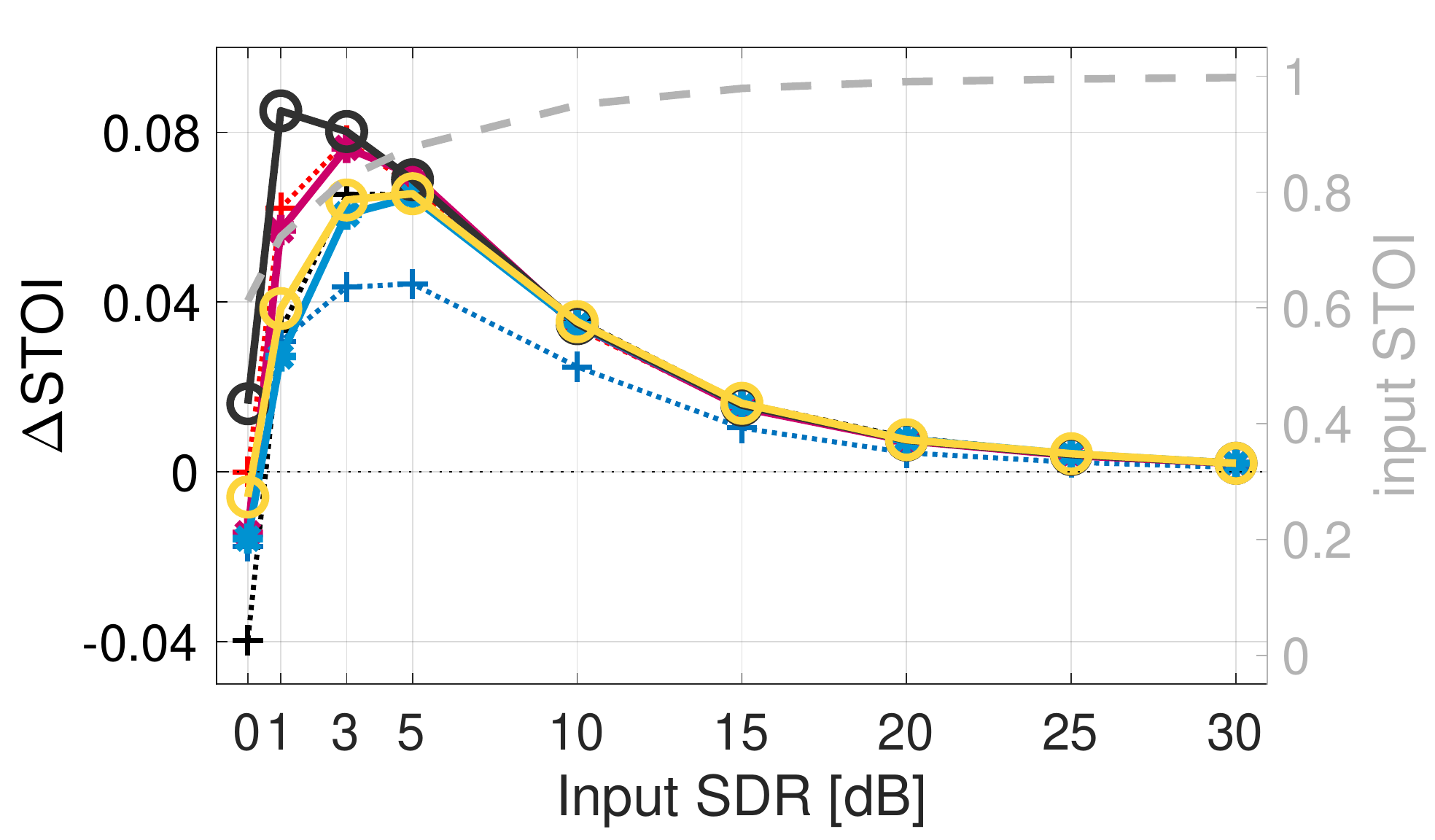}}
	\caption{State-of-the-art comparison (\ident{Signal-based} \& \ident{perceptually-motivated objective quality measures}) on SMALL dataset.\label{fig:SMALLSOADec}}
\end{figure*}

\begin{table*}%
	\centering
	\caption{State-of-the-art comparison on SMALL dataset (Mean $\pm$ std) of $\Delta$SDR }%
		\label{tab:SMALLSOAstd}%
	\ident{\subfloat[SMALL Music]{
	\addtolength{\tabcolsep}{-5.75pt}
	\resizebox{.5\textwidth}{!}{%
	\begin{tabular}{|l|c|c|c|c|c|c|c|c|}
		\hline
		\begin{tabular}[c]{@{}l@{}}Input\\ SDR\\ {[}dB{]}\end{tabular} & \multicolumn{1}{c|}{\begin{tabular}[c]{@{}c@{}}Adaptive\\ Interpolation\\ \cite{janssen1986adaptive}\end{tabular}} & \multicolumn{1}{c|}{\begin{tabular}[c]{@{}c@{}}C-IHT\\ \cite{kitic2013consistent}\end{tabular}} & \multicolumn{1}{c|}{\begin{tabular}[c]{@{}c@{}}A-SPADE\\ \cite{kitic2015sparsity}\end{tabular}} & \multicolumn{1}{c|}{\begin{tabular}[c]{@{}c@{}}Social\\ Sparsity\\ \cite{siedenburg2014audio}\end{tabular}} & \multicolumn{1}{c|}{\begin{tabular}[c]{@{}c@{}}Plain\\ Sparse\end{tabular}} & \multicolumn{1}{c|}{\begin{tabular}[c]{@{}c@{}}Plain\\ Cosparse\end{tabular}} & \multicolumn{1}{c|}{\begin{tabular}[c]{@{}c@{}}Adaptive\\ Social\\ Sparse\end{tabular}} & \multicolumn{1}{c|}{\begin{tabular}[c]{@{}c@{}}Adaptive\\ Social\\ Cosparse\end{tabular}} \\ \hline
		0 & -1.4$\pm$2.5 & 0.1$\pm$0.03 & 0.001$\pm$0.003 & 0.3$\pm$0.1 & 0.7$\pm$0.1 & 0.5$\pm$0.07 & 0.2$\pm$0.1 & 0.2$\pm$0.05 \\ \hline
		1 & -6.8$\pm$11.8 & 1.1$\pm$0.4 & 5.5$\pm$1.1 & 3.9$\pm$1.1 & 5.3$\pm$1.7 & 4.2$\pm$0.9 & 1.8$\pm$0.5 & 1.8$\pm$0.6 \\ \hline
		3 & -3.0$\pm$2 & 3.1$\pm$1.2 & 8.3$\pm$2.3 & 6.9$\pm$2.5 & 6.4$\pm$1.6 & 8.1$\pm$1.8 & 4.8$\pm$1.9 & 4.8$\pm$2 \\ \hline
		5 & -1.9$\pm$3.9 & 4.7$\pm$1.6 & 8.5$\pm$3.2 & 8.4$\pm$3.1 & 7.5$\pm$2.3 & 8.8$\pm$2.7 & 6.6$\pm$2.7 & 6.7$\pm$2.7 \\ \hline
		10 & 1.5$\pm$8.1 & 6.7$\pm$2.3 & 9.3$\pm$4.2 & 9.8$\pm$5.1 & 9.0$\pm$3.8 & 9.9$\pm$3.7 & 9.2$\pm$4.2 & 8.9$\pm$4 \\ \hline
		15 & 5.8$\pm$6.2 & 7.0$\pm$2.6 & 8.6$\pm$5.1 & 10.1$\pm$5.1 & 9.0$\pm$4.6 & 9.9$\pm$4.6 & 9.7$\pm$5.1 & 8.9$\pm$4.4 \\ \hline
		20 & 6.1$\pm$6.5 & 6.4$\pm$2.6 & 7.4$\pm$5.6 & 9.6$\pm$5 & 8.0$\pm$5.3 & 8.8$\pm$4.8 & 9.4$\pm$5.4 & 8.8$\pm$4.7 \\ \hline
		25 & 6.4$\pm$6.7 & 6.0$\pm$2.5 & 6.8$\pm$5.6 & 9.5$\pm$5.6 & 8.2$\pm$5.6 & 8.4$\pm$5 & 9.4$\pm$5.5 & 9.0$\pm$4.8 \\ \hline
		30 & 5.5$\pm$6.5 & 5.3$\pm$3.3 & 5.6$\pm$6.2 & 8.2$\pm$5.5 & 7.6$\pm$5.6 & 7.6$\pm$5.4 & 8.5$\pm$5.8 & 7.6$\pm$4.8 \\ \hline
	\end{tabular}%
	}
	}}%
	\subfloat[SMALL Speech]{
		\addtolength{\tabcolsep}{-5.75pt}
		\resizebox{.5\textwidth}{!}{
		\begin{tabular}{|l|c|c|c|c|c|c|c|c|}
			\hline
			\begin{tabular}[c]{@{}l@{}}Input\\ SDR\\ {[}dB{]}\end{tabular} & \begin{tabular}[c]{@{}c@{}}Adaptive\\ Interpolation\\ \cite{janssen1986adaptive}\end{tabular} & \begin{tabular}[c]{@{}c@{}}C-IHT\\ \cite{kitic2013consistent}\end{tabular} & \begin{tabular}[c]{@{}c@{}}A-SPADE\\ \cite{kitic2015sparsity}\end{tabular} & \begin{tabular}[c]{@{}c@{}}Social\\ Sparsity\\ \cite{siedenburg2014audio}\end{tabular} & \begin{tabular}[c]{@{}c@{}}Plain\\ Sparse\end{tabular} & \begin{tabular}[c]{@{}c@{}}Plain\\ Cosparse\end{tabular} & \begin{tabular}[c]{@{}c@{}}Adaptive\\ Social\\ Sparse\end{tabular} & \begin{tabular}[c]{@{}c@{}}Adaptive\\ Social\\ Cosparse\end{tabular} \\ \hline
			0 & -241.4$\pm$6.8 & 0.1$\pm$0.01 & 0.0$\pm$0 & 0.2$\pm$0.05 & 0.6$\pm$0.08 & 0.6$\pm$0.02 & 0.2$\pm$0.04 & 0.2$\pm$0.03 \\ \hline
			1 & -253.9$\pm$11.3 & 0.9$\pm$0.2 & 5.0$\pm$0.8 & 3.8$\pm$0.7 & 3.9$\pm$0.5 & 4.2$\pm$0.3 & 2.0$\pm$0.4 & 2.0$\pm$0.4 \\ \hline
			3 & -198.3$\pm$105.9 & 2.4$\pm$0.4 & 7.3$\pm$1.5 & 6.9$\pm$1.1 & 6.1$\pm$0.6 & 7.2$\pm$1.1 & 4.9$\pm$0.9 & 5.2$\pm$1.1 \\ \hline
			5 & -46.6$\pm$98.1 & 3.5$\pm$0.5 & 8.0$\pm$2.1 & 7.8$\pm$1.8 & 7.1$\pm$2.3 & 8.2$\pm$1.7 & 7.3$\pm$2 & 7.5$\pm$2 \\ \hline
			10 & 2.5$\pm$3.1 & 5.4$\pm$0.9 & 7.8$\pm$4.2 & 9.6$\pm$2.2 & 8.0$\pm$3.7 & 8.4$\pm$3.6 & 9.5$\pm$2.4 & 9.2$\pm$2.2 \\ \hline
			15 & 7.1$\pm$4.8 & 5.7$\pm$1.1 & 7.4$\pm$5.3 & 10.2$\pm$3.3 & 7.9$\pm$4.9 & 7.9$\pm$4.9 & 10.2$\pm$3.3 & 9.8$\pm$2.7 \\ \hline
			20 & 8.8$\pm$5 & 4.8$\pm$2.3 & 9.6$\pm$5.4 & 12.2$\pm$5.4 & 10.3$\pm$6.5 & 9.9$\pm$5.5 & 11.2$\pm$6.5 & 10.3$\pm$4.7 \\ \hline
			25 & 9.3$\pm$6.3 & 4.1$\pm$3.5 & 9.1$\pm$7.1 & 13.3$\pm$6.1 & 11.8$\pm$8.3 & 9.8$\pm$7.1 & 12.9$\pm$6.8 & 11.3$\pm$5.1 \\ \hline
			30 & 9.6$\pm$6.5 & 3.6$\pm$2.2 & 10.7$\pm$6.4 & 14.4$\pm$7.9 & 12.8$\pm$8.1 & 10.7$\pm$6 & 14.5$\pm$8.2 & 12.1$\pm$5.3 \\ \hline
		\end{tabular}%
		}	
	}
\end{table*}

State-of-the-art comparisons are first conducted on the 20 audio examples from the SMALL dataset described in the introduction, for its wide use in the literature, and as a preliminary tuning for subsequent experiments.

Then, the proposed large scale objective experiments are conducted on excerpts from the RWC Music Database \cite{goto2002RWC}. We use the ``Pop'' and ``Jazz'' genres as is, and subcategorize the ``Classic'' genre (Vocals, Chamber, Symphonies), leading to 5 subsets. All the tracks are sufficiently diverse to reflect the robustness of the approach on different audio content. We also perform experiments on excerpts of the TIMIT database \cite{garofolo1993darpa} for evaluation on speech content.

All the audio examples used for objective experiments are either natively sampled at 16 kHz or down-sampled to 16 kHz, and down-mixed (by simple averaging of the two channels) to provide single-channel data. Each excerpt was normalized in amplitude ($\norm{ \text{vec}(\mtrx{x})}{\infty} = 1$) then artificially clipped under the hard-clipping model in \eqref{eq:ClipModel} to meet a given input SDR. \autoref{tab:testData} summarizes the data used for these experiments.

\rev{
\begin{remark}
Since existing small scale benchmarks were conducted at 16 kHz, and given the substantial increase in computational resources that would be needed to conduct benchmarks on 44.1 kHz audio, we keep the 16 kHz format for the large-scale experiments, putting the focus on the diversity of considered musical genres. A different option is to consider less diverse audio files at 44.1 kHz. This is the complementary approach of the parallel and independent survey  \cite{Zaviska:2020ua}.
\end{remark}}

Additionally, we perform 
listening tests based on the MUltiple Stimuli with Hidden Reference and Anchor (MUSHRA) evaluation procedure \cite{MushraITU}. We use the webMUSHRA framework \cite{schoeffler2018webmushra}
and use saturated versions of some original \rev{44.1 kHz} stereo excerpts of the 5 RWC subsets, used also \rev{(in a 16 kHz monophonic version)} for objective \ident{experiments}. Further details are given in \autoref{subsec:Mushra}.

\subsection{Performance measure}

\rev{As mentioned before, in this work we use the SDR as an objective quality measure of a (de)clipped audio excerpt. Hence, an audio declipping algorithm can be considered performant if, when applied to a clipped signal with low input SDR, it recovers signal with significantly higher SDR.} This is naturally captured by the $\Delta$SDR performance measure.

On top of $\Delta$SDR as a first objective measure, 
\ident{we provide perceptually-motivated objective measures of speech quality based on}
MOS-PESQ scores~\cite{rix2001perceptual}, and \rev{analyze} a real-valued index predicting intelligibility through the Short-Time Objective Intelligibility (STOI)~\cite{taal2011algorithm}. For music, we 
\ident{compute the Objective Difference Grade (ODG) PEAQ scores~\cite{thiede2000peaq} on the entire collection of RWC excerpts.}
\ident{PEAQ scores are computed on 48 kHz upsampled version of the signals using the open-source implementation available with \cite{kabal2002peaq}.}
Finally, for all methods we compare computation times displaying the ratio to \rev{real-time} processing ($\times$RT).
\subsection{Small scale objective experiments}
In the following we compare the various instances of the proposed generic algorithmic framework with two baseline declipping methods: \emph{Adaptive interpolation} \cite{janssen1986adaptive} and \emph{Consistent Iterative Hard-Thresholding} (C-IHT,~\cite{kitic2013consistent}), as well as with state-of-the-art methods: the Social Sparsity Declipper \cite{siedenburg2014audio} and the \emph{Analysis SParse Audio DEclipper} (A-SPADE, \cite{kitic2015sparsity}). The \emph{plain cosparse} algorithm mainly differs from A-SPADE in terms of the relative norm criterion used to stop the procedure. This new choice provides substantial improvements for light to moderate clipping conditions (see \autoref{fig:SMALLMusicSDR}).

Common algorithm parameters are set as listed below.
\begin{itemize}
	\item Frame size \ident{$\ing{L}$ is the number of samples corresponding to the frame duration of $64$ ms for music, and $32$ ms for speech};
	\item \ident{Square-root periodic} Hamming windows, overlap: $ 75~\%$, for methods allowing frame-based processing;
	\item Maximum number of iterations $\ing{i}_{\max}=10^{6};$
	\item Analysis operator, $\mtrx{A} =$ twice redundant DFT;
	\item Synthesis operator, $\mtrx{D} =$ twice redundant inverse DFT ($\mtrx{D} = \htransp{\mtrx{A}}$).
\end{itemize}
\ident{As the temporal neighbourhood size is given by ${\ing{b} = 2\ing{T} + 1}$, setting  $\ing{T} = 5$ for music and $\ing{T} = 1$ for speech, yields the observation matrices $\mtrx{y}\in\Rset{L}{11}$ and $\mtrx{y}\in\Rset{L}{3}$, respectively.}

For each algorithm, we choose the stopping criterion $\beta \in \{10^{-1}, 10^{-2},10^{-3}, 10^{-4}\}$ that provides the best averaged SDR improvement. 
The value $\beta = 10^{-3}$ appeared as a good 
\ident{comon choice to allow all algorithms to be simulteanously close to their optimal performance. This value}
 will be re-used for all algorithms in the large-scale experiments described in the next section.

\begin{table*}[!ht]
	\centering
	\caption{Computational performance of declipping methods (Ratio to real-time processing $\times$RT)} 
	\label{tab:CompTimeMonodeclipSOA}
	\begin{tabular}{|l|c|c|c|c|c|c|c|c|c|}
		\hline
		\begin{tabular}[c]{@{}l@{}}Input\\ SDR\\ {[}dB{]}\end{tabular} & \begin{tabular}[c]{@{}c@{}}Adaptive\\ Interpolation\\ \cite{janssen1986adaptive} \end{tabular} & \begin{tabular}[c]{@{}c@{}}C-IHT\\ \cite{kitic2013consistent}\end{tabular} & \begin{tabular}[c]{@{}c@{}}A-SPADE\\ \cite{kitic2015sparsity}\end{tabular} & \begin{tabular}[c]{@{}c@{}}Social\\ Sparsity \cite{siedenburg2014audio}\\ LTFAT\end{tabular} & \begin{tabular}[c]{@{}c@{}}Social\\ Sparsity \cite{siedenburg2014audio}\\ Compiled LTFAT\end{tabular} & \begin{tabular}[c]{@{}c@{}}Plain\\ Sparse\end{tabular} & \begin{tabular}[c]{@{}c@{}}Plain\\ Cosparse\end{tabular} & \begin{tabular}[c]{@{}c@{}}Adaptive\\ Social\\ Sparse\end{tabular} & \begin{tabular}[c]{@{}c@{}}Adaptive\\ Social\\ Cosparse\end{tabular} \\ \hline
		0                                                              & 134.6                                                                      & 19.4                                                    & 0.07                                                      & 754.2                                                                     & 62.1                                                                               & 5.2                                                    & 10.1                                                     & 303.7                                                              & 296.7                                                                \\ \hline
		1                                                              & 82.2                                                                       & 19                                                      & 17.7                                                      & 753.3                                                                     & 62.2                                                                               & 6.9                                                    & 11                                                       & 236.7                                                              & 205.7                                                                \\ \hline
		3                                                              & 47                                                                         & 17.2                                                    & 24.9                                                      & 754.8                                                                     & 62.2                                                                               & 8.7                                                    & 12.8                                                     & 191.1                                                              & 165.4                                                                \\ \hline
		5                                                              & 30.5                                                                       & 14.9                                                    & 27.3                                                      & 754.5                                                                     & 62.1                                                                               & 9.5                                                    & 13.5                                                     & 302.9                                                              & 143.8                                                                \\ \hline
		10                                                             & 12.9                                                                       & 9.8                                                     & 24.4                                                      & 755.4                                                                     & 62.1                                                                               & 10.1                                                   & 13.3                                                     & 169.6                                                              & 121                                                                  \\ \hline
		15                                                             & 6.5                                                                        & 5.7                                                     & 16.5                                                      & 754.6                                                                     & 62.1                                                                               & 9.3                                                    & 11.4                                                     & 163.5                                                              & 180.5                                                                \\ \hline
		20                                                             & 3.8                                                                        & 3.4                                                     & 10.4                                                      & 756.1                                                                     & 62.2                                                                               & 7.6                                                    & 9.1                                                      & 70.2                                                               & 91.7                                                                 \\ \hline
		25                                                             & 2.2                                                                        & 2.2                                                     & 6.6                                                       & 754.6                                                                     & 62.1                                                                               & 2.2                                                    & 6.2                                                      & 56                                                                 & 47.4                                                                 \\ \hline
		30                                                             & 1.3                                                                        & 1.3                                                     & 4.2                                                       & 756.4                                                                     & 62.2                                                                               & 3.4                                                    & 4.1                                                      & 25.3                                                               & 22.4                                                                 \\ \hline
	\end{tabular}		
\end{table*}

\begin{figure*}[ht]
	\centering
	\subfloat[Plain sparse \label{fig:ITERPlainSparseDec}]{\includegraphics[trim=1cm 7.5cm 2cm 8.5cm,clip,width=0.5\columnwidth]{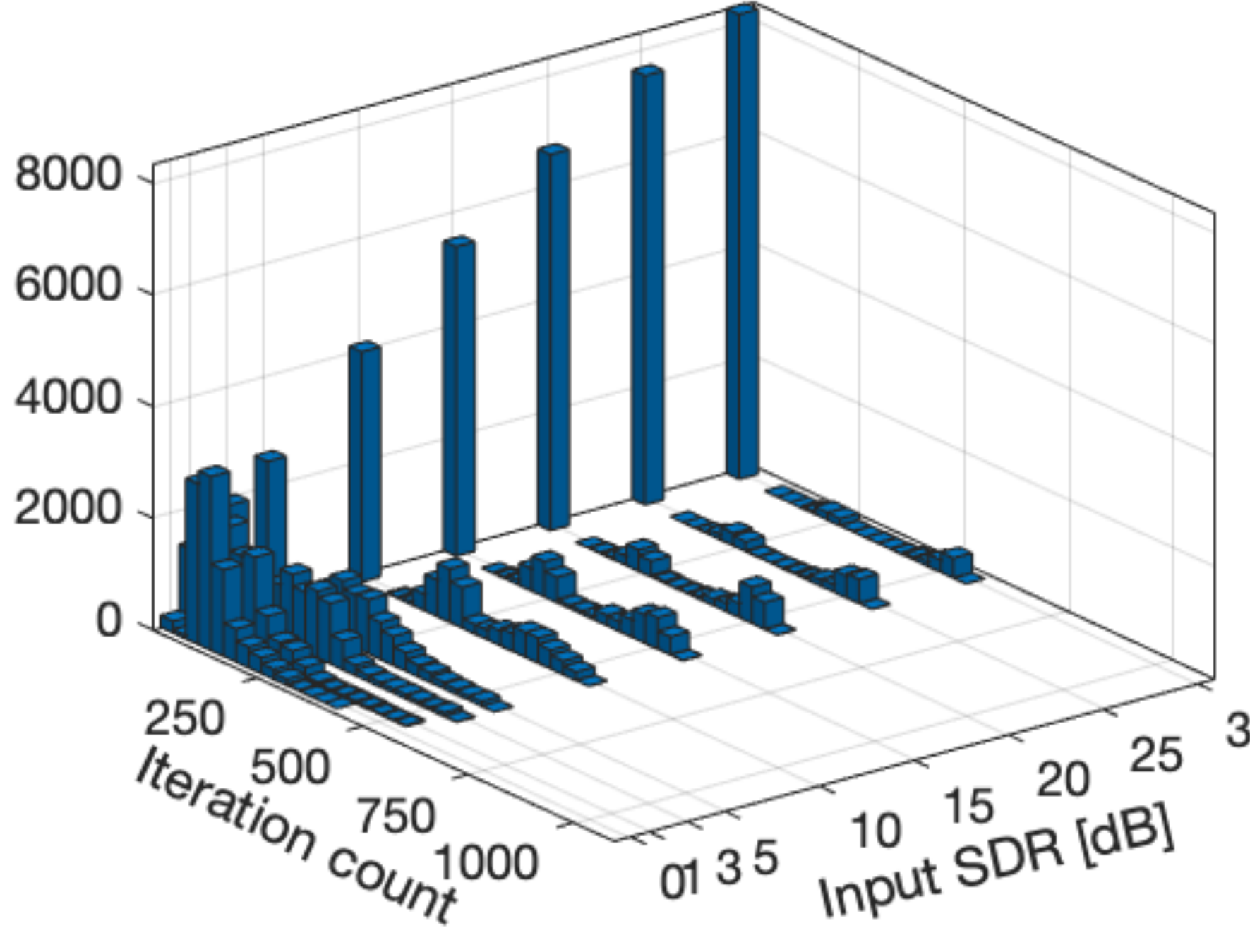}}
	~
	\subfloat[Plain Cosparse \label{fig:ITERPlainCosparseDec}]{\includegraphics[trim=1cm 7.5cm 2cm 8.5cm,clip,width=0.5\columnwidth]{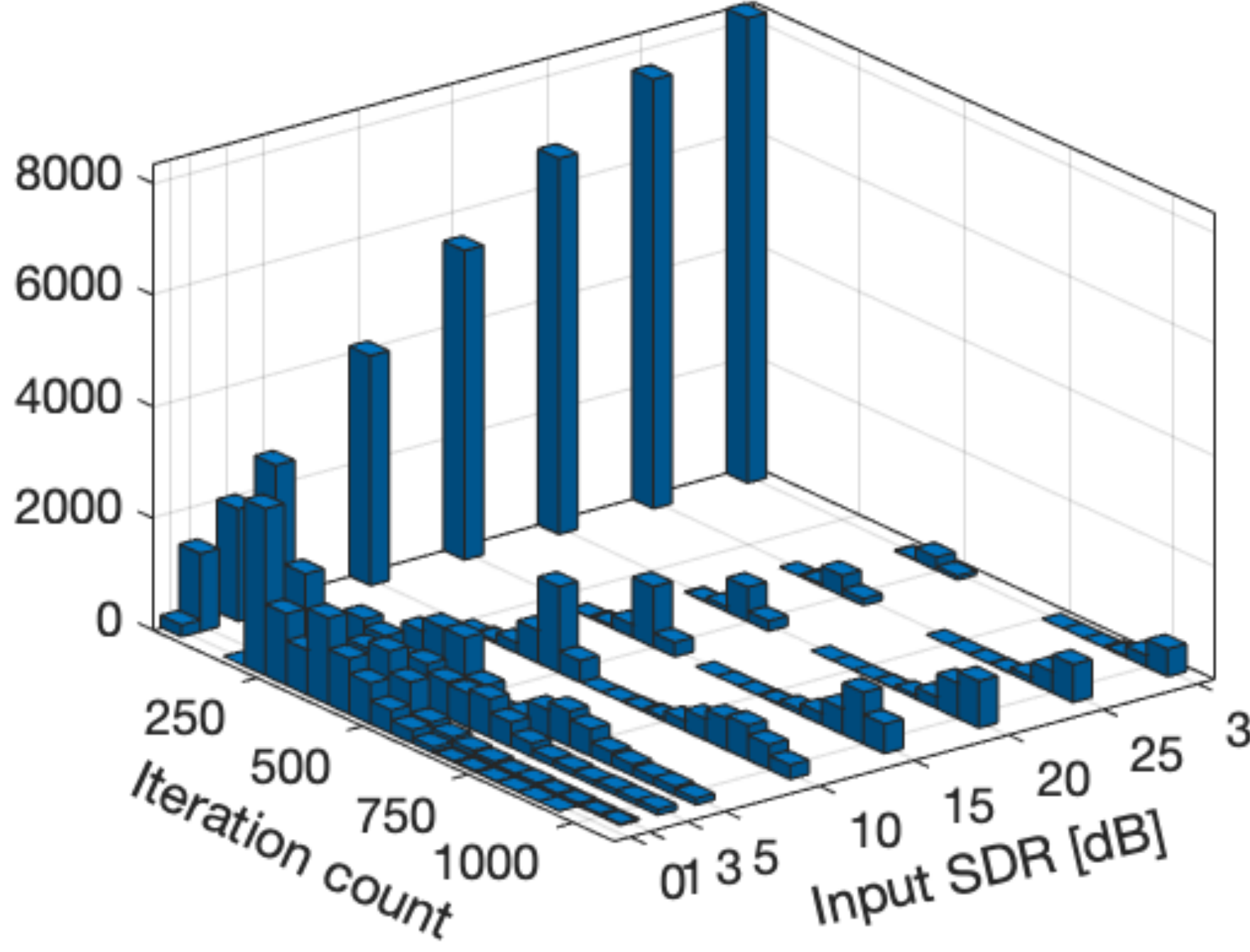}}
	~
	\subfloat[Adaptive social sparse \label{fig:ITERPEWSparseDec}]{\includegraphics[trim=1cm 7.5cm 2cm 8.5cm,clip,width=0.5\columnwidth]{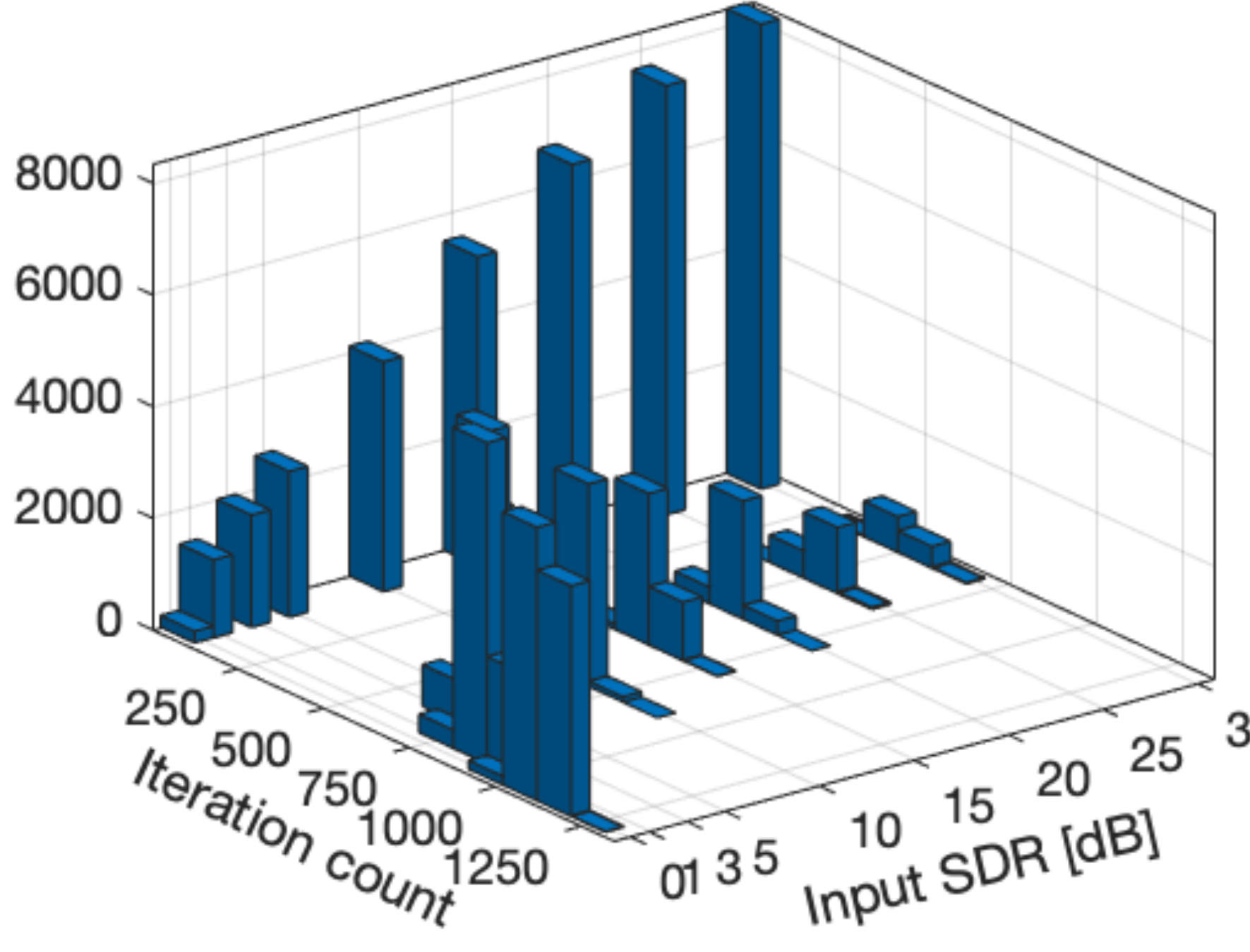}}
	~
	\subfloat[Adaptive social cosparse \label{fig:ITERPEWCosparseDec}]{\includegraphics[trim=1cm 7.5cm 2cm 8.5cm,clip,width=0.5\columnwidth]{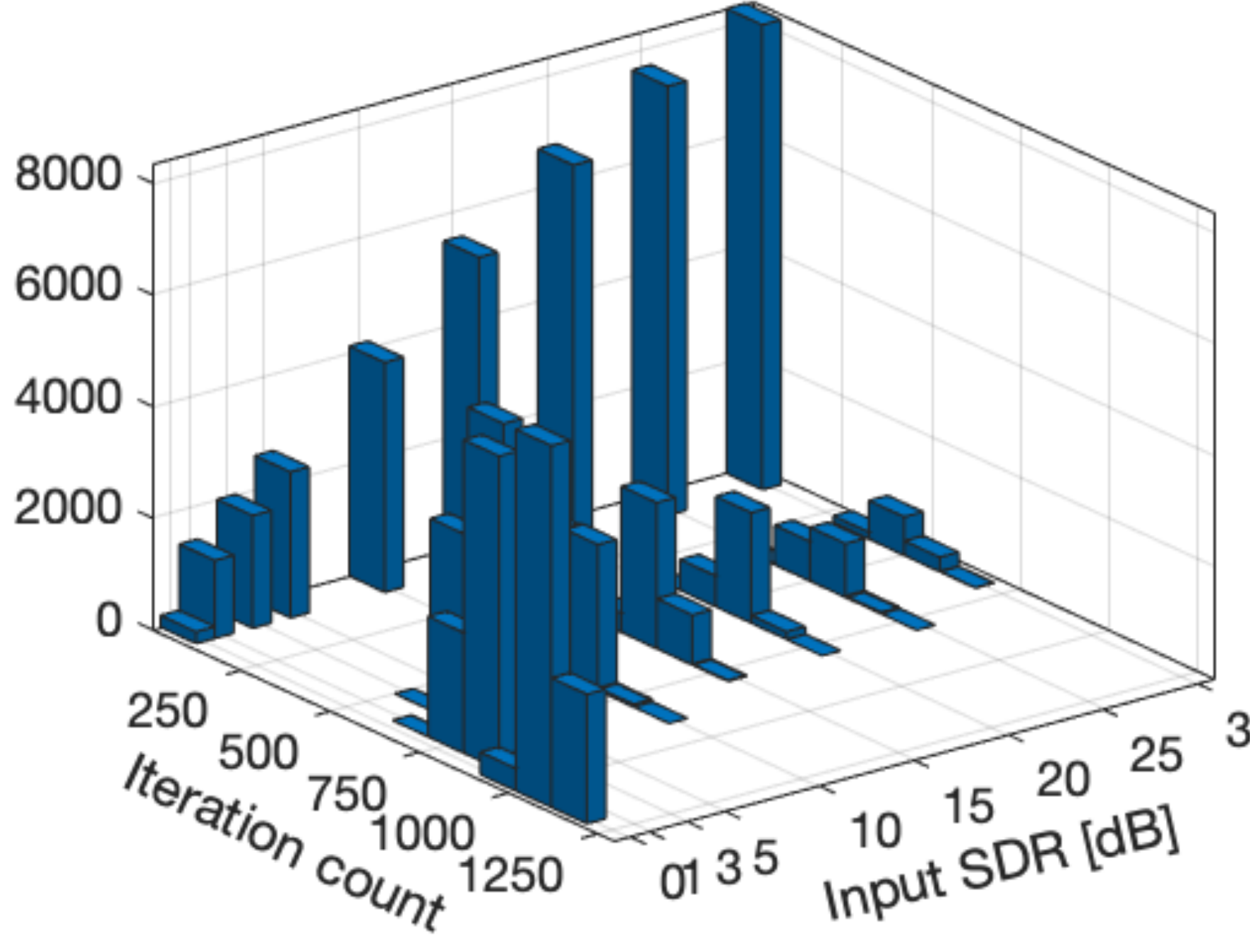}}
	\caption{Total iteration count distribution on SMALL dataset.}
	\label{fig:ITERDeclip}
\end{figure*}

The pattern $\Gamma$ for \ident{the non-adaptive social sparse method}
 is fixed as recommended in \cite{siedenburg2014audio} to obtain the best results. \ident{Namely, the patterns extend only in the temporal dimension ($\Gamma \in \Rset{1}{(2T+1)}$), where $\ing{T}=1$ for speech, and $\ing{T}=3$ for music signals.}
Considering the adaptive social (co)sparse declippers, we set the collection of time-frequency patterns $\left\{ \Gamma_{\ing{k}}\right\}_{\ing{k}=1}^{\ing{K}}$ to match the one presented on \autoref{fig:Stencils} for music and \autoref{fig:SpeechStencils} for speech. 

For each pattern, the initial shrinkage strength in \autoref{alg:AUDASCITYDeclip} is set as $\iter{\mu_{\ing{k}}}{0} := \norm{\Gamma_{\ing{k}}}{0}\times\left(1-\norm{\text{vec}(\mtrx{Y})}{\infty}\right)$. With this choice, the more severely clipped is the signal $\mtrx{Y}$, the lower its magnitude $\norm{\text{vec}(\mtrx{Y})}{\infty}$, hence the stronger the shrinkage. 
Moreover, given the expression \eqref{eqPEW} of PEW shrinkage, setting $\mu$ proportional to the sparsity degree $ \norm{\Gamma}{0}$ of the considered time-frequency neighborhood makes it act as a threshold on the \emph{average time-frequency power} in this neighborhood, rather than a threshold on the total energy in $\Gamma$.

Once the proper $\Gamma_{\ing{k}^{\star}}$ is selected, we obtained the best declipping results with $\mu$ following a geometric progression of common ratio $\alpha$ with $\alpha = 0.99$.


\autoref{fig:SMALLSOADec} displays average SDR improvements, predicted quality improvements ($\Delta$PEAQ, $\Delta$PESQ) and intelligibility improvements ($\Delta$STOI) for all the aforementioned methods on the SMALL examples. \ident{To supplement the graphical representation, the average and standard deviation of the $\Delta$SDR measure are given in Table~\ref{tab:SMALLSOAstd} . Based on these results, we} note that for severe \ident{clipping threshold}s (low input SDR), plain (co)sparse models provide better SDR improvements than social sparse models. \ident{In contrast}, we notice superiority of methods that include social sparse models for lighter degradation (input SDR $\geq 15$~dB) \ident{and for perceptually-motivated objective quality measure improvements ($\Delta$PEAQ)}.

\autoref{tab:CompTimeMonodeclipSOA}~presents computational efficiency corresponding to the results obtained in \autoref{fig:SMALLSOADec}. All the experiments were performed using a \Matlab implementation of the algorithms on a workstation equipped with a 2.4 GHz \ProcBluesy processor and 32 GB of RAM. We note that the method provided in \cite{siedenburg2014audio} uses the Structured Sparsity Toolbox\footnote{\tiny \url{https://homepage.univie.ac.at/monika.doerfler/StrucAudio.html}} relying on the Large Time-Frequency Analysis Toolbox\footnote{\tiny \url{http://ltfat.github.io}} \cite{ltfatnote015}. We note that for almost all methods, at input SDR of 5~dB or more, the processing time seems to be monotonically decreasing with the input SDR. The exception is for \cite{siedenburg2014audio} whose computational efficiency seems to be independent from the input SDR. The plausible explanation is that \ident{contrary} to the other methods, it only relies on an upper bound on the iteration count to stop the algorithm. Hence, the corresponding higher computational time could certainly be drastically reduced by lowering the maximum iteration count.

\begin{remark}
Tuning the parametrization of the plain (co)sparse methods (in particular increasing $\beta$, reducing the number of iterations or increasing the rate of decay of the shrinkage strength) makes it possible to achieve real-time processing on a regular laptop computer, possibly at the cost of tradeoffs with the resulting declipping performance. We also note that for \cite{siedenburg2014audio} some code optimization (\rev{\emph{i.e.}}, C backend for the LTFAT toolbox) can drastically improve the computational performance (see. the $\text{4}^\text{th}$column of \autoref{tab:CompTimeMonodeclipSOA}).
For the four methods presented here, the maximum number of iteration $\ing{i}_{\max}=10^{6}$ is never reached anyway as shown in \autoref{fig:ITERDeclip} for a given $\beta = 10^{-3}$.

\end{remark}
\subsection{Large scale objective experiments}

\begin{figure*}[t]
	\centerline{
		\subfloat[RWC: \emph{Pop}\label{fig:PopDec}]{\includegraphics[trim=1cm 8.5cm 2cm 9.5cm,clip,width=0.6\columnwidth]{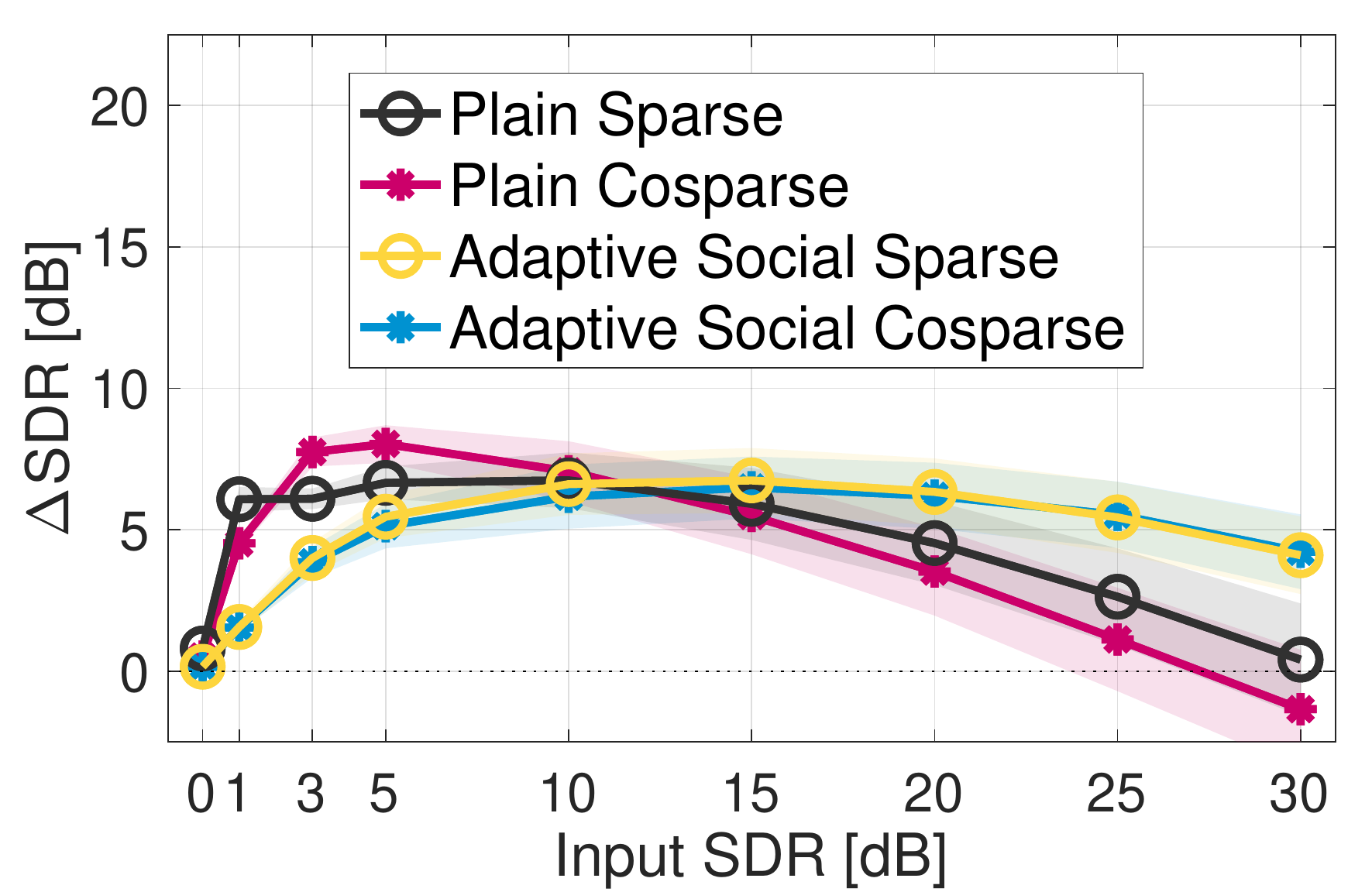}}
		\hfill
		\subfloat[RWC: \emph{Jazz}\label{fig:JazzDec}]{\includegraphics[trim=1cm 8.5cm 2cm 9.5cm,clip,width=0.6\columnwidth]{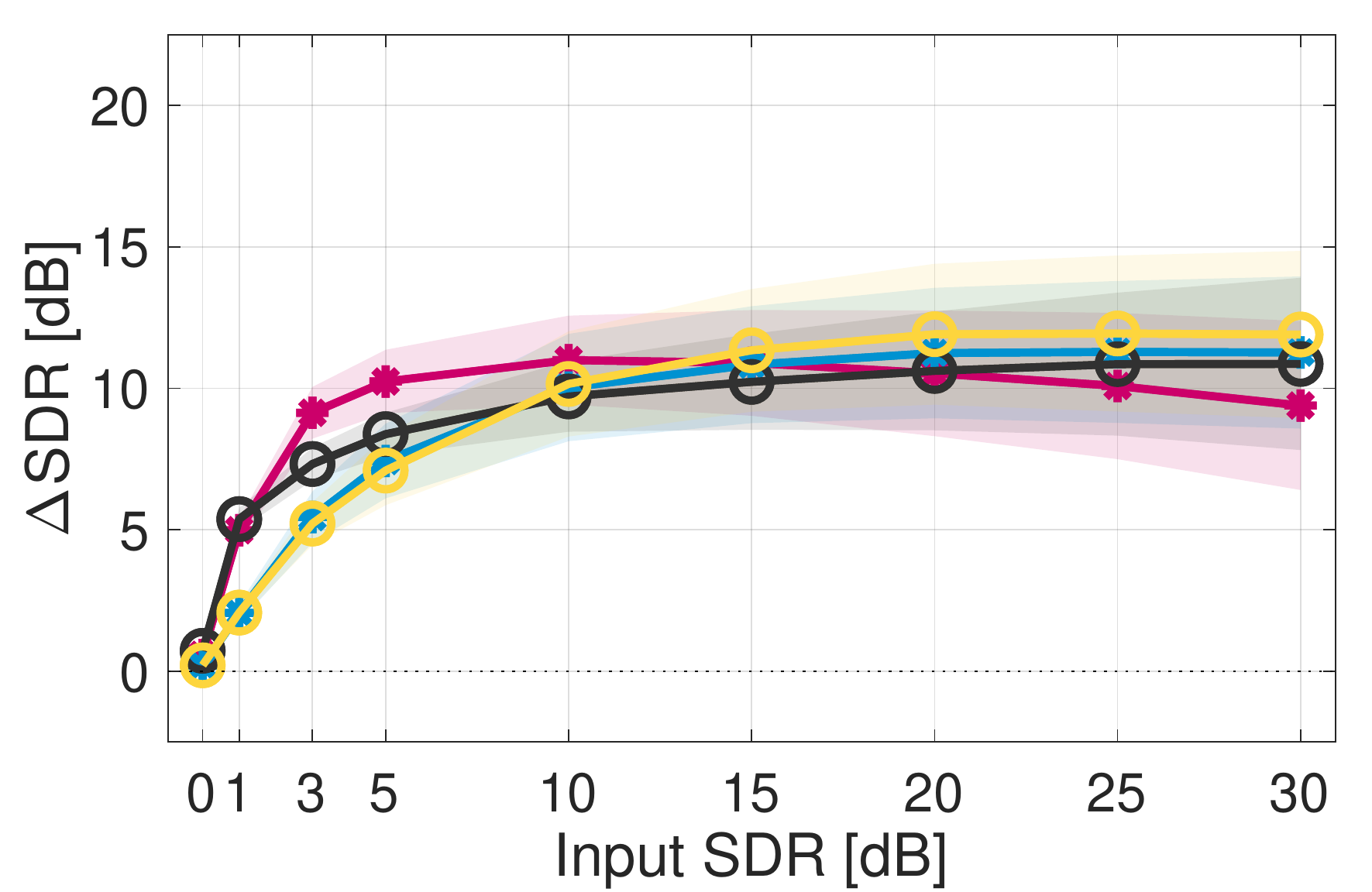}}
		\hfill
		\subfloat[RWC: \emph{Orchestra}\label{fig:SymphoniesDec}]{\includegraphics[trim=1cm 8.5cm 2cm 9.5cm,clip,width=0.6\columnwidth]{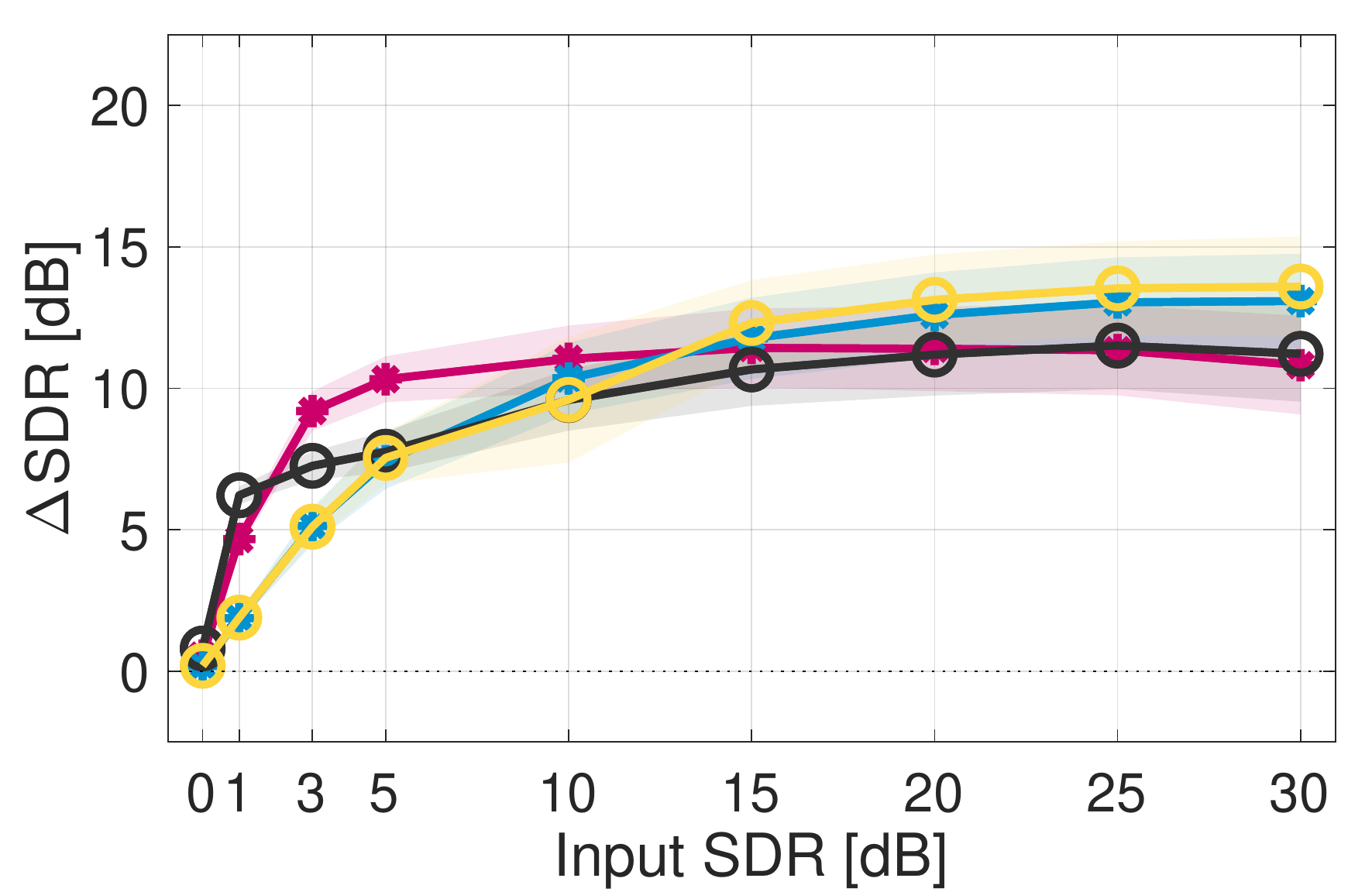}}
	}
	~\\
	\centerline{
		\subfloat[RWC: \emph{Vocals}\label{fig:VocalsDec}]{\includegraphics[trim=1cm 8.5cm 2cm 9.5cm,clip,width=0.6\columnwidth]{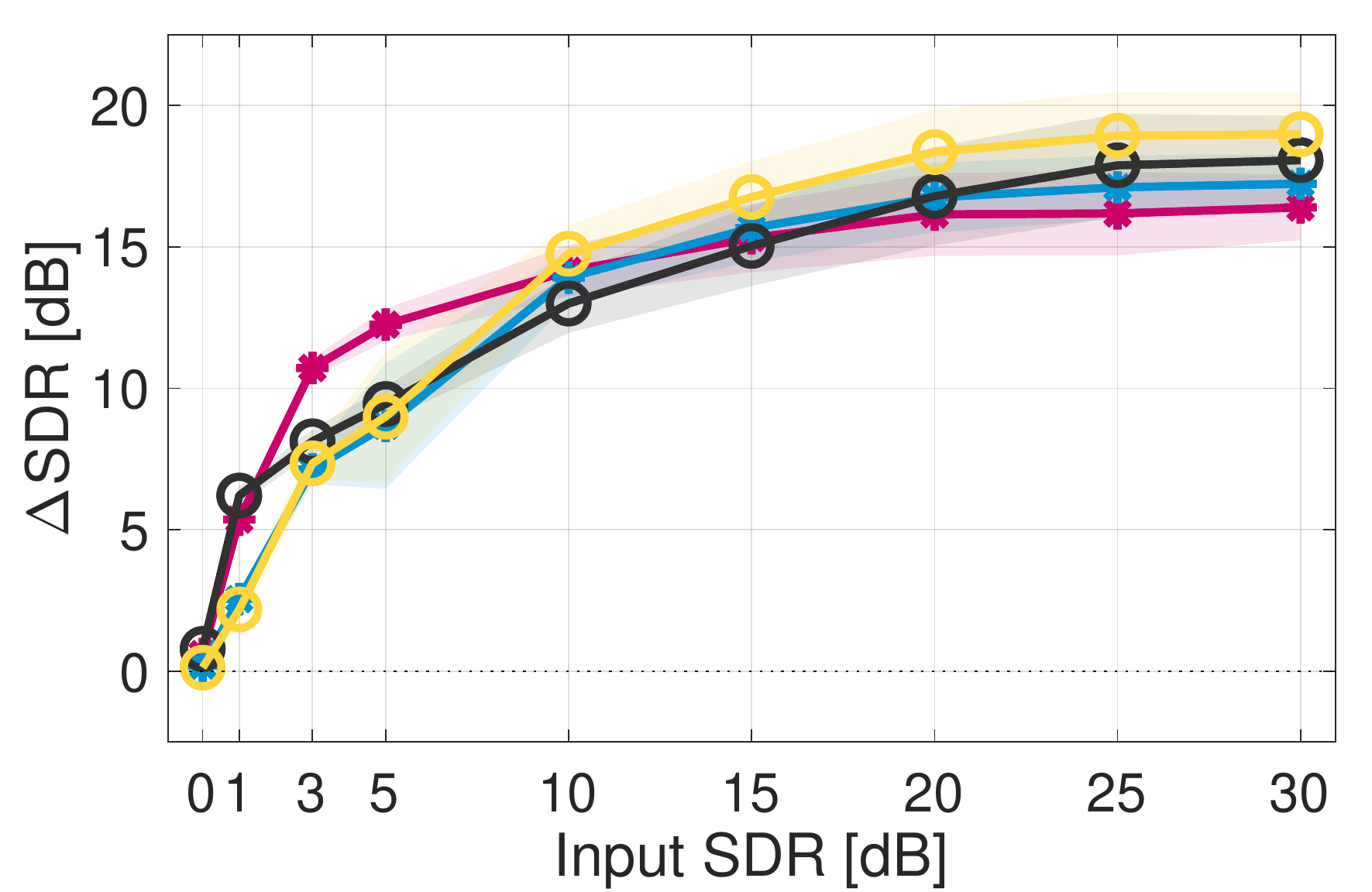}}
		\hfill
		\subfloat[RWC: \emph{Chamber}\label{fig:ChamberDec}]{\includegraphics[trim=1cm 8.5cm 2cm 9.5cm,clip,width=0.6\columnwidth]{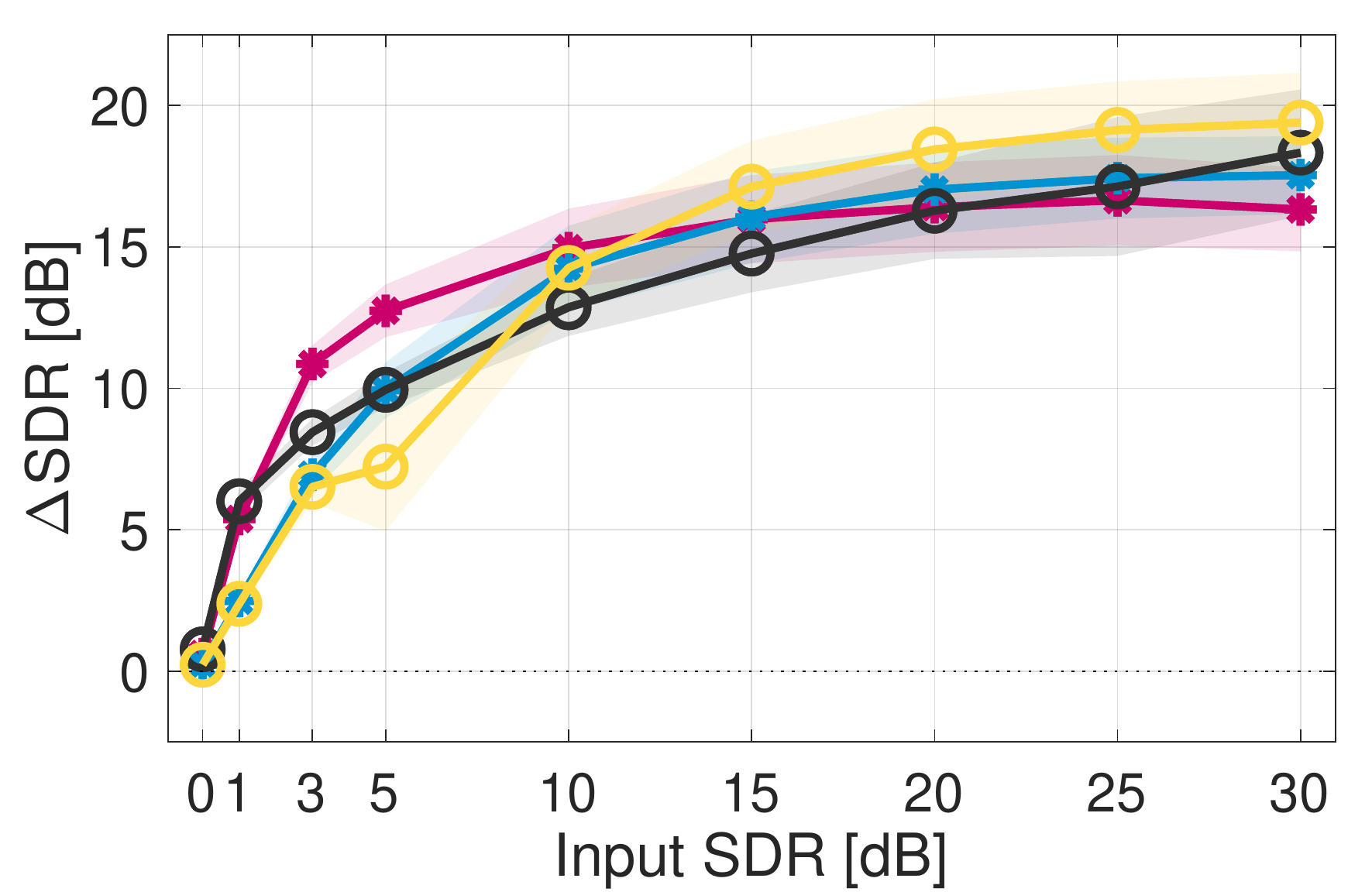}}
		\hfill
		\subfloat[TIMIT: Speech\label{fig:SpeechDec}]{\includegraphics[trim=1cm 8.5cm 2cm 9.5cm,clip,width=0.6\columnwidth]{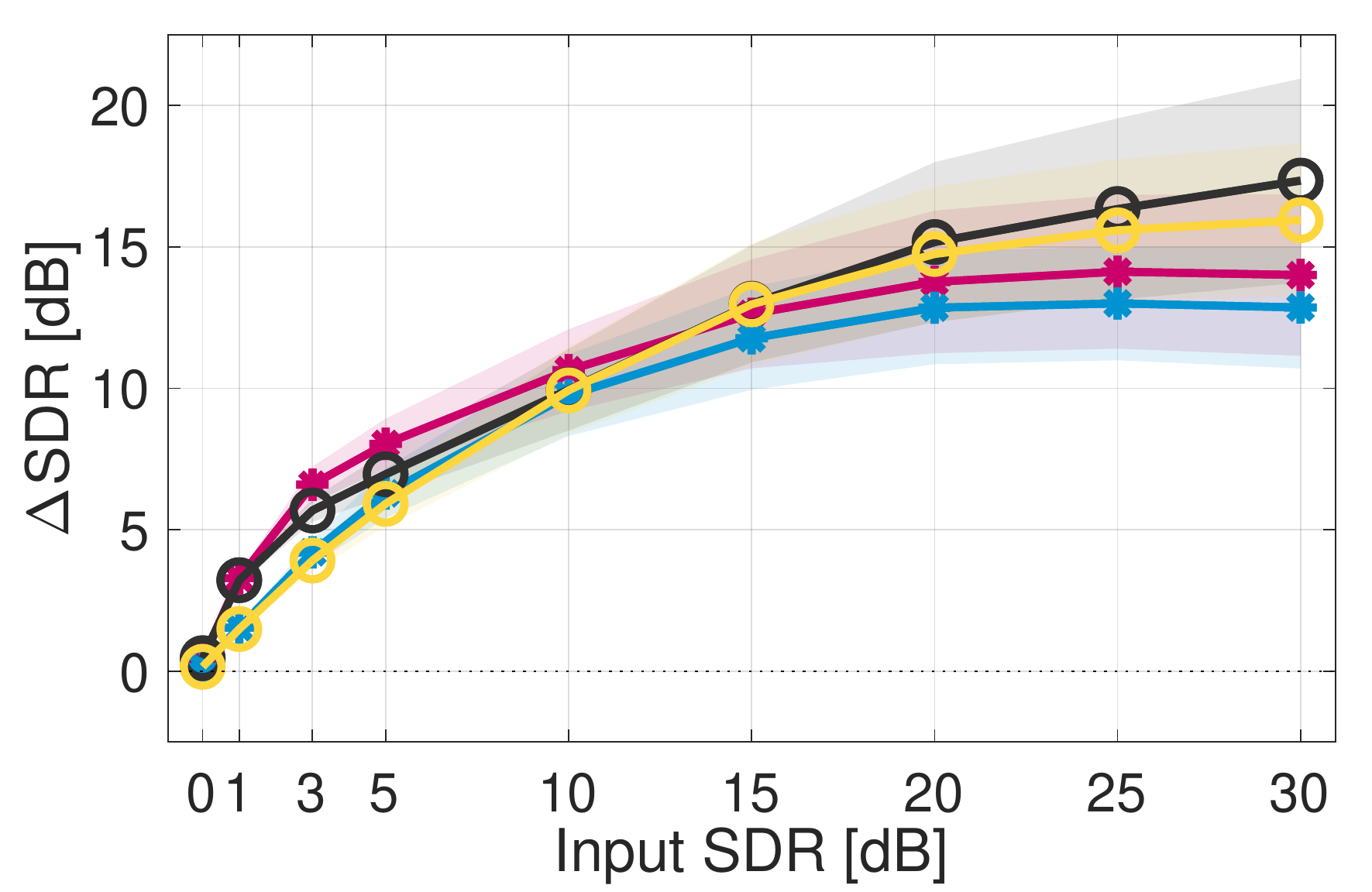}}
	}
	\caption{Performance on large-scale dataset: $\Delta$SDR {[}dB{]}}
	\label{fig:ResNumClip}
\end{figure*}
\begin{figure*}[ht]
	\centerline{
		\subfloat[TIMIT: $\Delta$STOI\label{fig:STOITimitDec}]{\includegraphics[trim=0.5cm 8.5cm 0.5cm 9.5cm,clip,width=0.6\columnwidth]{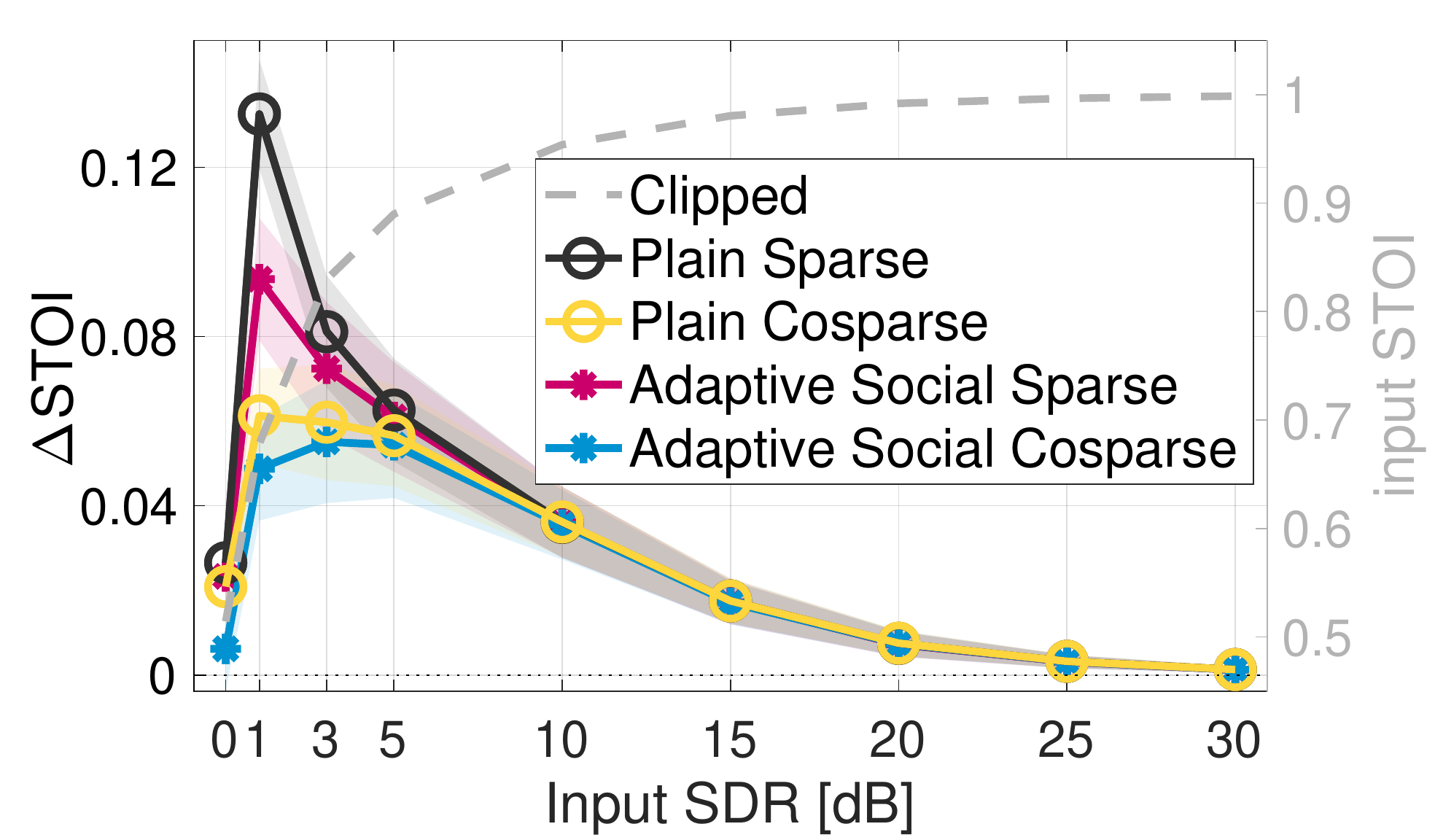}}
		\hfill
		\subfloat[TIMIT: $\Delta$PESQ\label{fig:PESQTimitDec}]{\includegraphics[trim=0.5cm 8.5cm 0.5cm 9.5cm,clip,width=0.6\columnwidth]{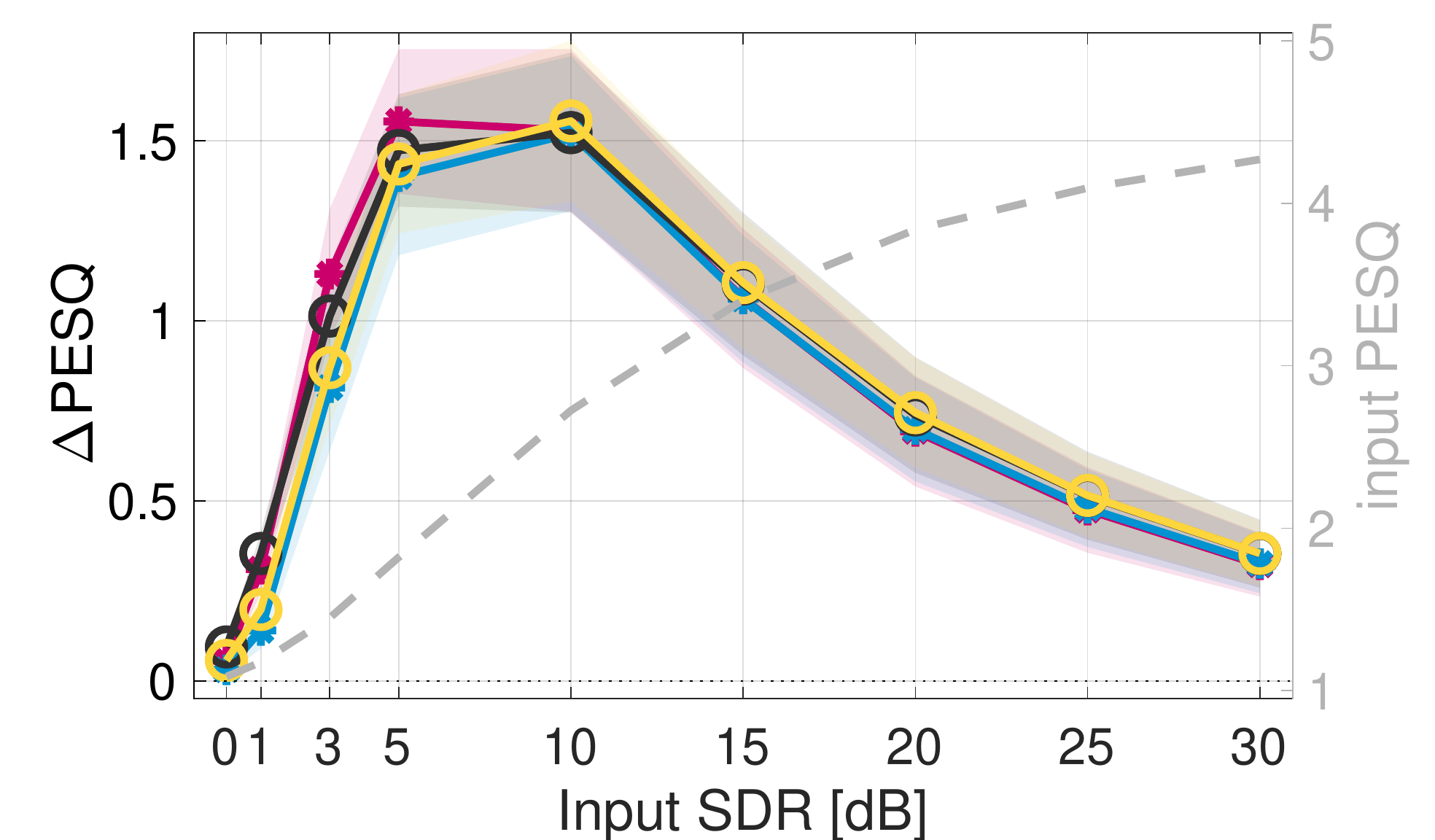}}
		\hfill
		\subfloat[RWC: $\Delta$PEAQ\label{fig:PEAQDecRWC}]{\includegraphics[trim=0.5cm 8.5cm 0.5cm 9.5cm,clip,width=0.6\columnwidth]{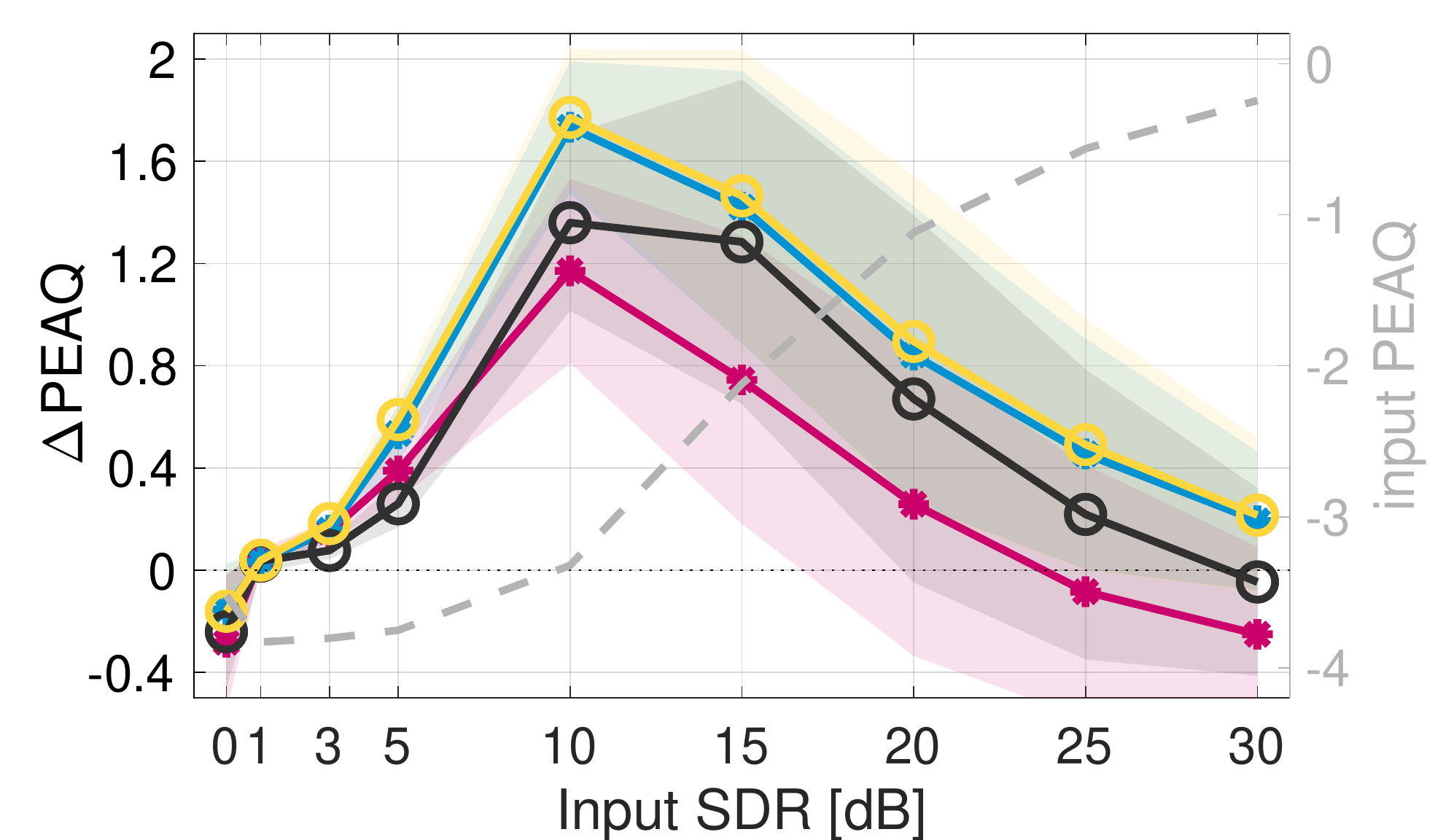}}
	}
	\caption{Performance on large-scale dataset: \ident{$\Delta$STOI, $\Delta$PESQ, $\Delta$PEAQ}}
	\label{fig:ResObjClip}
\end{figure*}

In order to accurately study the influence of the (social) (co)sparse models, we extend here the study to a wide-range comparison on RWC database excerpts. To the best of our knowledge, it is the first time such a large scale validation is performed. Results presented in \autoref{fig:ResNumClip} and \autoref{fig:ResObjClip} show averaged measurements over all available sounds. In order to assess the effect of different signal models only, we choose a single stopping criterion $\beta$. In light of the experiments on the SMALL dataset, we retained $\beta = 10^{-3}$ as a good compromise for the performance of all considered algorithms.
\autoref{fig:ResNumClip} shows the behavior of the four methods as a function of the input degradation level. 
For all the considered datasets, all declipping methods provide significant SDR improvements (often more than $8$~dB) at (almost) all considered input SDRs. This remains the case even for relatively high input SDRs, with one exception: the Pop category, for which the Plain Cosparse brings some degradation at very high input SDR, and the overall improvement never exceeds $8$~dB. This may be due to the fact that most of the 100 unclipped excerpts in this category are mixes containing one or more tracks of dynamically compressed drums, and that at least 21 of them contain saturated guitar sounds. \ident{Such sound tracks are intrinsically saturated (on purpose) and it is worth emphasizing that, while the objective of the discussed declipping algorithms is to restore a clipped {\em mix}, it is not to restore, e.g., unsaturated guitar {\em sound tracks}. This being said, saturating a sound track tends to enrich its spectral content, so that the corresponding mix fails to satisfy the main regularization hypothesis\rev{---sparsity in the time-frequency plane---}allowing the algorithms to work well.}


The benefit of social (co)sparse modeling in terms of these objective measures is clear for high to moderate input SDR ($>10~\text{dB}$, mild clipping), and vice-versa, there is also a distinct superiority of the plain cosparse method for low input SDRs (strong clipping). These tendency will also be observed in listening tests at a 3~dB input SDR. Actually, the plain approaches perform 2 to 4~dB better than the adaptive social methods for input SDRs ranging from 1 to 5~dB on audio content from the RWC database. On the opposite, the trend tends to reverse above 10~dB input SDR as the social methods features improvements between 1 and 4~dB (even 7~dB for the Pop category) above the plain (co)sparse techniques. For speech content, the difference is less obvious yet \autoref{fig:SpeechDec}, \autoref{fig:STOITimitDec} and \autoref{fig:PESQTimitDec} displays better improvements either in terms of SDR, 
\ident{STOI or PESQ}
 for the plain sparse declipper.\\

Standard deviation results (showed on shaded colors) indicate that the social cosparse declipper produces less variable results across examples: we tested 54 configurations of sound categories and input SDRs, and this declipper produced the smallest standard deviation in 67~\% of these configurations. We also observed that, for any of the considered algorithms, the variability of the improvement seems to be higher for higher SDR. \\

The difference in performance between the plain and social cosparse declippers on music at low input SDR might come from the nature of the degradation. Indeed, \ident{contrary} to additive noise, \ident{clipping} 
adds broadband stripes in the time-frequency plane due to discontinuities of the derivative in the time domain. 
This way, the signal's underlying structure (embodied by a time-frequency pattern $\Gamma$) is not only hidden as it would be in the case of additive noise, but also possibly distorted: during the initialization loop of the adaptive social approaches, it is possible that a ``wrong'' pattern $\Gamma^{*}$ is selected. In contrast, the plain cosparse declipper cannot be affected by this type of behaviour. Another interesting result which could support this hypothesis is that for higher SDR, the social methods are actually benefiting from the time-frequency structure identification as the adaptive approach performs better.

\rev{
To ease the comparison of the listening tests that come next (conducted on 44.1 kHz audio material) with the PEAQ results of Figure~\ref{fig:ResObjClip} (conducted on 16 kHz material), we conduct additional experiments: for each channel of each 44.1 kHz stereo sample used in the listening test, we computed an objective performance using PEAQ, both on: a) the original 44.1 kHz channel and its clipped+declipped counterpart 44.1 kHz channel; and b) a 16 kHz version of the channel and its clipped+declipped (all at 16 kHz) counterpart. The average results 
show differences on the order of $0.01$ to $0.1$ on input values of PEAQ and $\Delta$PEAQ.
}
\subsection{\ident{Listening tests}} 
\label{subsec:Mushra}

This listening test is designed \ident{using the ITU-R BS.1534-3 recommendation MUSHRA method \cite{MushraITU},} to provide insights on global audio quality of several declipping methods. To avoid possible difficulties in rating speech quality with non-native speakers participants and as speech intelligibility listening tests are out of the scope of this study, we focus on music excerpts. We evaluate a subset of the conditions presented for the numerical experiments. For the test not to last longer than 30 minutes, we choose a single input SDR of 3~dB. For the sound material,   for each genre of the RWC database described in \autoref{subsec:Datasets}, 2 excerpts  of 5 seconds each were randomly picked. 
\ident{The same excerpts were presented to all participants.}

\begin{figure}[!t]
	\includegraphics[trim=0cm 9.6cm 1cm 11cm,clip,width=\columnwidth]{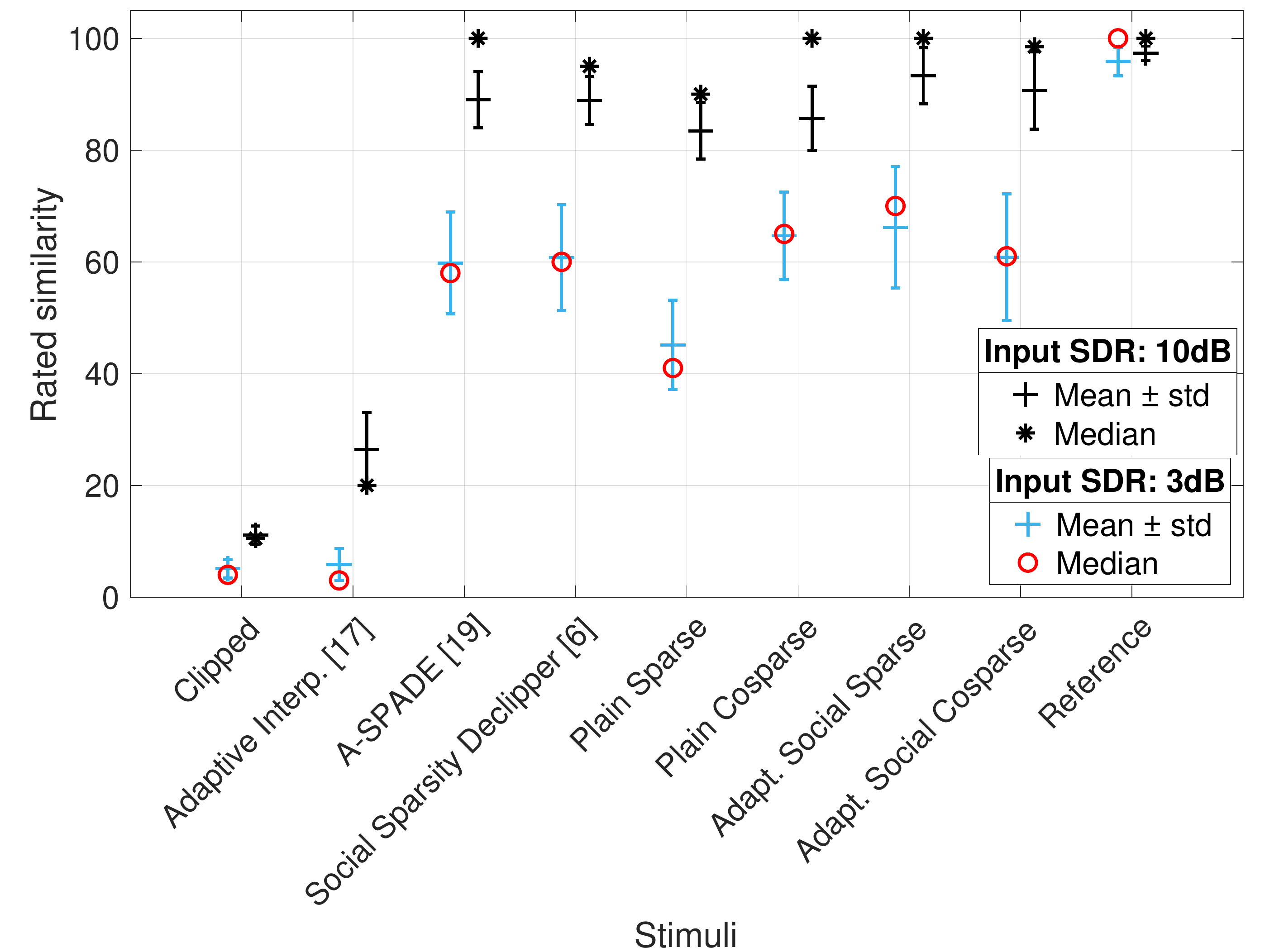}
	\caption{\ident{Audio rating via listening tests} 
	\label{fig:MUSHRA}}
\end{figure}

We use here the original stereo excerpts sampled at 44.1 kHz to preserve original quality. Each channel is first artificially saturated to reach \ident{an identical input SDR level on each channel. Two independant listening tests are performed featuring different initial degradation conditions: $3$~dB and $10$~dB of input SDR. 

Each channel of the tested items is} processed independently with the 6 best performing sparsity-based methods of \autoref{fig:SMALLMusicSDR} \ident{and two anchors}. The compared methods are here A-SPADE \cite{kitic2015sparsity}, Social sparsity declipper \cite{siedenburg2014audio}, and the 4 
\ident{variants} of the framework presented earlier: plain cosparse, plain sparse, adaptive social cosparse, adaptive social sparse. As a lower anchor, we use the algorithm of \cite{janssen1986adaptive} based on auto-regressive modeling and interpolation. As \autoref{fig:ITERDeclip} suggests that the four algorithms presented here stop far below $10^6$ iterations, we set here \ident{$\ing{i}_{\text{max}} = 1300$}, the other parameters match those used described earlier. 
Participants are asked to rate the similarity between the clean reference and the processed signals on a scale from 0 to 100. \ident{They remotely access the test through a web-browser based interface thanks to the web-MUSHRA framework \cite{schoeffler2018webmushra}. 26 participants aged between 18 and 48 years old volunteered for this listening study. 3 of them reported themselves as expert listeners for audio quality ratings.
23 volunteered for the $3$~dB input SDR condition and 14 volunteered for the $10$~dB input SDR testing condition.}
\autoref{fig:MUSHRA} displays results for this listening experiment.

\ident{For the $3$~dB input SDR condition,} 
despite the variability between participants, \ident{a repeated measures} analysis of variance (ANOVA) and t-paired tests \cite{cohen2013statistical} reveal that the average results for \ident{the lower anchor} \cite{janssen1986adaptive} and the \emph{plain sparse} algorithm are significantly different from \ident{(and lower than)} the 5 other methods (\emph{p-values} $\leq 9\cdot 10^{-8}$). Even if these results do not show statistically significant different means between \ident{A-SPADE} \cite{kitic2015sparsity}, \ident{the Social sparsity declipper} \cite{siedenburg2014audio}, adaptive social (co)sparse and plain cosparse algorithms, individual results shows the same ranking trend from one participant to another.

\ident{For the second testing condition ($10$~dB input SDR), results show very similar ratings between all sparsity-based methods and the reference. Only the lower anchor \cite{janssen1986adaptive} shows significant differences with the other methods (\emph{p-values} $\leq 8\cdot 10^{-8}$).
Results for this lighter input degradation condition also show comparable median results between the reference and methods using adaptive-social and/or analysis sparse models.
This suggests that it might be difficult to discriminate between the restored and reference signals for such input degradation model. Hence, further SDR improvements in this type of testing condition might not be so important when targeting perceived audio quality.}

\section{Conclusion}
Recent progress in audio declipping has made it possible to address severe degradation levels with unprecedented performance. The versatile algorithmic framework proposed here, which handles transparently both the analysis and synthesis 
\ident{variants} of sparse time-frequency models as well as the popular ``social'' time-frequency constraint, allowed us to conduct a systematic evaluation of many popular algorithms, in addition to a novel variant (adaptive social cosparse). The systematic coverage of algorithmic options and of a large ballpark of 
\ident{clipping thresholds}, evaluated on a wide range of data and performance measures, lead us to draw conclusions and guidelines for the practice of declipping as well as its performance evaluation.

The first message to be drawn is the importance of experimental methodology to truly assess declipping performance, which should consider: i) measuring the initial degradation and its improvement in terms of Signal-to-Distortion Ratio; ii) choosing a perceptually relevant range of degradation levels, more ambitious than the one used in early work on the topic, down to only a few decibels; iii) as in many other fields (if not all), gathering large test sets, covering a wide scope of musical genres and speech; iv) conducting listening tests.

In terms of algorithmic choices, we can summarize our observations as follows. The first expected result is the supremacy of sparsity-based methods (no matter the \rev{variant}) over the 
\ident{well-established autoregressive} interpolation method, in all \ident{clipping} 
regimes but the inaudible ones. For light \ident{clipping} 
regimes (Input SDR above 5~dB), recent sparse methods consistently bring improvements of 8~dB and more for various types of speech and music. Significant improvements, typically of a few decibels and up to 10~dB, are also obtained by these methods in severe clipping cases (input SDR from 0 to 5~dB). The only notable exception is pop music, possibly due to the presence of dynamically compressed drums and saturated guitar sounds. Similar trends are observed with \ident{perceptually-motivated} objective quality measures.


The plain and social 
\ident{variants} have comparable quantitative restoration performance, although the social constraint brings more benefit at mild \ident{clipping} 
regimes; plain versions could be preferable in more challenging conditions. This could be due to the difficulty to detect social sparsity patterns heavily corrupted by strong \ident{clipping}, 
and calls for further studies to guide the choice of such patterns, possibly based on learning techniques. However, from a perceptual point of view, adaptive social algorithms were slightly preferred by listeners, even at 3~dB. 

Improvements brought by the social constraint, however, comes with an increased computational cost. While cosparsity has been previously thought to be faster in the declipping task \cite{kitic2015sparsity}, better implementations and a fair tuning of stopping criteria suggest here that synthesis models are still on the go. Finally, at the moment, only plain versions show an immediate potential for real-time applications.

The algorithmic framework for audio restoration proposed in this paper is versatile from many points of view. In particular, it can handle several types of audio distortion models: demonstrated here on declipping, it can act as well on denoising and dereverberation problems \cite{gaultier2019design}. Further work could include extensions of the framework to these degradations, and multichannel scenarios.

\appendix[Generalized projection for declipping\label{app:ProxDeclipping}]

For the synthesis case, solving \eqref{eq:ConstrainedProjection}
with $\Theta = \Theta_{\textrm{time-freq}}(\mtrx{Y})$
%
%
%
%
and $\mtrx{M}=\mtrx{I}$
%
%
can be recast as:
\begin{equation}
\label{eq:synthprojmodif}
\hat{\mtrx{w}}=\argmin_{\mtrx{w}}\norm{\mtrx{w}-\mtrx{z}}{\text{F}}^2 \subjto \mtrx{D}\mtrx{w}\in\tilde{\Theta}
\end{equation}
\noindent and 
$$\tilde{\Theta} := \left\{ 
\mtrx{V}\ |
\begin{array}{l}
\mtrx{V}_{\Omega_r} = \mtrx{Y}_{\Omega_r};\\
\mtrx{V}_{\Omega_+} \succcurlyeq \mtrx{Y}_{\Omega_+}; \\ 
\mtrx{V}_{\Omega_-} \preccurlyeq \mtrx{Y}_{\Omega_-}.\end{array} \right\}.$$

\noindent As shown in \cite{Rajmic:2019dd,sorel2016efficient}, when $\mtrx{D}\htransp{\mtrx{D}} = \Id$ and $\tilde{\Theta}$ embodies a multidimensional interval constraint, the closed-form solution for~\eqref{eq:synthprojmodif} writes: 

\begin{equation}
\hat{\mtrx{w}} = \mtrx{Z} - \htransp{\mtrx{D}}(\mtrx{D}\mtrx{Z} - \Pi(\mtrx{Z})),
\end{equation}

\noindent with
$$[\Pi(\mtrx{Z})]_{\ing{nt}}~=~\left\{ 
\begin{array}{l l}
(\mtrx{D}\mtrx{Z})_{\ing{nt}} & \text{if~}\left\{\begin{array}{l}\ing{nt}\in\Omega_+,(\mtrx{D}\mtrx{Z})_{\ing{nt}} \geq \tau;\\ \text{or}\\\ing{nt}\in\Omega_-,(\mtrx{D}\mtrx{Z})_{\ing{nt}} \leq -\tau;\end{array}\right.\\
\mtrx{Y}_{\ing{nt}}
 & \text{otherwise.}\\
\end{array} \right.$$


%

%
%

\section*{Acknowledgment}

The authors thank Matthieu Kowalski for providing his implementation of the algorithm described in \cite{siedenburg2014audio}.

\ifCLASSOPTIONcaptionsoff
  \newpage
\fi



%

\Urlmuskip=0mu plus 1mu\relax
\bibliographystyle{IEEEtran}
\bibliography{bibliography}

%

%
%
%





\end{document}